\documentclass[12pt]{article}

\usepackage{graphicx}
\usepackage{tabularx}
\usepackage{caption2}
\usepackage{color}
\usepackage{slashed}
\usepackage{parskip}
\usepackage{latexsym,gensymb,amsmath,amssymb,amsthm}
\usepackage{url,hyperref}
\usepackage{times}

\usepackage[utf8]{inputenc}

\author{Aleksandra A. Drozd, University Of Warsaw}
\title{RGE and the Fine-Tuning Problem}
\date{MSc Thesis, Warsaw, August 2010}

\begin{document}
\maketitle

\renewcommand{\abstractname}{Abstract}
\begin{abstract}
In this work we study the fine-tuning problem in a general gauge theory with scalars and fermions. Then we apply our results to the Standard Model and its extension with additional singlet scalar field. The correlation between the Higgs mass and the scale at which new physics is expected to occur, is studied based on a fine-tuning arguments such as the Veltman condition. 
\end{abstract}

\newpage
\renewcommand{\contentsname}{Table Of Contents}
\tableofcontents


\newpage
\section*{Introduction}

The aim of this work is to investigate the fine-tuning problem. We will start with the RGE and quadratic divergences for a generic gauge theory and then apply our results to the Standard Model (SM) and the minimal SM extension. Then we will adopt the fine-tuning argument to estimate the range of allowed Higgs boson mass as a function of the UV cut-off $\Lambda$.\\

\section*{Renormalization group equations}

The idea of the renormalization group is based on the arbitrariness of renormalization prescription. 
Renormalization procedure is based on expressing the parameters of our theory with help of physical quantities obtained from experiments. Unfortunately, Quantum Field Theory does not properly describe physics at very short distances, which  results in divergences at almost every step of calculations at higher orders of perturbative expansion. To interpret such theory, one can introduce a procedure for regularization of divergences. There are very different renormalization and regularization schemes which give the same results, up to the specific order of perturbative calculations.

A particularly useful type of changing the renormalization prescription is changing the mass scale parameter $\mu$.
For example, the parameter could be the renormalization point at which we define the value of the 1PI Green's function. As a consequence of RGE, we have for a given physical theory, a definite values of coupling parameters as functions of the energy scale $\mu$. These are called running coupling constants and can be derived from specific differential equations (see section \ref{Calculating_beta_functions}).

Results of calulations of renormalization group functions are very useful and can be found in literature up to  several loops order.

\section*{Standard Model and its problems}
Physicists are able to describe the fundamental particles and their mutual interactions, with increasing accuracy. As for today, the Standard Model of particle physics is the best theory we have. It has passed almost all of the experimental challenges (except for neutrino oscillations) and is an excellent description of fundamental particles. It has been verified for example in LEP and SLC experiments.

But the SM also contains very important gaps and problems. The main issue is the very existence of Higgs boson. It is not proven yet, but there is a lot of hope towards experiments in Large Hadron Collider, Geneva. Higgs boson existence would explain a fundamental problem of masses. Higgs mechanism, which is based on generating masses through a non-zero vacuum expectation value of a specific field, is a very simple and beautiful way of obtaining massive vector particles through symmetry breaking. There exists lots of variations to this idea, but the beauty of this basic concept challenges many scientists to look for a Higgs or Higgs-like particle in experiments.

Higgs mass $m_H$ is the most commonly pointed out unknown parameter of Standard Model, but not the only one. If we assume that Standard Model is only a low-energy limit of a more fundamental theory (which does not necessarily have to be a quantum field theory) that could for example explain why the electroweak symmetry is broken.

Other problems with the SM are the combined issues of fine-tuning and naturalness. In theoretical physics, fine-tuning refers to circumstances when the parameters of a model must be adjusted very precisely in order to agree with observations~\footnote{There are some discussions in the literature over the definition of fine-tuning and the degrees of precision in adjustments of parameters. The definition of fine-tuning adopted here will be specified later}. The requirement of a fine-tuning in a theory is generally unwelcome by physicists, permissible with a presence of a mechanism to explain the precisely needed values. 
A so called, \textit{little hierarchy problem} is a problem of fine-tuning of the Higgs boson mass corrections. For the SM energy scale much larger than the W boson mass, $\Lambda \gg m_{W}$, corrections to the Higgs mass should cancel each other to a very high precision in order for the mass to be in order of electroweak scale.

\section*{A simple extension of Standard Model}

Standard Model is known to be a good approximation of fundamental interactions, but there are many attempts to extend this theory and get rid of the aforementioned problems.

The very simplest extension of Standard Model is an addition of singlet scalar particle. Assuming interactions of $N_{\phi}$ singlet scalar $\phi_{n}$ particles and Higgs, a potential with a discrete $Z_{2}$ symmetry $\phi \rightarrow - \phi$ can be introduced:
\begin{eqnarray}
V(H, \phi_{n}) = -\mu^2 H ^\dagger H + \lambda (H ^\dagger H)^2 + \sum _{i} ^{N_{\phi}} \frac{\mu ^{i}_{\phi}}{2}   \phi _ i ^2 + 
\sum _{i,j}^{N_{\phi}} \lambda_{\phi}^{ij} \phi ^2 _{i} \phi ^2 _{j} +
\sum _{i}^{N_{\phi}} \lambda_{x}^{i} (H ^\dagger H) \phi _{i} ^ 2 
\label{scalar_potential}
\end{eqnarray}
If $N_{\phi} = 1$ then this extension leaves us with three additional parameters $\lambda_{x}$, $\lambda_{\phi}$ and the additional particle mass. \\

In this work we will discuss theoretical constraints on $m_H$ and $\Lambda$ due to the fine-tuning argument. The letter is organized as follows.

The first chapter is about 1-loop renormalization of a general gauge theory with fermions and scalars. We will use the dimensional regularization scheme and calculate the RGE beta functions of such theory. In the second chapter we will concentrate on 1-loop corrections in cut-off regularization scheme to the general gauge theory. Third chapter is to present higher order corrections of the perturbative expansion using previously obtained results. In fourth chapter we will concentrate on the Higgs mass corrections and estimation of this parameter using the 'Veltman condition' and the 2-loop fine-tuning. The fifth chapter presents results in a presence of an additional singlet scalar field in the model.

\section{Derivation of beta functions in a generic gauge
   theory with fermions and scalars}
\subsection{Lagrangian and the counterterms}\label{general_theory}
 
We will start our calculation with analysing the most general case: a gauge invariant Lagrangian of a theory with a number of real scalar fields $ \phi _{i}, i = 1,..., N_{\phi} $ and spin-$\frac{1}{2}$ fields $ \psi _{n} , n = 1,..., N_{\psi}$, with a single gauge symmetry and corresponding hermitian gauge fields $ A^{a}_{\mu} $. We adopt the $R_\xi$ gauge and $\eta_{a}$ stands for the ghost fields. Everywhere summation over repeated indices is assumed. 
\begin{eqnarray}
\mathcal{L} & = & - \frac{1}{4} F^{a}_{\mu \nu} F^{a \mu \nu} 
- \frac{1}{2\xi} \left( \partial_{\mu} A^{\mu}_{a} \right)^{2}
+ \frac{1}{2} (D_{\mu} \phi)_{i}(D^{\mu} \phi)^{i} + 
\overline{\psi} ( \imath \gamma^{\mu} D_{\mu} - M ) \psi - \nonumber\\
& & + \partial _{\mu} \eta^{*}_{a}  \left( \delta_{a b} \partial ^{\mu} \eta_{b} + g f_{a b c} \eta_{b} A_{c}^{\mu} \right) 
-  \overline{\psi} \kappa^{i} \psi \phi_{i} - V(\phi)
\label{lagrangian_general}
\end{eqnarray}
where
\begin{equation}
F^{a}_{\mu \nu} = \partial_{\mu} A^{a}_{\nu} - \partial_{\nu} A^{a}_{\mu} - g f_{a b c} A^{b}_{\mu} A^{c}_{\nu}
\end{equation} 
The covariant derivative of a field $ \phi $ and $\psi$ can be written as
\begin{equation}
D_{\mu}\phi = \left( \partial_{\mu}+ \dot{\imath} g \textbf{T}^{a} A^{a}_{\mu} \right) \phi
\end{equation}
\begin{equation}
D_{\mu}\psi = \left( \partial_{\mu}+ \dot{\imath} g \overline{\textbf{T}}^{a} A^{a}_{\mu} \right) \psi
\end{equation}
where $ g $ is the group constant. In general, scalar and fermion fields can transform under different representations of the gauge group, so there are two different sets of generators $\textbf{T}^{a}$ and $\overline{\textbf{T}}^{b}$ for scalar and fermion fields respectively. For each $\phi _{i}$ scalar field and $\psi _{n}$ fermion field one can write the covariant derivative in form of:
\begin{eqnarray}
& & (D_{\mu}\phi ) _{i} = \partial_{\mu} \phi _{i} + \dot{\imath} g \textbf{T}^{a}_{i j} A^{a}_{\mu}  \phi _{j} \\
& & (D_{\mu}\psi )_{n} =  \partial_{\mu} \psi _{n} + \dot{\imath} g \overline{\textbf{T}}^{a}_{n m} A^{a}_{\mu} \psi_{m}
\end{eqnarray}

Generators fulfil the following relations:
\begin{eqnarray}
& & [ \textbf{T}^{a}, \textbf{T}^{b} ] =  i f_{a b c} \textbf{T}^{c} \\
& & C_{1} \delta_{a b} = f_{acd} f_{cdb}
\label{C1}\\
& & \textbf{T}^{a}_{i j} \textbf{T}^{b}_{j i} = \textbf{Tr} (\textbf{T}^a \textbf{T}^b) = C_{2} (R) \delta_{a b}  \label{C2} \\
& & (\textbf{T}^{a} \textbf{T}^{a})_{m n} = C_{3} \delta_{m n}
\label{C3}
\end{eqnarray} 
where $f_{a b c}$ are the structure constants, group factors $C_{1}$ and $C_{3}$ depends only on the group we consider, while $C_{2}$ depends on specific representation $R$. All above equations can be simply written in terms of $\overline{\textbf{T}}^{a}$ generators. \\
There are also some constraints on the couplings and generators which result from the hermiticity and gauge invariance of the Lagrangian, some of them will be discussed later. \\

The potential $V(\phi)$ to consider is no more than quartic in $\phi$. We omit cubic and quadratic terms as they are not relevant hereafter.
\begin{equation}
V(\phi) = \frac{1}{4!} h_{i j k l} \ \phi_{i} \phi_{j} \phi_{k} \phi_{l} + \textit{lower-order terms}
\end{equation} 
We will only consider real scalar fields case. For complex scalars it is always possible to rewrite the Lagrangian in terms of real degrees of freedom and re-evaluate the result.

To proceed with the renormalization we write for the bare Lagrangian
\begin{equation}
\mathcal{L}_{B} = \mathcal{L} + \Delta \mathcal{L}
\end{equation} 
where the $ \Delta \mathcal{L} $ is for the counter terms. We assume the form of the bare Lagrangian to be the same as in the renormalised Lagrangian but with the bare fields (like $\psi _{B}$ or $\phi _{B}$ ) and coupling constants (like $g_{B}$) replaced by the corresponding renormalised quantities. \\

We will assume such relationships between bare and renormalised fields:
\begin{eqnarray}
(A^{a}_{\mu})_{B} & = & \left( 1+ \Delta Z_{A} \right) ^{1/2} A^{a}_{\mu} = Z^{1/2}_{A} A^{a}_{\mu} \label{niediagonalne_wektory}\\
(\eta^{a})_{B} & = & \left( 1+ \Delta Z_{\eta} \right)^{1/2} \eta^{a} = Z^{1/2}_{\eta} \eta^{a} \label{niediagonalne_duchy}\\
(\psi _{n})_{B} & = & \left( (1+ \Delta Z _{\psi})^{1/2}  \right) _{n m} \psi _{m} =
\left( Z^{1/2}_{\psi} \right) _{n m} \psi _{m} \label{niediagonalne_fermiony}\\
(\phi _{i})_{B} & = & \left((1+ \Delta Z_{\phi})^{1/2}\right)_{i j} \phi_{j}  =
\left( Z^{1/2}_{\phi} \right) _{i j} \phi _{j} \label{niediagonalne_scalary}
\end{eqnarray}

Because Yukawa couplings in our considerations are hermitian, renormalization constant $Z_{\psi}$ is generally a complex matrix. $Z_{\phi}$ and $Z_{A}$ must be real for real scalar and real vector fields.

Even at the 1-loop order renormalization, one needs to consider that there can be non-diagonal corrections to the propagators (see \cite{bouzas}). It was done in (\ref{niediagonalne_fermiony}) and (\ref{niediagonalne_scalary}). In the case of the gauge field, as one can see in later discussions, it happens that 1-loop corrections are purely diagonal and we assumed this in (\ref{niediagonalne_wektory}). We will not discuss later the corrections for the ghost field, but they are diagonal too, as in (\ref{niediagonalne_duchy}).

Now we can write the counter terms for the Lagrangian (below only terms important for our calculations):
\begin{eqnarray}
 \Delta \mathcal{L} & = & 
- \frac{1}{4} \Delta Z_{A} 
(\partial_{\mu} A_{a \nu} - \partial_{\nu} A_{a \mu}) (\partial^{\mu} A_{a}^{\nu} - \partial^{\nu} A_{a}^{\mu}) 
- \frac{K_{\xi}}{2\xi} \left( \partial_{\mu} A^{\mu}_{a} \right)^{2} 
\nonumber\\
& & 
+   \overline{\psi} \Delta Z_{\psi} \ i \gamma^{\mu} \partial_{\mu} \psi
-  \overline{\psi} \gamma^{\mu} \Delta g \overline{\textbf{T}}^{a}  \psi A^{a}_{\mu}
-  \overline{\psi} \Delta \kappa _{i}  \psi  \phi_{i}
+ \frac{1}{2} (\partial_{\mu} \phi) \Delta Z_{\phi} (\partial^{\mu} \phi) 
\nonumber\\
& & 
+ \Delta Z_{\eta} \partial _{\mu} \eta^{*}_{a} \partial ^{\mu} \eta_{a}
- \frac{1}{4!} \Delta h_{i j k l} \ \phi_{i} \phi_{j} \phi_{k} \phi_{l} + \ldots
\label{counter}
\end{eqnarray}

$\Delta g$, $\Delta \kappa_{i}$ and $\Delta h_{i j k l}$ will be specified further while $K_{\xi}$ is defined by the following relation:
\begin{eqnarray}
(\xi)_{B}^{-1} & = & \left( 1 + K_{\xi} \right) Z^{-1}_{A} \xi ^{-1} \label{Zetxi}
\end{eqnarray}

Bare coupling constants dependence on the renormalized quantities and the renormalization constants in general have a complicated form, because of the previously mentioned non-diagonality of the corrections to the propagator. There are different expressions for $g_{B}$, depending on the vertex we consider, and they result in some relationships between renormalization constants. In later discussion we will consider only the $\overline{\psi} \psi A_{\mu} $ vertex to calculate the beta function of $g$ coupling. 

The formula for the bare coupling in terms of renormalized quantities is:
\begin{eqnarray}
g_{B} \overline{T}_{nm} ^{a} & = & (Z^{-1/2}_{\psi})^{\dagger}_{n n'} 
\left( (g + \Delta g)\overline{T}^{a} \right)_{n' m'}
(Z^{-1/2}_{\psi})_{m' m}Z^{-1/2}_{A}
\label{counter_g}
\end{eqnarray}
If we expand the formula using (\ref{niediagonalne_wektory}) and (\ref{niediagonalne_fermiony}) we can get a relation as follows
\begin{eqnarray}
g_{B} \overline{T}_{nm} ^{a} &=& g \overline{T}_{n m} ^{a}
- \frac{1}{2} (\Delta Z^{\dagger}_{\psi})_{n n'} g \overline{T}_{n' m} ^{a}
- \frac{1}{2} g \overline{T}_{n n'} ^{a} (\Delta Z _{\psi})_{n' m} \nonumber\\
& & - \frac{1}{2} g \overline{T}_{n m} ^{a} \Delta Z_{A}
+ \left( \Delta g \overline{T}^{a} \right)_{n m} + \ldots \label{gB_all}
\end{eqnarray}

As one can see, the general and complete relation between the bare and renormalized coupling is complicated - it includes not only the non-diagonal propagator corrections, but also the group generators. In section \ref{renormalization_of_ffb} we show that after specific calculations all the non-diagonal contributions cancel each other at the 1-loop accuracy.
We can use this fact to simplify our result.

We define $\Delta Z_{\psi}$ as the non-cancelling part of the $\left( \Delta Z_{\psi} \right)_{m n}$, which (as we can see in section \ref{renormalization_of_ffb}) happens to be a number multiplying the generator $\overline{T}^{a}$.
Similarly, $\Delta \tilde{g} \, \overline{T}^{a}$ is the non-cancelling diagonal part of $\left( \Delta g \overline{T}^{a} \right)_{n m}$ from $\overline{\psi} \psi A_{\mu}$ vertex  renormalization. And while the non-diagonal contributions cancel, the simplified equation for $g_{B}$ takes the form
\begin{eqnarray}
g_{B} = g - \frac{1}{2} g \, \Delta Z_{A} - g \, \Delta Z_{\psi}
+  \Delta \tilde{g} + \ldots \label{easy_gB}
\end{eqnarray}
where $\Delta \tilde{g} = K_2 \, g$ with the renormalization constant $K_2$. For the beta function $\beta(g)$ calculations in section \ref{Calculating_beta_functions} we will use the simplified formula.

For the Yukawa coupling constant and quadrilinear couplings, non-diagonal terms are present in the calculations and don't vanish.
\begin{eqnarray}    
(\kappa^{i} _{n m})_{B} &=& 
\sum _{i' n' m'} 
\left(  (Z^{\dagger}_{\psi})^{-1/2} \right)_{n n'}
\left(  Z^{-1/2}_{\psi} \right)_{m' m}
\left(  Z^{-1/2}_{\phi} \right)_{i' i} (\kappa^{i'}_{n' m'} + \Delta \kappa^{i'}_{n' m'})\label{ZetKappa1}
\\
(h_{i j k l})_{B} &=& 
\sum _{i' j' k' l'} 
\left(  Z^{-1/2}_{\phi} \right)_{i i'}
\left(  Z^{-1/2}_{\phi} \right)_{j j'}
\left(  Z^{-1/2}_{\phi} \right)_{k k'}
\left(  Z^{-1/2}_{\phi} \right)_{l l'}
(h_{i' j' k' l'} + \Delta h_{i' j' k' l'}) \label{Zeth} \nonumber\\
\end{eqnarray}
where we can express the $\Delta \kappa^{i} _{n m}$ and $\Delta h_{i j k l}$ as follows:
\begin{eqnarray}    
\Delta \kappa^{i} _{n m} = \sum _{i' j' k' l'} K_{i n m}^{i' n' m'} \kappa^{i'} _{n' m'} 
\label{ZetKappa} \\
\Delta h_{i j k l} = \sum _{i' j' k' l'}
L_{i j k l}^{i' j' k' l'} h_{i' j' k' l'} 
\label{ZetH}
\end{eqnarray}

\subsection{Renormalization of propagators}

We will regularise the divergent integrals adopting the dimensional regularization (we set the number of dimensions to be $d = 4 - \epsilon $). Feynmann diagrams for a general gauge theory can be found in the Appendix A and remarks on calculating the symmetry factors can be found in \cite{palmer}. At every step of the calculation we mention only the diagrams that contribute in dimensional regularization.

\subsubsection{Two point function for a gauge field}

\begin{figure}
  \centering
  \includegraphics[height = 3.5 cm]{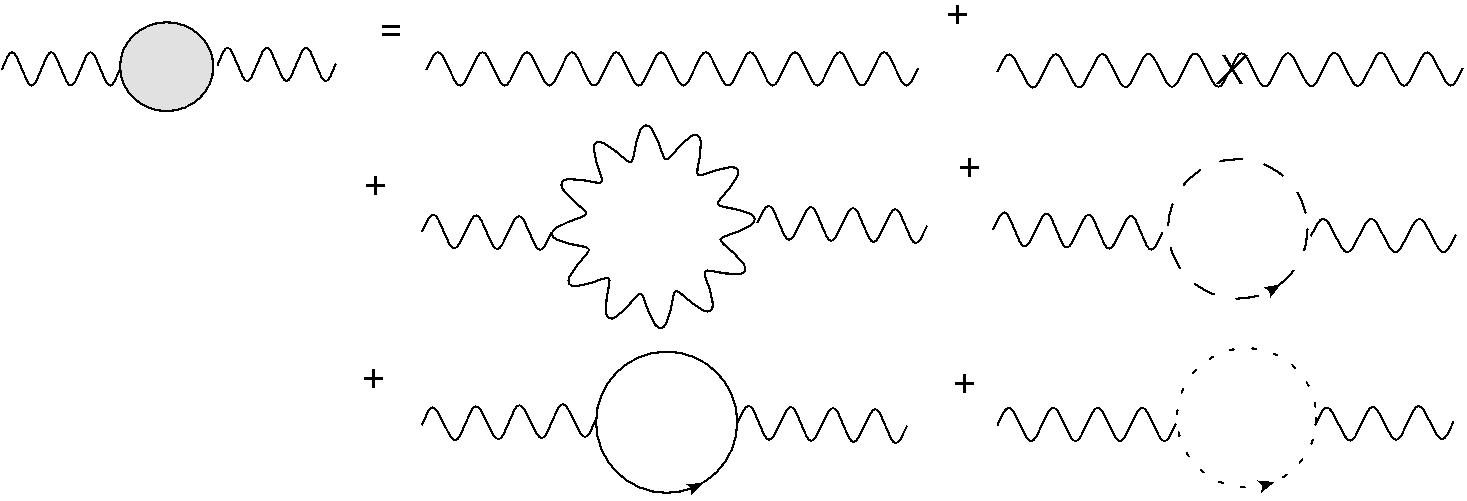}
  \caption{OPI self energy corrections to the vector boson propagator that contribute in dimensional regularization}
  \label{OPI_gauge_propagator}
\end{figure}

One can draw the one particle irreducible (OPI) diagrams contributing in the dimensional regularization to the gauge boson propagator as in fig. \ref{OPI_gauge_propagator}.\\
\renewcommand{\figurename}{Diagram}
\begin{figure}
  \centering
  \includegraphics[height = 2 cm]{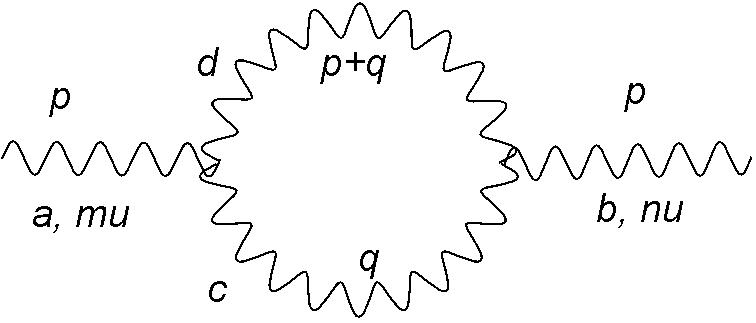} 
  \caption{}
\label{2point_boson_1}
\end{figure}
We will do all the calculations step by step starting with the gauge fields loop (fig. \ref{2point_boson_1}). Here  the symmetry factor is $\frac{1}{2}$, so the boson self-energy contribution takes the form:
\begin{equation}
(\texttt{diagram \ref{2point_boson_1}}) = \frac{1}{2} \mu^{\epsilon} g^{2} f_{acd} f_{bcd} \int \frac{d^{d} q}{(2\pi)^{d}} 
\tilde{D}_{F}^{\sigma \tau}(p+q) \tilde{D}^{\lambda \rho}_{F}(q) J_{\sigma \mu \rho \tau \lambda\nu} (p,q) \label{2point_boson_integral}
\end{equation}
where $\tilde{D}^{\lambda \rho}_{F}$ is a gauge boson propagator and
\begin{eqnarray}
J_{\sigma \mu \rho \tau \lambda \nu}  (p,q) & = & [(p-q)_{\sigma} g_{\mu \rho} + (2q+p)_{\mu} g_{\rho \sigma} -(2p + q)_{\rho} g_{\mu} \sigma] 
\times \nonumber\\
& & [(p-q)_{\tau} g_{\lambda \nu} - (2p+q)_{\lambda} g_{\tau \nu} + (2q + p)_{\nu} g_{\tau \lambda}] 
\end{eqnarray}
From (\ref{2point_boson_integral}) after some calculations one can get the divergent term.
\begin{equation}
(\texttt{diagram \ref{2point_boson_1}})  = \frac{i g^2 C_{1} \delta_{ab}}{16 \pi^2 \epsilon} \left[
 \left( -\frac{11}{3} - 2\eta \right) p_{\mu} p_{\nu} + \left( \frac{19}{6} + \eta \right) p^2 g_{\mu \nu} 
 \right]
\end{equation}
where the group theory factor is defined in (\ref{C1}) and $\eta = 1 - \xi $.

\begin{figure}
  \centering
  \includegraphics[height = 2 cm]{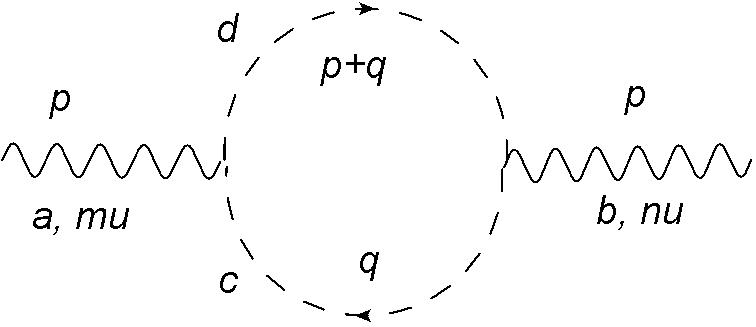} 
  \caption{}
  \label{2point_boson_2}
\end{figure}

For calculating the ghost fields loop, the symmetry factor is 1 and there is a minus sign because of the closed loop of Grassmann fields. \\
\begin{equation}
(\texttt{diagram \ref{2point_boson_2}}) = (-1) g^{2} f_{dca} f_{cdb} \int \frac{d^{d} q}{(2\pi)^{d}} 
(p+q)_{\mu} q_{\nu} \tilde{G}_{F} (p+q) \tilde{G}_{F}(q)
\end{equation}
where $\tilde{G}_{F}$ is a ghost propagator. Using (\ref{C1}) the contribution from fig. \ref{2point_boson_2} takes a form
\begin{equation}
(\texttt{diagram \ref{2point_boson_2}}) = - g^{2} C_{1} \delta_{a b} \int \frac{d^{d} q}{(2\pi)^{d}} 
\frac{(p+q)_{\mu} q_{\nu}}{(p+q)^2 q^2}
\end{equation}
with the final result of
\begin{equation}
(\texttt{diagram \ref{2point_boson_2}}) = \frac{i g^2 C_{1} \delta_{ab}}{16 \pi^2 \epsilon} \left(
 \frac{1}{3} p_{\mu} p_{\nu} + \frac{1}{6} p^2 g_{\mu \nu} 
 \right)
\end{equation}\\

\begin{figure}
  \centering
  \includegraphics[height = 2 cm]{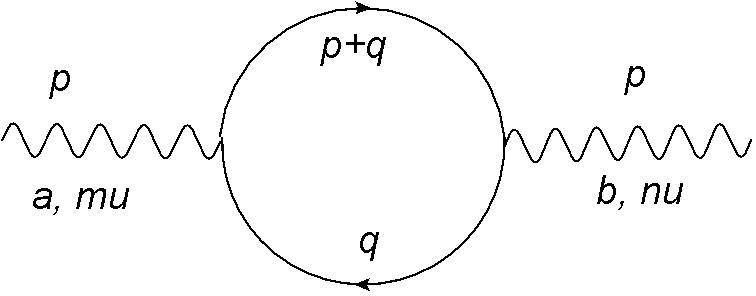} 
  \caption{}
\label{2point_boson_3}
\end{figure}

For calculating the contribution from fermion fields, one has as before loop symmetry factor 1 and a minus sign for the closed loop of Grassmann fields. \\
\begin{equation}
(\texttt{diagram \ref{2point_boson_3}}) = - \mu^{\epsilon} g^{2} \overline{C}_{2} \delta_{a b} \int \frac{d^{d} q}{(2\pi)^{d}} 
\textbf{Tr} \left( 
\gamma_{\mu} \tilde{S}_{F}(q) \gamma_{\nu} \tilde{S}_{F} (p+q)
\right)
\end{equation}
where (\ref{C2}) was adopted and $\tilde{S}_{F}$ denotes the fermion propagator. To extract the pole term the easiest way, one can put fermion masses to zero and then obtain
\begin{equation}
(\texttt{diagram \ref{2point_boson_3}}) = - \frac{i g^2 \overline{C}_{2} \delta_{ab}}{16 \pi^2 \epsilon}
 \frac{8}{3} \left(
- p_{\mu} p_{\nu} + p^2 g_{\mu \nu} 
 \right)
\end{equation}
\begin{figure}
  \centering
  \includegraphics[height = 2 cm]{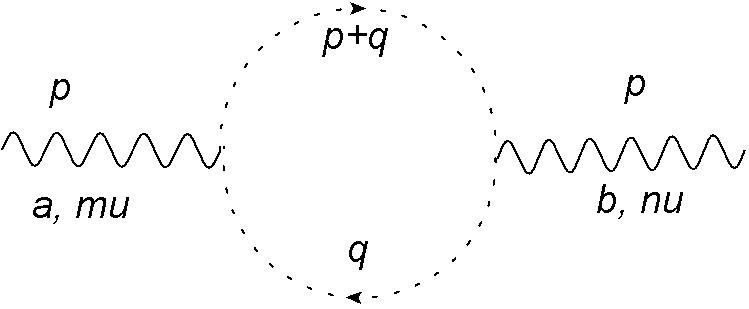} 
  \caption{}
\label{2point_boson_4}
\end{figure}
Calculating the scalar fields loop include symmetry factor $\frac{1}{2}$.
\begin{equation}
(\texttt{diagram \ref{2point_boson_4}}) = - \frac{1}{2} g^{2} \textbf{T}^{a}_{i j} \textbf{T}^{b}_{j i} \int \frac{d^{d} q}{(2\pi)^{d}} 
(2q+p)_{\mu} (2q+p)_{\nu} \tilde{D}_{F}(q)\tilde{D}_{F}(p+q)
\end{equation}
Using the group theory factor from (\ref{C2}) and simplifying, one can get the formula
\begin{equation}
(\texttt{diagram \ref{2point_boson_4}}) =  \frac{1}{2} g^{2} C_{2} \delta_{ab} \int \frac{d^{d} q}{(2\pi)^{d}} 
\frac{(2q+p)_{\mu} (2q+p)_{\nu}}{(p+q)^2 q^2}
\end{equation}
and a final result of:
\begin{equation}
(\texttt{diagram \ref{2point_boson_4}}) = - \frac{i g^2 C_{2} \delta_{ab}}{16 \pi^2 \epsilon}
\frac{1}{3}
 \left(
-p_{\mu} p_{\nu} + p^2 g_{\mu \nu} 
 \right)
\end{equation}

From those results we can calculate the $\Delta Z_{A}$ and $K_{\xi}$ renormalization constants:
\begin{eqnarray}
&& \dot{\imath}\Delta Z_{A} (-p^2 g_{\mu \nu}  +  p_{\mu} p_{\nu}) - \dot{\imath}  \frac{1}{\xi} K_{\xi} p_{\mu} p_{\nu} =
\frac{- \dot{\imath}  g^2}{16 \pi^2 \epsilon}  \times \nonumber\\ && \times 
 \left[(- p^2 g_{\mu \nu} + p_{\mu} p_{\nu})\left[\frac{8}{3}\overline{C}_{2}+(-\frac{10}{3} - \eta) C_{1} + 
\frac{1}{3} C_{2} \right] -(\eta C_{1} + 4 C_{2} ) p_{\mu} p_{\nu} \right] \nonumber
\end{eqnarray}
\begin{eqnarray}
\Delta Z_{A} 
  &=&  \frac{- g^2 }{16 \pi^2 \epsilon} 
\left[(-\frac{10}{3} - \eta) C_{1} +\frac{8}{3}\overline{C}_{2} + 
\frac{1}{3} C_{2} \right] 
\label{ZA} \\
K_{\xi} &=& - \frac{\xi g^2}{16 \pi^2 \epsilon} \left( \eta C_{1} + 4 C_{2} \right)
\end{eqnarray}

\subsubsection{Two point function for a fermion field}
\renewcommand{\figurename}{Figure}
\begin{figure}
  \centering
  \includegraphics[height = 2cm]{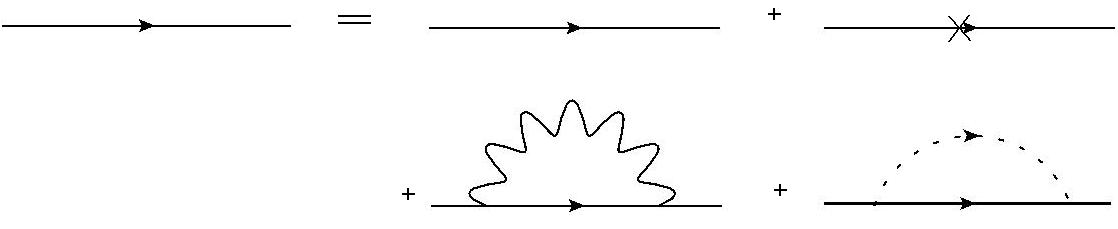}
  \caption{OPI self energy corrections to the fermion propagator that contribute in dimensional regularization}
  \label{OPI_fermion_propagator}
\end{figure}

Only two diagrams contribute to the renormalised propagator at the 1-loop accuracy. They are shown in fig.\ref{OPI_fermion_propagator}.
\renewcommand{\figurename}{Diagram}
\begin{figure}
  \centering
  \includegraphics[height = 2 cm]{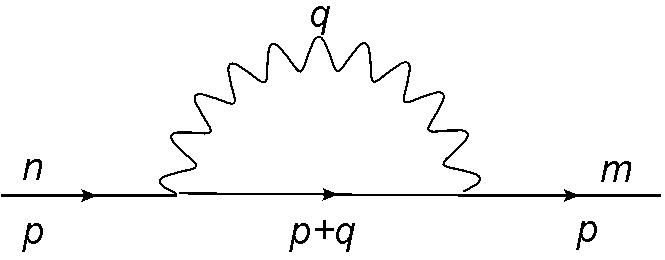} 
  \caption{}
\label{2point_fermion_1}
\end{figure}

\begin{equation}
(\texttt{diagram \ref{2point_fermion_1}}) = - g^{2}  \overline{C}_{3} \delta_{m n} \int \frac{d^{d} q}{(2\pi)^{d}} 
\left( \gamma_{\mu}  \tilde{S}_{F}(p+q) \gamma_{\nu} \right)_{ \beta \alpha} \tilde{D}_{F}^{\mu \nu}(q)
\end{equation}
To calculate the pole term in (diagram \ref{2point_fermion_1}) we put $m = 0$ in denominator and use the following identities:
\begin{eqnarray}
\gamma^{\mu}  \gamma _{\mu} = d \textbf{I}   \\
\gamma^{\mu} \gamma _{\rho} \gamma_{\mu} = (2-d) \gamma_{\rho}
\end{eqnarray}
And after some simple calculations one can obtain the result of:
\begin{equation}
(\texttt{diagram \ref{2point_fermion_1}}) = \frac{2 i g^2  \overline{C}_{3} \delta_{m n}(1 - \eta) }{16 \pi^2 \epsilon} 
\left( \slashed{p} \right) _{\beta \alpha}
\end{equation}

\begin{figure}
  \centering
  \includegraphics[height = 2.5cm]{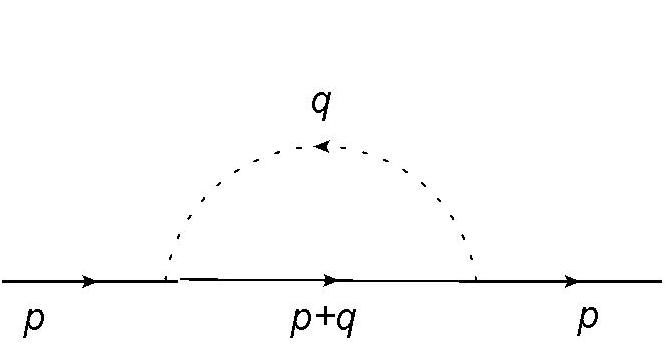} 
  \caption{}
\label{2point_fermion_2}
\end{figure}

For the scalar contribution we have no additional factors, so the pole term can be calculated from:
\begin{equation}
(\texttt{diagram \ref{2point_fermion_2}}) = - \kappa^{i}_{m n'} \kappa^{i}_{n' n} \int \frac{d^{d} q}{(2\pi)^{d}} 
\left(  \tilde{S}_{F}(p+q) \right)_{ \beta \alpha} \tilde{D}_{F}(q)
\end{equation}
with the result of:
\begin{equation}
(\texttt{diagram \ref{2point_fermion_2}}) = \frac{i (\kappa^{i} \kappa^{i})_{m n}}{16 \pi^2 \epsilon} 
\left( \slashed{p}  \right)_{\beta \alpha}
\end{equation}

Hence, using the Feynman rules for the counterterms from the Appendix \ref{fermion_propagator_counter}, one can evaluate the fermion propagator counterterm:
\begin{eqnarray}
\frac{1}{2} \left( \Delta Z_{\Psi}^{\dagger} + \Delta Z_{\Psi}\right)_{m n} =
- \frac{2  g^2  \overline{C}_{3} \delta_{m n}(1 - \eta) }{16 \pi^2 \epsilon}  
- \frac{ (\kappa ^{i} \kappa ^{i})_{m n}}{16 \pi^2 \epsilon} 
\label{Zpsi}
\end{eqnarray}

\subsubsection{Two point function for a scalar field}

\renewcommand{\figurename}{Figure}
\begin{figure}
  \centering
  \includegraphics[height = 2.5 cm]{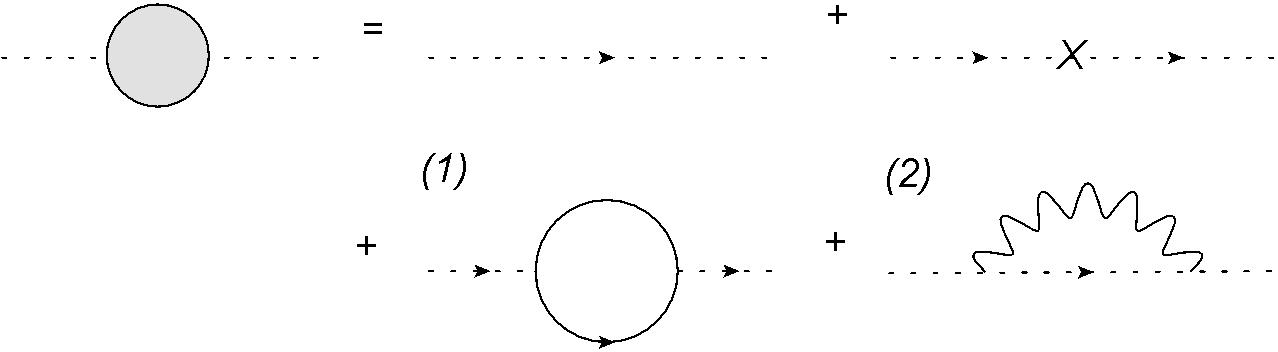}
  \caption{OPI self energy corrections to the scalar field that contribute in dimensional regularization}
  \label{OPI_scalar_propagator}
\end{figure}

\renewcommand{\figurename}{Diagram}
\begin{figure}
  \centering
  \includegraphics[height = 2 cm]{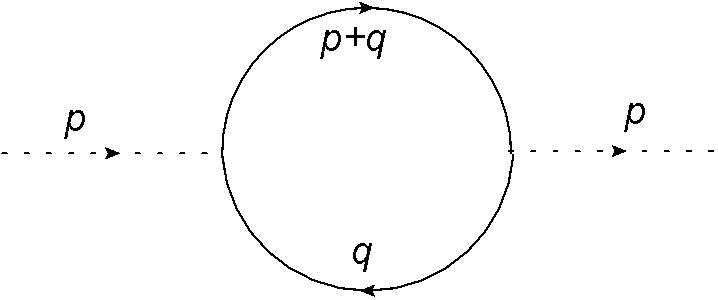} 
  \caption{}
\label{2point_scalar_1}
\end{figure}

In diagram \ref{2point_scalar_1} one has to include a $(-1)$ factor from the fermion closed loop.
\begin{equation}
(\texttt{diagram \ref{2point_scalar_1}}) = \textbf{Tr} \left( \kappa^{i} \kappa^{j} \right) \int \frac{d^d q}{(2 \pi)^d} \textbf{Tr} \left( 
\tilde{S}_{F}(p+q) \tilde{S}_{F}(q)
\right)
\end{equation}
The important term for beta function calculations is the pole term proportional to the $p^2$, so by putting $m$ equal to zero, one can evaluate the pole term simpler and get the result of:
\begin{equation}
(\texttt{diagram \ref{2point_scalar_1}}) = \frac{4 i p^2 \, \textbf{Tr} \left( \kappa^{i} \kappa^{j} \right)}{16 \pi ^2 \epsilon} 
\end{equation}

For the gauge boson contribution one gets:
\begin{figure}
  \centering
  \includegraphics[height = 2 cm]{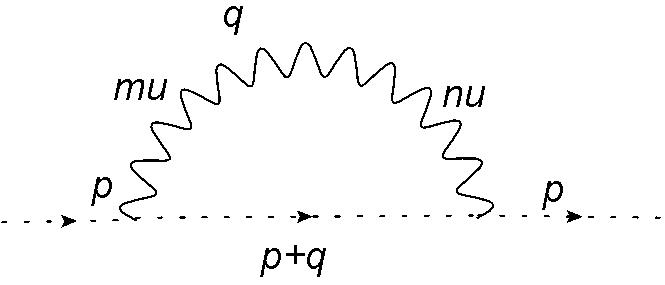}
  \caption{}
\label{2point_scalar_3}
\end{figure}
\begin{equation}
(\texttt{diagram \ref{2point_scalar_3}}) = - g^2 \textbf{T}^a_{ii'} \textbf{T}^a_{i'j}  \int \frac{d^d q}{(2 \pi)^d} 
\left( 2p+q \right)_{\mu} \left( 2p+q \right) _{\nu}
\tilde{D}_{F}(p+q) \tilde{D}^{\mu \nu}_{F}(q)
\end{equation}
Using (\ref{C3}) and simplifying one can get the result of:
\begin{equation}
(\texttt{diagram \ref{2point_scalar_3}}) = - \frac{\dot{\imath} g^2 p^2}{16 \pi^2 \epsilon}  \left( 4  + 2 \eta \right) C_{3} \delta_{ij}
\end{equation}

Using those results one can calculate the $\Delta Z_{\phi}$ renormalization constants.
\begin{eqnarray}
\left( \Delta Z_{\phi} \right) _{i j} = 
- \frac{4 \textbf{Tr} \left( \kappa^{i} \kappa^{j} \right)}{16 \pi ^2 \epsilon}  +
\frac{ g^2 }{16 \pi^2 \epsilon} \left( 4 + 2 \eta \right) C_{3} \delta_{i j}
\label{Zphi}
\end{eqnarray}

\subsection{Renormalization of fermion-fermion-vector boson coupling} \label{renormalization_of_ffb}

\renewcommand{\figurename}{Figure}
\begin{figure}
  \centering
  \includegraphics[height = 4.5 cm]{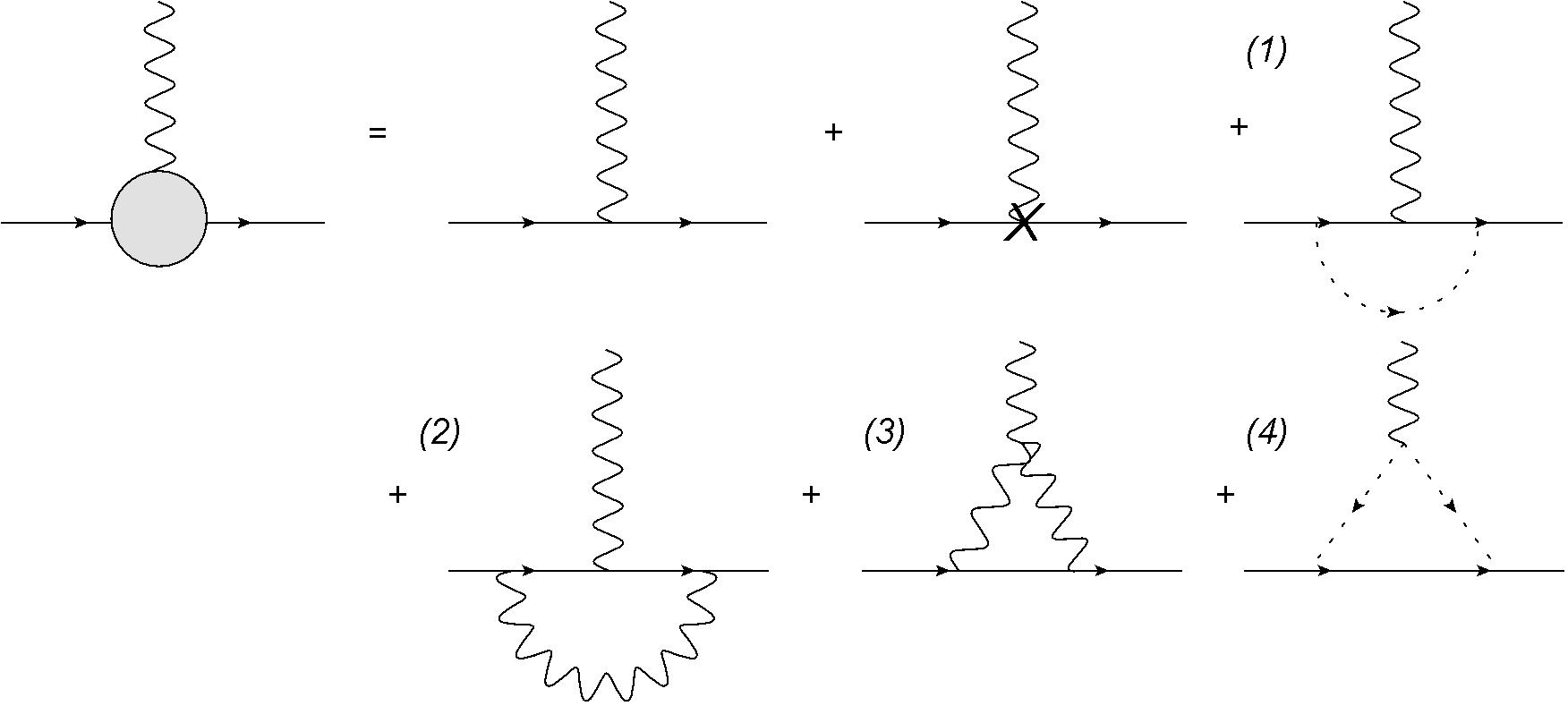}
  \caption{Corrections to the $\overline{\psi} \psi A_{\mu}$ coupling }
  \label{3point_ffA}
\end{figure}

\renewcommand{\figurename}{Diagram}
\begin{figure}
  \centering
  \includegraphics[height = 2.5 cm]{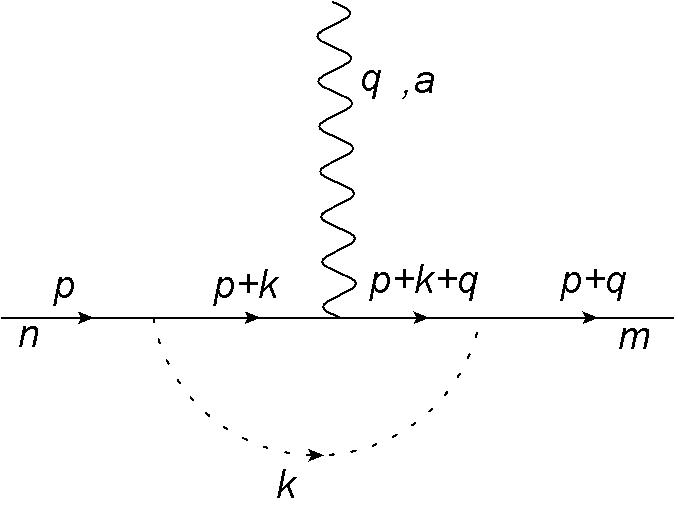} 
  \caption{}
\label{3point_1}
\end{figure}

For all diagrams in fig. \ref{3point_ffA} there are no additional symmetry factors. In most cases, the evaluation of the pole term can be done the easiest way with masses and the momentum carried by the gauge boson equal to zero (it can be done when the counter terms for vertices have no momentum or mass dependence). The full expresion for the first diagram is: 
\begin{equation}
(\texttt{diagram \ref{3point_1}}) = i \sum _{n' m'} g \, \kappa ^{i}_{m m'}\overline{\textbf{T}}^{a}_{m' n'} \kappa ^{i}_{n' n}  \int \frac{d^{d} k}{(2\pi)^{d}} 
\left(  \tilde{S}_{F}(p+k + q) \gamma_{\mu} \tilde{S}_{F}(p+k) \right) \tilde{D}_{F}(k)
\end{equation}
After some simplifications, the integral we are interested in, takes the form:
\begin{equation}
(\texttt{diagram \ref{3point_1}}) = g \, \sum _{n' m'} g \, \kappa ^{i}_{m m'} \overline{\textbf{T}}^{a}_{m' n'} \kappa ^{i}_{n' n}
 \int \frac{d^{d} k}{(2\pi)^{d}} 
\frac{\left( p+k  \right)^{\alpha} \left( p+k  \right)^{\beta} }{k^2 (p+k)^2 (p+k)^2} 
\gamma_{\alpha} \gamma_{\mu} \gamma_{\beta}
\end{equation}
Using the equality
\begin{eqnarray}
\gamma_{\mu} \gamma_{\alpha} \gamma^{\mu} = (2 - d) \gamma_{\alpha} 
\end{eqnarray}
one can get the final result
\begin{equation}
(\texttt{diagram \ref{3point_1}}) = - \frac{\dot{\imath} g \, (\kappa ^{i} \overline{\textbf{T}}^{a} \kappa ^{i})_{m n} }{16 \pi^2 \epsilon}  \gamma_{\mu}
\end{equation}

\begin{figure}
  \centering
  \includegraphics[height = 2.5 cm]{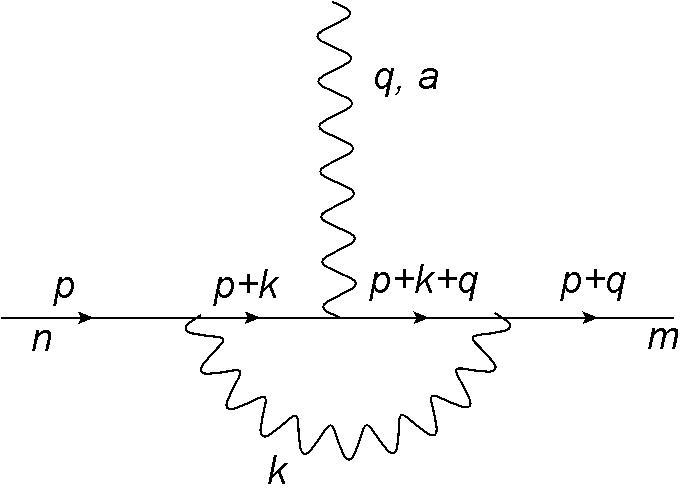} 
  \caption{}
\label{3point_2}
\end{figure}
To evaluate contribution from diagram \ref{3point_2} one needs to simplify the group theory factor.
\begin{equation}
(\texttt{diagram \ref{3point_2}}) = g^2 g (\overline{\textbf{T}}^{b} \overline{\textbf{T}}^{a} \overline{\textbf{T}}^{b})_{m n} 
\int \frac{d^{d}k}{(2\pi)^d} \tilde{D}_{F}^{\nu \rho} (k) \left[ \gamma_{\rho} \tilde{S}_{F}(p+q+k) \gamma_{\mu}
\tilde{S}_{F}(p+k) \gamma _{\nu} \right]
\end{equation}
\begin{eqnarray}
(\overline{\textbf{T}}^{b} \overline{\textbf{T}}^{a} \overline{\textbf{T}}^{b})_{n m} & = & 
\frac{1}{2} \left( \overline{\textbf{T}}^{b} \overline{\textbf{T}}^{b} \overline{\textbf{T}}^{a} + \dot{\imath} f_{abc} \overline{\textbf{T}}^{b} \overline{\textbf{T}}^{c}\right)_{n m} + 
\frac{1}{2} \left( \overline{\textbf{T}}^{a} \overline{\textbf{T}}^{b} \overline{\textbf{T}}^{b} - \dot{\imath}  f_{abc} \overline{\textbf{T}}^{c} \overline{\textbf{T}}^{b}\right)_{n m} = \nonumber\\
& = &  \left( \overline{C}_{3} - \frac{1}{2} C_{1}\right) \overline{\textbf{T}}^{a}_{n m} 
\end{eqnarray}
We use the previously mentioned simplification to calculate the pole term. With some help of the identity
\begin{equation}
\gamma_{\rho} \gamma_{\lambda} \gamma_{\mu} \gamma_{\nu} \gamma^{\rho} = -2 \gamma_{\mu} \gamma_{\nu} \gamma_{\lambda}  + (2-d)\gamma_{\lambda} \gamma_{\mu} \gamma_{\nu}
\end{equation}\\
one can obtain the result
\begin{equation}
(\texttt{diagram \ref{3point_2}}) = 
- \frac{2 \dot{\imath} g^3 (\overline{C}_{3} - \frac{1}{2}C_{1})(1-\eta) }{16 \pi^2} \overline{\textbf{T}}^{a}_{m n} \gamma _{\mu}
\end{equation}

\begin{figure}
  \centering
  \includegraphics[height = 2.5 cm]{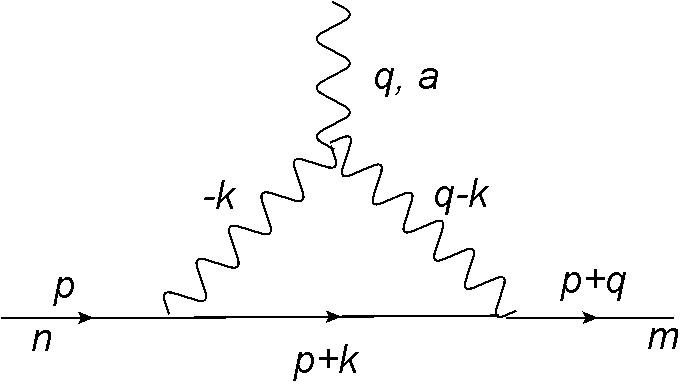} 
  \caption{}
\label{3point_3}
\end{figure}
\newpage
To evaluate contribution from diagram \ref{3point_3} we need to calculate the following expression:
\begin{eqnarray}
(\texttt{diagram \ref{3point_3}}) &=& \dot{\imath} g^3 \sum _{b, c} f_{abc}(\overline{\textbf{T}}^{c} \overline{\textbf{T}}^{b})_{m n} 
\int \frac{d^{d}k}{(2\pi)^d} 
\tilde{D}_{F}^{\tau \rho} (q - k) \tilde{D}_{F}^{\nu \sigma} (k) \left[ \gamma_{\tau} \tilde{S}_{F}(p+k) \gamma_{\sigma} \right] \times \nonumber\\
& &\times 
\left[
(2k-q)_{\mu}g_{\nu \rho} - (q+k)_{\rho}g_{\mu \nu} +(2q - q)_{\nu} g_{\rho \mu}
\right]
\end{eqnarray}
One can express the group theory factor using $C_{1}$ as follows:
\begin{eqnarray}
\sum _{b, c} f_{abc} \overline{\textbf{T}}^{c} \, \overline{\textbf{T}}^{b} = 
- i \sum _{b} [\overline{\textbf{T}}^{a}, \overline{\textbf{T}}^{b}] \, \overline{\textbf{T}}^{b} = -\frac{\dot{\imath}}{2} C_{1} \overline{\textbf{T}}^{a}
\end{eqnarray}
With the same procedure as before we find the pole term:
\begin{eqnarray}
(\texttt{diagram \ref{3point_3}}) = - \frac{3}{2} \frac{\dot{\imath} g^3 C_{1}(1+\xi)}{16 \pi ^2 \epsilon} (\overline{\textbf{T}}^{a})_{m n} \gamma_{\mu}
\end{eqnarray}

\begin{figure}
  \centering
  \includegraphics[height = 2.5 cm]{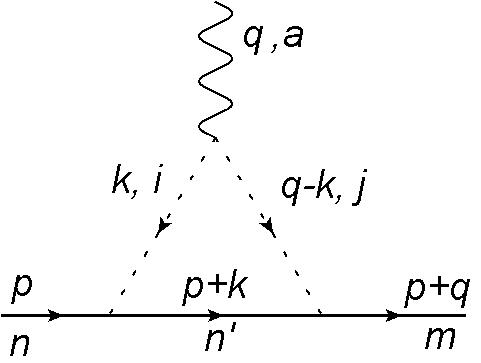} 
  \caption{}
\label{3point_4}
\end{figure}

To evaluate contribution from diagram \ref{3point_4} one needs to extract the pole term from the following expression:
\begin{eqnarray}
(\texttt{diagram \ref{3point_4}})  &=& i g \textbf{T}^{a} _{i j} \kappa ^{j} _{m n'} \kappa ^{i} _{n' n}
\times \nonumber\\
& & \times \int \frac{d^{d}k}{(2\pi)^d} 
\tilde{S}_{F} \left( p+k \right)
\tilde{D}_{F} \left( k \right)
\tilde{D}_{F} \left( q-k \right)
\left( 2k - q \right) ^{\mu}
\end{eqnarray}
which simplifies to the form
\begin{eqnarray}
(\texttt{diagram \ref{3point_4}}) &=& \textbf{T}^{a} _{i j} \kappa ^{j} _{m n'} \kappa ^{i} _{n' n}
\frac{i g }{16 \pi ^2 \epsilon} \gamma ^{\mu}
\label{row2}
\end{eqnarray}

\begin{figure}
  \centering
  \includegraphics[height = 2.3 cm]{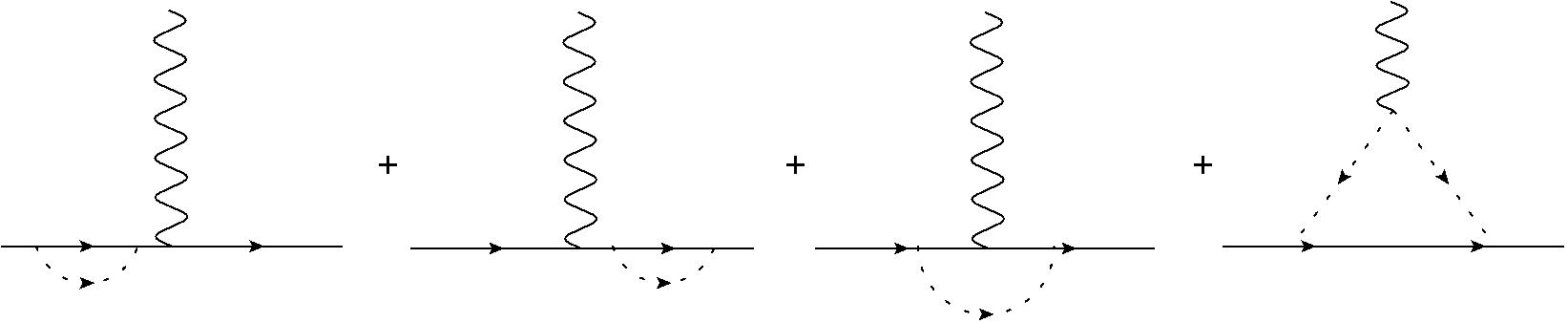} 
  \caption{One particle reductible corrections to  $\overline{\psi} \psi A_{\mu}$  coupling that cancel each other.}
  \label{non_OPI}
\end{figure}

As we have previously mentioned in section \ref{general_theory}, all the 1-loop non diagonal corrections that occur in the equation (\ref{gB_all}) cancel each other out. Cancelling diagrams are shown in fig. \ref{non_OPI}.
\renewcommand{\figurename}{Figure}

We will write partially the equation (\ref{gB_all}) - only with contributions from diagrams in fig. (\ref{non_OPI}). Assuming that the renormalization matrix $Z_{\psi}$ is hermitian we get 
\begin{eqnarray}
g_{B} \overline {\textbf{T}}^{a} _{n m} &=& g \overline {\textbf{T}}^{a} _{n m} + 
\frac{1}{16 \pi^2 \epsilon} (
\frac{1}{2}  \kappa ^{i} _{m m'} \kappa ^{i} _{m' n'} \overline {\textbf{T}}^{a} _{n' n} 
+ \frac{1}{2} \overline {\textbf{T}}^{a} _{m m'} \kappa ^{i} _{m' n'} \kappa ^{i} _{n' n} 
- \kappa ^{i} _{m m'} \overline {\textbf{T}}^{a} _{m' n'} \kappa ^{i} _{n' n} \nonumber\\
& &+ \textbf{T}^{a} _{i j} \kappa ^{j} _{m n'} \kappa ^{i} _{n' n}   ) + \ldots
\label{kappy}
\end{eqnarray}
To show the cancellation, one needs to consider how fermion and scalar fields change under infinitesimal gauge transformation.
\begin{eqnarray}
\psi _{n}' = \psi _{n} - i g \overline{\textbf{T}}^{a}_{n m } \Lambda ^{a} \psi _{m} \\
\overline{\psi} _{n}' = \overline{\psi} _{n} + i g \overline{\psi} _{m} \overline{\textbf{T}}^{a}_{m n} \Lambda ^{a}
\\
\phi _{i}' = \phi _{j} - i g \textbf{T}^{a}_{i j } \Lambda ^{a} \phi _{j}
\end{eqnarray}

From the invariance of the Yukawa term under gauge symmetry one can get a relation between the Yukawa coupling and gauge transformation generators. 
\begin{eqnarray}
\textbf{T}^{a} _{j i} \kappa ^{j} _{n m} =
 \overline {\textbf{T}}^{a} _{n n'} \kappa ^{i} _{n' m} 
- \kappa ^{i} _{n n'} \overline {\textbf{T}}^{a} _{n' m} =
[  \overline {\textbf{T}}^{a} , \kappa ^{i}
] _{n' m} 
\label{tozsamosc_kapp}
\end{eqnarray}
which guarantees that
\begin{eqnarray}
\frac{1}{2}  \kappa ^{i} _{m m'} \kappa ^{i} _{m' n'} \overline {\textbf{T}}^{a} _{n' n} 
+ \frac{1}{2} \overline {\textbf{T}}^{a} _{m m'} \kappa ^{i} _{m' n'} \kappa ^{i} _{n' n} 
- \kappa ^{i} _{m m'} \overline {\textbf{T}}^{a} _{m' n'} \kappa ^{i} _{n' n} + \textbf{T}^{a} _{i j} \kappa ^{j} _{m n'} \kappa ^{i} _{n' n}  = 0
\end{eqnarray}

Now we can simplify the equation (\ref{gB_all}) to the form (\ref{easy_gB}), where $K_{2}$ and $\Delta Z_{\psi}$ are as follows:
\begin{eqnarray}
K_{2} &=& - \left(\frac{3}{2} C_{1} + \frac{1}{2} C_{1} \xi + 2 \overline{C} _{3} \xi \right) \frac{\dot{\imath} g^2}{16 \pi ^2 \epsilon}
\label{Z2} \\
\Delta Z_{\psi} &=&
\frac{ g^2 }{16 \pi^2 \epsilon} \left( 4 + 2 \eta \right) C_{3}
\label{Zpsi2}
\end{eqnarray}

\subsection{Renormalization of $\phi^4$ interaction}

\renewcommand{\figurename}{Figure}
\begin{figure}
  \centering
  \includegraphics[height = 9 cm]{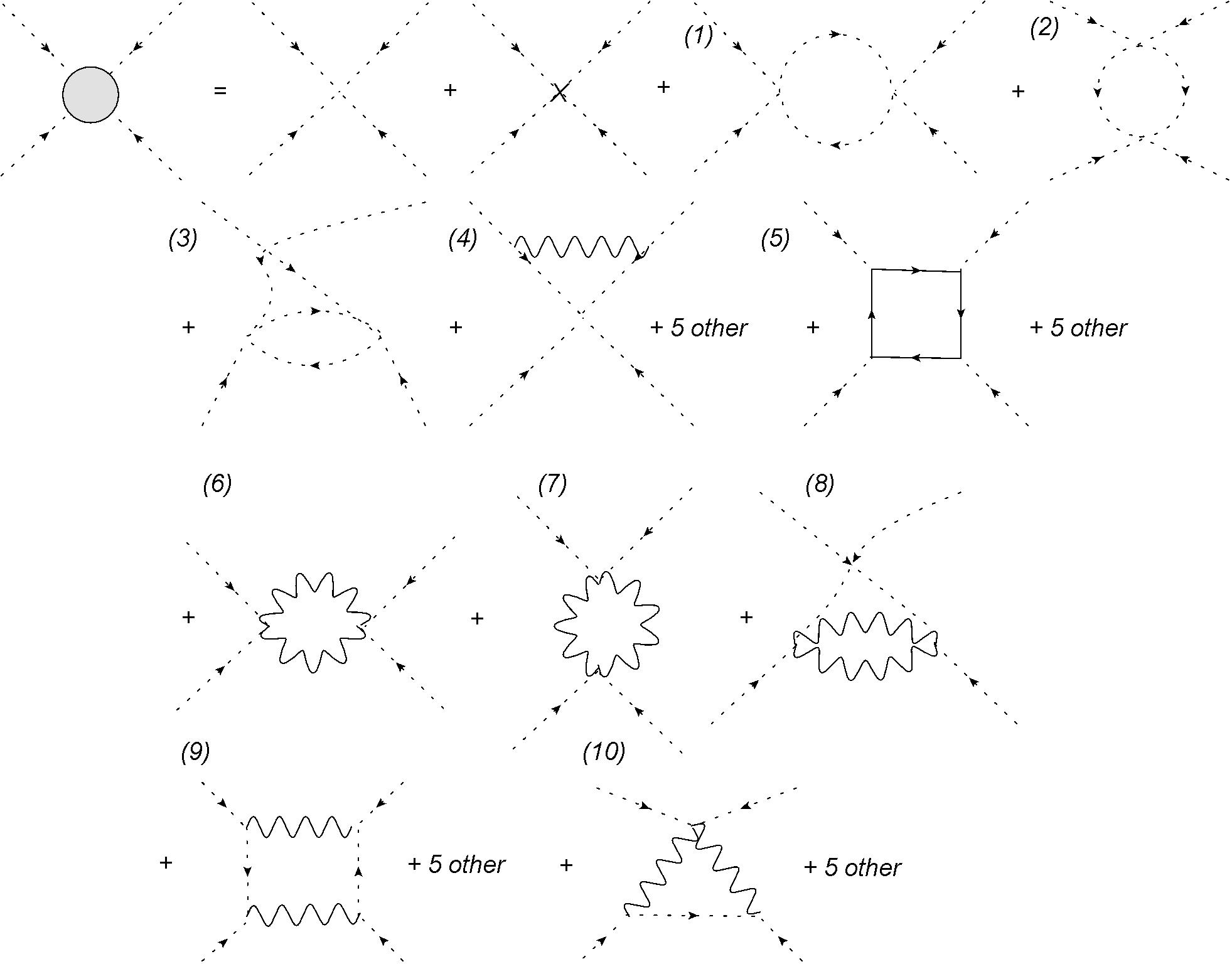}
  \caption{ Quadrilinear scalar coupling corrections that contribute in dimensional regularization}
  \label{4point_scalar}
\end{figure}

\renewcommand{\figurename}{Diagram}
\begin{figure}
  \centering
  \includegraphics[height = 2.5 cm]{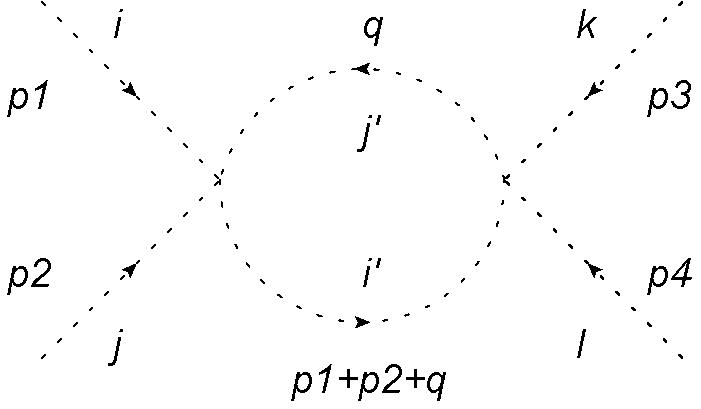} 
  \caption{}
\label{4point_1}
\end{figure}
All contributing diagrams to the 1-loop renormalization of $\phi^4$ interaction are shown in fig.~\ref{4point_scalar}. 
In diagram \ref{4point_1} there is a symmetry factor $\frac{1}{2}$, and one should sum over all $i', j'$ scalar fields.

\begin{eqnarray}
(\texttt{diagram \ref{4point_1}}) = - \frac{1}{2} \sum _{i' j'} 
h_{i j i' j'} h_{i' j' k l}
\int \frac{d^d q}{(2 \pi)^d} 
\tilde{D}_{F} (p_{1} + p_{2} + q) \tilde{D}_{F} ( q)
\end{eqnarray}

For two similar diagrams, but with differently connected scalar lines, the expressions are analogous. Summing them together result in:
\begin{eqnarray}
(\texttt{diagram \ref{4point_1} + 2 other})  =  \frac{ \dot{\imath} }{16 \pi ^2 \epsilon } \sum _{i' j'} 
\left( 
h_{i j i' j'} h_{i' j' k l} + h_{i k i' j'} h_{i' j' j l}+ h_{k j i' j'} h_{i' j' i l}
\right)
\end{eqnarray}

\begin{figure}
  \centering
  \includegraphics[height = 3 cm]{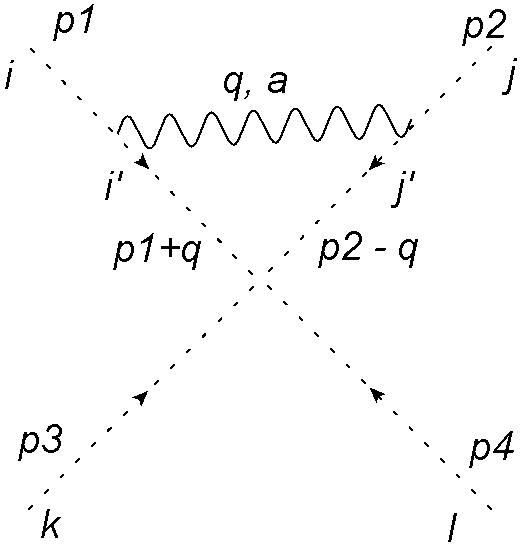} 
  \caption{}
\label{4point_3}
\end{figure}

There are 6 diagrams of the type shown on \ref{4point_3}. Evaluating the contribution from \ref{4point_3} one can get:
\begin{eqnarray}
(\texttt{diagram \ref{4point_3}}) &=& i g^2 \sum_{i' j'} \sum_{a}  \textbf{T} ^{a} _{i i'} \textbf{T} ^{a} _{j j'} h_{i' j' k l}
\int \frac{d^d q}{(2 \pi)^d} 
(2 p_{1} + q)_{\mu} (2 p_{2} - q)_{\nu} 
\times \nonumber\\
& & \times
\tilde{D}_{F} ^{\mu \nu} ( q) 
\tilde{D}_{F} ( p_{2} - q) \tilde{D}_{F} (p_{1} + q) 
\end{eqnarray}
\begin{eqnarray}
(\texttt{diagram \ref{4point_3}}) &=&   \sum_{i', j'} \sum_{a} \textbf{T} ^{a} _{i i'} \textbf{T} ^{a} _{j j'} h_{i' j' k l}
 \frac{2 i g^2 (1 - \eta)}{16 \pi^2 \epsilon}
\end{eqnarray}
For a full contribution we sum all the diagrams of this type.
\begin{eqnarray}
(\texttt{diagram \ref{4point_3} + 5 other}) &=&  \frac{2 i g^2 (1 - \eta)}{16 \pi^2 \epsilon} 
\times \nonumber\\ & & \times 
\sum_{a} \sum_{b, c} (
\textbf{T} ^{a} _{i b} \textbf{T} ^{a} _{j c} h_{b c k l} +
\textbf{T} ^{a} _{i b} \textbf{T} ^{a} _{k c} h_{b j c l} +
\textbf{T} ^{a} _{i b} \textbf{T} ^{a} _{l c} h_{b j k c} +
\nonumber\\ & & +
\textbf{T} ^{a} _{j b} \textbf{T} ^{a} _{k c} h_{i b c l} +
\textbf{T} ^{a} _{j b} \textbf{T} ^{a} _{l c} h_{i b k c} +
\textbf{T} ^{a} _{k b} \textbf{T} ^{a} _{l c} h_{i j b c}
) 
\end{eqnarray}
To simplify this expression we will use an identity obtained from the quadrilinear term invariance under infinitesimal gauge transformation.
\begin{eqnarray}
\textbf{T} ^{a} _{i i'}  h_{i' j k l} +
\textbf{T} ^{a} _{j j'}  h_{i j' k l} +
\textbf{T} ^{a} _{k k'}  h_{i j k' l} +
\textbf{T} ^{a} _{l l'}  h_{i j k l'} = 0
\label{tozsamosc_h}
\end{eqnarray}
and write the factor containing generators in a form: 
\begin{eqnarray}
& &
\textbf{T} ^{a} _{i i'} \textbf{T} ^{a} _{j j'} h_{i' j' k l} +
\textbf{T} ^{a} _{i i'} \textbf{T} ^{a} _{k k'} h_{i' j k' l} +
\textbf{T} ^{a} _{i i'} \textbf{T} ^{a} _{l l'} h_{i' j k l'} +
 \nonumber\\ & & 
\textbf{T} ^{a} _{j j'} \textbf{T} ^{a} _{k k'} h_{i j' k' l} +
\textbf{T} ^{a} _{j j'} \textbf{T} ^{a} _{l l'} h_{i j' k l'} +  
\textbf{T} ^{a} _{k k'} \textbf{T} ^{a} _{l l'} h_{i j k' l'}
= 
\\ & & 
= \frac{1}{2} \left( 
\textbf{T} ^{a} _{i i'} \textbf{T} ^{a} _{j j'} h_{i' j' k l} +
\textbf{T} ^{a} _{i i'} \textbf{T} ^{a} _{k k'} h_{i' j k' l} +
\textbf{T} ^{a} _{i i'} \textbf{T} ^{a} _{l l'} h_{i' j k l'} +
\textbf{T} ^{a} _{j j'} \textbf{T} ^{a} _{i i'} h_{i' j' k l} + \right.
\nonumber\\ & & +
\textbf{T} ^{a} _{j j'} \textbf{T} ^{a} _{k k'} h_{i j' k' l} +
\textbf{T} ^{a} _{j j'} \textbf{T} ^{a} _{l l'} h_{i j' k l'} +
\textbf{T} ^{a} _{k k'} \textbf{T} ^{a} _{i i'} h_{i' j k' l} +
\textbf{T} ^{a} _{k k'} \textbf{T} ^{a} _{j j'} h_{i j' k' l} +
\nonumber\\ & & \left. +
\textbf{T} ^{a} _{k k'} \textbf{T} ^{a} _{l l'} h_{i j k' l'} +
\textbf{T} ^{a} _{l l'} \textbf{T} ^{a} _{i i'} h_{i' j k l'} +
\textbf{T} ^{a} _{l l'} \textbf{T} ^{a} _{j j'} h_{i j' k l'} +
\textbf{T} ^{a} _{l l'} \textbf{T} ^{a} _{k k'} h_{i j k' l'} 
\right)
\\ & & = - \frac{1}{2} 
\textbf{T} ^{a} _{i i'} \textbf{T} ^{a} _{i' i''} h_{i'' j k l} - \frac{1}{2} 
\textbf{T} ^{a} _{j j'} \textbf{T} ^{a} _{j' j''} h_{i j'' k l}
- \frac{1}{2} 
\textbf{T} ^{a} _{k k'} \textbf{T} ^{a} _{k' k''} h_{i j k'' l} 
- \frac{1}{2} 
\textbf{T} ^{a} _{l l'} \textbf{T} ^{a} _{l' l''} h_{i j k l''}
\nonumber\\ & &
= - 2 C_{3} h_{i j k l}
\end{eqnarray}
Now we can write the result in a simpler form:
\begin{eqnarray}
(\texttt{diagram \ref{4point_3}} + \textit{5 others} ) &=&   - \frac{4 i g^2 (1 - \eta) C_{3} \delta_{i j}}{16 \pi^2 \epsilon} 
\end{eqnarray}

\begin{figure}
  \centering
  \includegraphics[height = 3 cm]{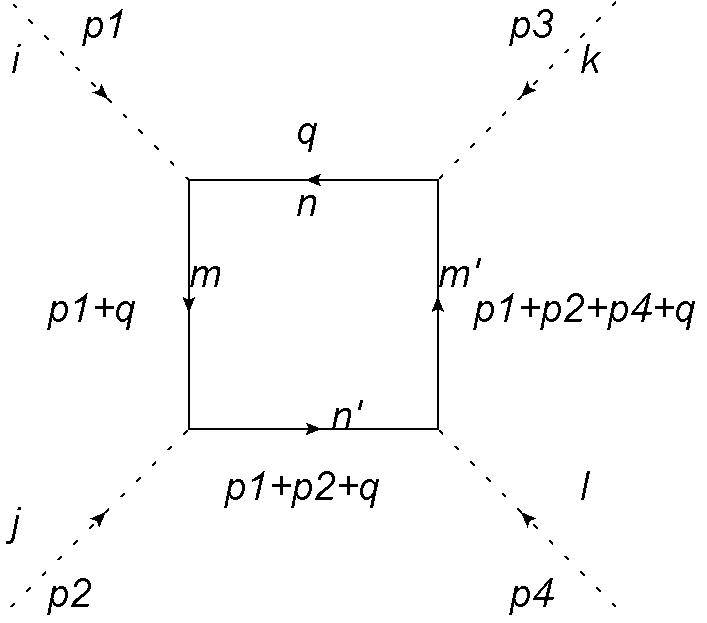} 
  \caption{}
\label{4point_4}
\end{figure}

Diagram \ref{4point_4} has to be considered with $(-1)$ factor from a closed fermion loop.
\begin{eqnarray}
(\texttt{diagram \ref{4point_4}}) &=& - \textbf{Tr} (\kappa _j \kappa _i \kappa_k \kappa _l) 
\int \frac{d^d q}{(2 \pi)^d}  \times \nonumber\\
&\times & 
\textbf{Tr} 
\left(
\tilde{S}_{F} (p_{1} + p_{2} + p_{4} + q) \, \tilde{S}_{F} (p_{1} + p_{2} + q) \,
\tilde{S}_{F} (p_{1} + q) \, \tilde{S}_{F} (q)
\right) \nonumber\\
\end{eqnarray}
To extract the pole term from this integral one can use the following identity.
\begin{eqnarray}
\textbf{Tr} (\gamma_{\alpha} \gamma_{\beta} \gamma_{\mu} \gamma_{\nu}) = 
4 \left( g^{\alpha \beta} g^{\mu \nu} - g^{\alpha \mu } g^{\beta \nu} + g^{\alpha \nu } g^{\mu \beta}
 \right)
\end{eqnarray}

There are 5 other diagrams similar to \ref{4point_4}. To simplify the result including all of them, we will introduce such quantity:
\begin{eqnarray}
A_{ijkl} = \textbf{Tr} \left( 
\kappa_{i} \kappa_{j} \lbrace \kappa_{l}, \kappa_{k}  \rbrace + \kappa_{i} \kappa_{k} \lbrace \kappa_{j}, \kappa_{l}  \rbrace + \kappa_{i} \kappa_{l} \lbrace \kappa_{j}, \kappa_{k}  \rbrace
 \right)
\end{eqnarray}
Then the final contribution is:
\begin{eqnarray}
(\texttt{diagram \ref{4point_4} + other})= - \frac{8 \dot{\imath} A_{ijkl} }{16 \pi ^2 \epsilon}
\end{eqnarray}

\begin{figure}
  \centering
  \includegraphics[height = 2.5 cm]{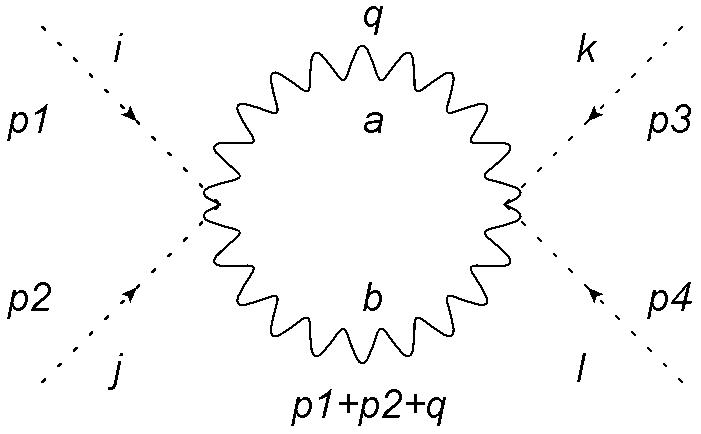} 
  \caption{}
\label{4point_5}
\end{figure}

Diagram \ref{4point_5} has a symmetry factor $\frac{1}{2}$.
\begin{eqnarray}
(\texttt{diagram \ref{4point_5}}) = & &  - \frac{1}{2} g^4 
\left(
\textbf{T}^a_{ni}\textbf{T}^b_{nj} + \textbf{T}^a_{nj}\textbf{T}^b_{ni}
\right)
\left(
\textbf{T}^a_{mk}\textbf{T}^b_{ml} + \textbf{T}^a_{ml}\textbf{T}^b_{mk}
\right) \times \nonumber\\
& &\times \int \frac{d^d q}{(2 \pi)^d} 
\tilde{D}^{\alpha \beta}_{F} (p_{1} + p_{2} + q) \tilde{D}^{\mu \nu}_{F} ( q) g_{\mu \alpha} g_{\beta \nu}
\end{eqnarray}

To calculate the contribution from this kind of diagrams one needs to perform the following integration:
\begin{eqnarray}
\int \frac{d^d q}{(2 \pi)^d} 
\tilde{D}^{\mu \nu}_{F} (p + q) \tilde{D}_{\mu \nu \, F} ( q) = 
- \frac{2 \dot{\imath} (4 - 2 \eta + \eta ^2) }{16 \pi^2 \epsilon}
\end{eqnarray}
To simplify the result including other similar diagrams, it is convenient to introduce the following constant
\begin{eqnarray}
B_{ijkl} &=& \lbrace \textbf{T}^a , \textbf{T}^b \rbrace _{ij} \lbrace \textbf{T}^a , \textbf{T}^b \rbrace _{kl} +
 \lbrace \textbf{T}^a , \textbf{T}^b \rbrace _{ik} \lbrace \textbf{T}^a , \textbf{T}^b \rbrace _{jl} + \nonumber\\
& &  \lbrace \textbf{T}^a , \textbf{T}^b \rbrace _{il} \lbrace \textbf{T}^a , \textbf{T}^b \rbrace _{jk}
\end{eqnarray}
where 
\begin{eqnarray}
\textbf{T}^a_{ni}\textbf{T}^b_{nj} + \textbf{T}^a_{nj}\textbf{T}^b_{ni}
 = - \lbrace \textbf{T}^a , \textbf{T}^b \rbrace _{ij}
\end{eqnarray}
Using this notation one can write the result as follows
\begin{eqnarray}
(\texttt{diagram \ref{4point_5} + other}) =  \frac{\dot{\imath} g^4 B_{ijkl}}{16 \pi ^2 \epsilon} 
\left(4 - 2 \eta + \eta^2 \right) = \frac{\dot{\imath} g^4 B_{ijkl}}{16 \pi ^2 \epsilon} 
\left(3 + (1 - \eta )^2 \right)
\end{eqnarray}

\begin{figure}
  \centering
  \includegraphics[height = 3 cm]{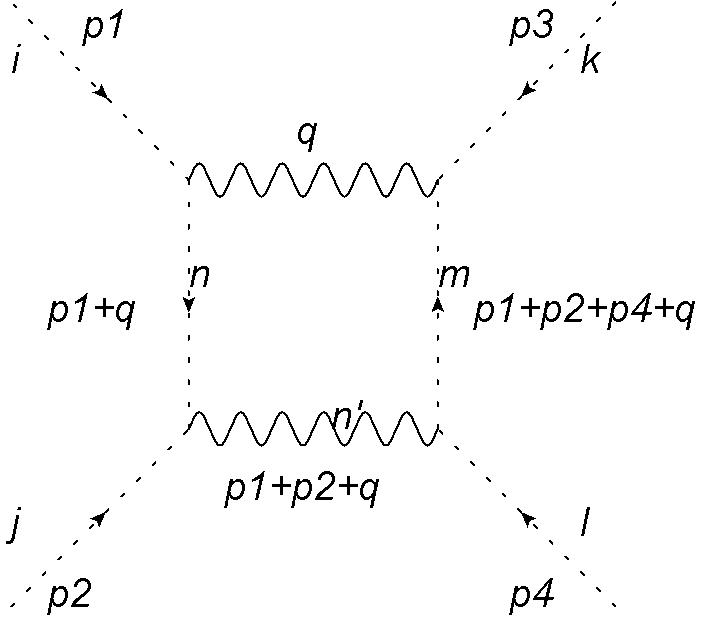}
  \caption{}
\label{4point_6}
\end{figure}

There are 6 diagrams of type \ref{4point_6} to include in our calculations. Contribution from diagram \ref{4point_6} takes the form:
\begin{eqnarray}
& & (\texttt{diagram \ref{4point_6}}) =   g^4 
\left(
\textbf{T}^a_{in}\textbf{T}^b_{jn} \textbf{T}^a_{km}\textbf{T}^b_{lm}
\right)
 \times \nonumber\\
& & \times \int \frac{d^d q}{(2 \pi)^d} 
\tilde{D}^{\alpha \beta}_{F} (p_{1} + p_{2} + q) \tilde{D}^{\mu \nu}_{F} ( q) 
\tilde{D}_{F} (p_{1} + p_{2} + p_{4} + q) \tilde{D}_{F} (p_{1} + q)
 \times \nonumber\\
& &
\times
\left(
2 p_{1} + q
\right)_{\mu}
\left(
p_{3} - p_{1} - p_{2} - p_{4} - q
\right)_{\nu}
\left(
2p_{4} + p_{1} + p_{2} + q
\right)_{\beta}
\left(
p_{2} - p_{1} - q
\right)_{\alpha} \nonumber\\ 
\end{eqnarray}
Being interested only in extracting the pole of this integral one can get after some simplifications the following form
\begin{eqnarray}
(\texttt{diagram \ref{4point_6}}) &=&  - g^4 (1-\eta)^2
\left(
\textbf{T}^a_{in}\textbf{T}^b_{jn} \textbf{T}^a_{km}\textbf{T}^b_{lm}
\right)
 \times \nonumber\\
& & \times \int \frac{d^d q}{(2 \pi)^d} 
\tilde{D}_{F} (p_{1} + p_{2} + p_{4} + q) \tilde{D}_{F} (p_{1} + q)
\end{eqnarray}
With the result
\begin{eqnarray}
(\texttt{diagram \ref{4point_6}}) = 
\left(
\textbf{T}^a_{in}\textbf{T}^b_{jn} \textbf{T}^a_{km}\textbf{T}^b_{lm}
\right)
\frac{2 i  g^4 (1-\eta)^2 }{16 \pi^2 \epsilon}
\end{eqnarray}

One needs to consider other similar diagrams with permutations of the $i,j,k,l$ indices. The final result reads:
\begin{eqnarray}
(\texttt{diagram \ref{4point_6} + other}) &= \frac{i g^4 B_{ijkl}}{16 \pi ^2 \epsilon} 
\left( 1 - \eta \right)^2
\end{eqnarray}

\begin{figure}
  \centering
  \includegraphics[height = 3 cm]{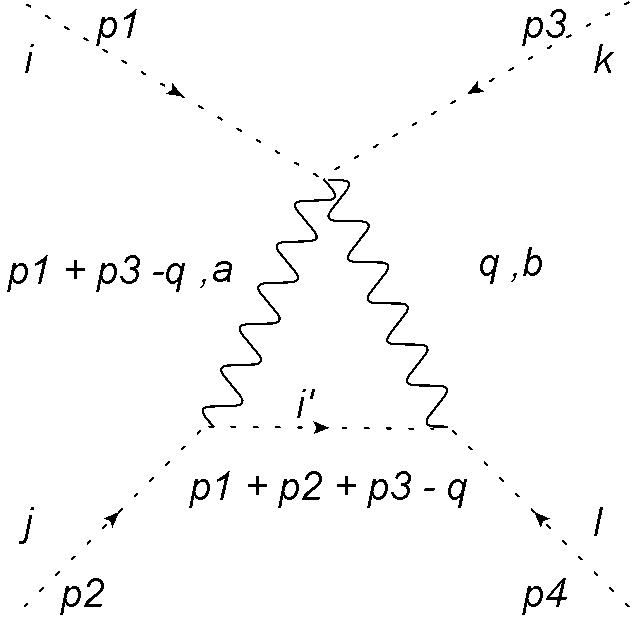} 
  \caption{}
\label{4point_7}
\end{figure}

There are 6 diagrams of type \ref{4point_7} to include in the calculations. Symmetry factor for these diagrams is $1$.
\begin{eqnarray}
(\texttt{diagram \ref{4point_7}}) &=& -i g^4 g^{\alpha \mu}
 \lbrace \textbf{T}^a , \textbf{T}^b \rbrace _{i k} \textbf{T}^a_{j i'}\textbf{T}^b_{i' l}
 \int \frac{d^d q}{(2 \pi)^d} 
\tilde{D}^{\mu \nu}_{F} ( q) 
\tilde{D}_{F} (p_{1} + p_{2} + p_{3} - q) 
 \nonumber\\
&\times &  
\tilde{D}^{\alpha \beta}_{F} (p_{1} + p_{2} - q) 
(p_{1} + 2 p_{2} + p_{3} - q) _{\beta}
(p_{1} + p_{2} + p_{3} - p_{4} - q) _{\nu} \nonumber\\
\end{eqnarray}
After considering other similar diagrams with permutations of the $i,j,k,l$ indices we have
\begin{eqnarray}
(\texttt{diagram \ref{4point_7} + other})= - \frac{2 i g^4 B_{ijkl}}{16 \pi ^2 \epsilon} 
( 1 - \eta )^2
\end{eqnarray}

The final result for the renormalization constant is below. As we can see, the gauge fixing parameter cancels within 6th to 10th diagram (see figure \ref{4point_scalar}) and only the term proportional to $C_{3}$ (originating from the 4th diagram) depends on the gauge choice.
\begin{eqnarray}
& & i L_{i j k l} ^{i' j' k' l'} h_{i' j' k' l'} = 
\frac{3 \dot{\imath} g^4 B_{ijkl}}{16 \pi ^2 \epsilon} 
- \frac{8 \dot{\imath} A_{ijkl}}{16 \pi ^2 \epsilon} 
+ \frac{ \dot{\imath} }{16 \pi ^2 \epsilon } \sum _{i' j'} ( h_{i j i' j'} h_{i' j' k l} +
\nonumber\\ & &
+ h_{i k i' j'} h_{i' j' j l}+ h_{k j i' j'} h_{i' j' i l} )
- \frac{4 i g^2 (1 - \eta) C_{3} \delta_{i j}}{16 \pi^2 \epsilon} 
\end{eqnarray}

\newpage
\subsection{Renormalization of Yukawa interaction} \label{renormalization_of_yukawa}
\renewcommand{\figurename}{Figure}
\begin{figure}
  \centering
  \includegraphics[height = 5.5 cm]{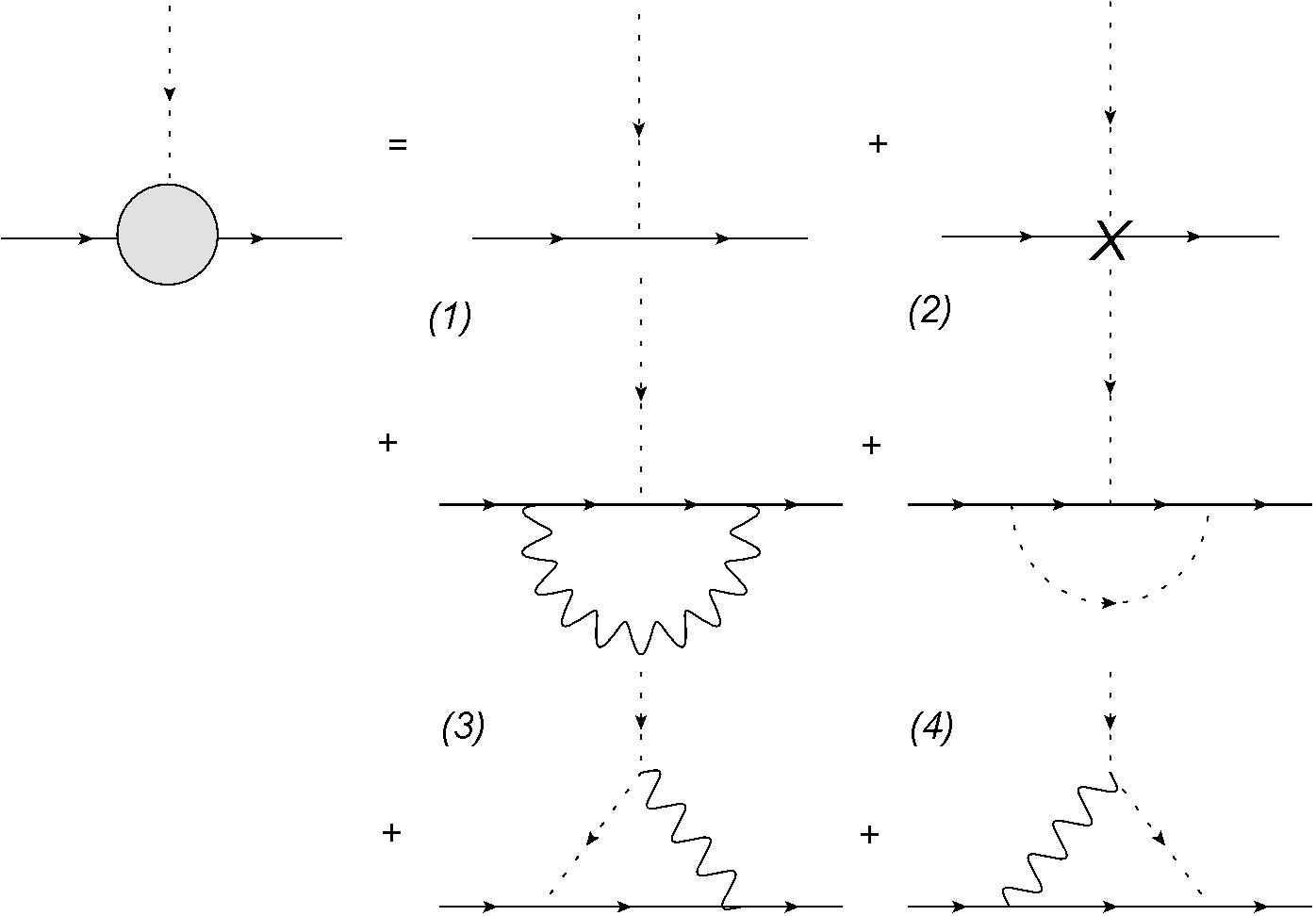}
  \caption{Corrections to the Yukawa that contribute in dimensional regularization}
  \label{3point_Yukawa}
\end{figure}

Diagrams contributing to the renormalization of Yukawa interaction are shown in fig. \ref{3point_Yukawa}.

\renewcommand{\figurename}{Diagram}
\begin{figure}
  \centering
  \includegraphics[height = 2.5 cm]{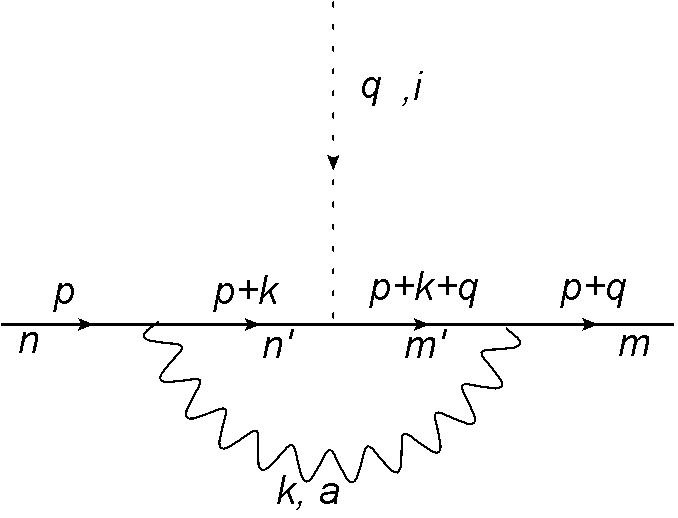}
  \caption{}
\label{yukawa_1}
\end{figure}
In diagram \ref{yukawa_1} the symmetry factor is equal to 1 and one should sum over all $n'$ and $m'$ indices for fermion fields and over $a$ for gauge fields.
\pagebreak
\begin{eqnarray}
(\texttt{diagram \ref{yukawa_1}}) &=&  i g^2 \sum _{n', m', a}
\left(
\overline{\textbf{T}}^a _{m m'} \kappa ^{i} _{m' n'}   \overline{\textbf{T}}^a _{n' n} 
\right)
 \times \nonumber \\
& & \times \int \frac{d^d k}{(2 \pi)^d} 
\tilde{D} ^{\mu \nu}_{F} (k) \gamma_{\mu} \tilde{S}_{F} (p + k + q) \tilde{S}_{F} (p + k) \gamma_{\nu}
\end{eqnarray}
After performing the integral one can write the pole term as follows
\begin{eqnarray}
(\texttt{diagram \ref{yukawa_1}}) =  \sum _{a}
\left(
\overline{\textbf{T}}^a  \kappa ^{i}   \overline{\textbf{T}}^a 
\right) _{m n}
\frac{2 i g^2 (-4 + \eta) }{16 \pi^2 \epsilon}
\end{eqnarray}

\begin{figure}
  \centering
  \includegraphics[height = 2.5 cm]{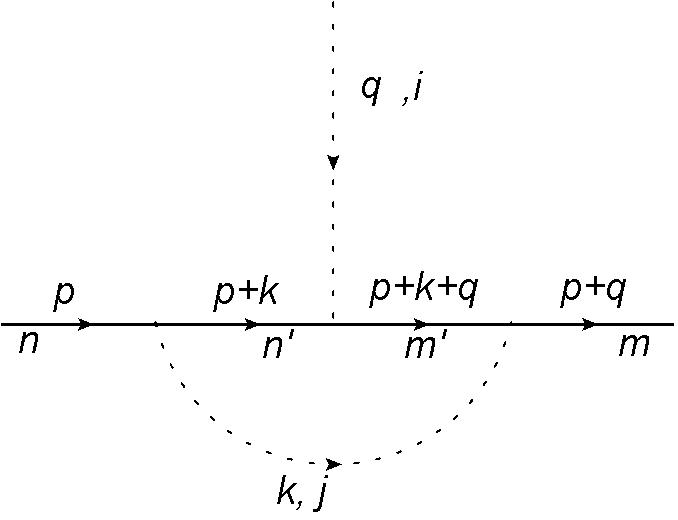}
  \caption{}
\label{yukawa_2}
\end{figure}

In diagram \ref{yukawa_2} one also sums over $n'$ and $m'$ indices of the fermion fields and $i$ index for the scalar field.
\begin{eqnarray}
(\texttt{diagram \ref{yukawa_2} })= i\sum _{n', m', j}
\left(
\kappa ^{j} _{m m'} \kappa ^{i} _{m' n'}  \kappa ^{j} _{n' n}
\right)
 \int \frac{d^d k}{(2 \pi)^d} 
\tilde{D} _{F} (k) \tilde{S}_{F} (p + k + q) \tilde{S}_{F} (p + k) \nonumber\\
\end{eqnarray}
The pole term contribution reads
\begin{eqnarray}
(\texttt{diagram \ref{yukawa_2} })=  \sum _{ j}
\left(
\kappa ^{j}  \kappa ^{i}   \kappa ^{j} 
\right) _{m n}
\frac{2 i}{16 \pi ^2 \epsilon}
\end{eqnarray}

\begin{figure}
  \centering
  \includegraphics[height = 2.5 cm]{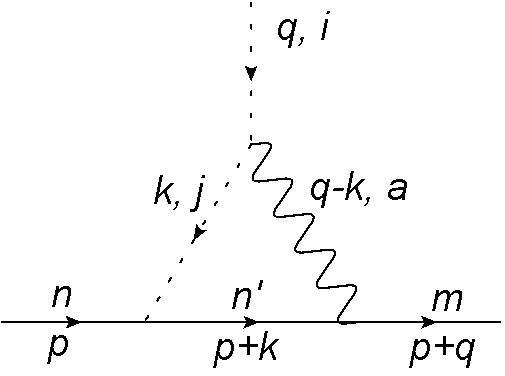}
  \caption{}
\label{yukawa_3}
\end{figure}

The last two diagrams to consider are very similar to each other and do not require any new calculation tricks.
\begin{eqnarray}
(\texttt{diagram \ref{yukawa_3}})= 
- i g^2 
\sum _{n', j, a} \overline{\textbf{T}}^{a} _{m n'} \kappa ^{j} _{n' n}  \textbf{T}^{a}_{ij}
 \int \frac{d^d k}{(2 \pi)^d} 
\gamma _{\nu} \tilde{S}_{F} (p + k) \tilde{D} _{F} (k)  (q + k)_{\mu} \tilde{D} _{F} ^{\mu \nu} (q-k) \nonumber\\
\end{eqnarray}
The result is:
\begin{eqnarray}
(\texttt{diagram \ref{yukawa_3}})= 
-  g^2 \sum _{n', j, a}  \overline{\textbf{T}}^{a} _{m n'} \kappa ^{j} _{n' n}  \textbf{T}^{a}_{ij}
\frac{i (-2 + 2 \eta)}{16 \pi^2 \epsilon}
\end{eqnarray}
Now we can add contribution from the second look-alike diagram, receiving:
\begin{eqnarray}
(\texttt{diagram \ref{yukawa_3} + other}) = 
g^2 \sum _{n', j, a}  \left(
-   \overline{\textbf{T}}^{a} _{m n'} \kappa ^{j} _{n' n}  \textbf{T}^{a}_{ij} +
 \kappa ^{j} _{m n'} \overline{\textbf{T}}^{a} _{n' n} \textbf{T}^{a}_{ij}
 \right)
\frac{i (-2 + 2 \eta)}{16 \pi^2 \epsilon} \nonumber\\
\end{eqnarray}

This way one can calculate $\Delta \kappa$ (see (\ref{ZetKappa1}) and (\ref{ZetKappa}))
\begin{eqnarray}
i K_{i m n}^{i' m' n'} \kappa^{i'} _{m' n'} &=& 
\sum _{a} \left( \overline{\textbf{T}}^a  \kappa ^{i}   \overline{\textbf{T}}^a  \right) _{m n}
\frac{2 i g^2 (-4 + \eta) }{16 \pi^2 \epsilon}
+ \sum _{j} \left( \kappa ^{j}  \kappa ^{i}   \kappa ^{j} \right) _{m n}
\frac{2 i}{16 \pi ^2 \epsilon} \nonumber\\
& & + g^2 \sum _{n', j, a} 
\left( \kappa ^{j} _{m n'} \overline{\textbf{T}}^{a} _{n' n} - \overline{\textbf{T}}^{a} _{m n'} \kappa ^{j} _{n' n} 
\right) \textbf{T}^{a}_{ij}
\frac{i (-2 + 2 \eta)}{16 \pi^2 \epsilon}
\end{eqnarray}

\subsection{Calculating beta functions}\label{Calculating_beta_functions}

To calculate beta functions in our generic gauge theory, we need relations between bare and renormalized coupling constants (see equations (\ref{Zetxi}), (\ref{easy_gB}), (\ref{ZetKappa1}) or (\ref{Zeth})). We will start from finding the expression for the beta function of the $g$-coupling. We repeat the relation between $g_{B}$ and renormalized coupling $g$ from equation (\ref{easy_gB}), also including the previously omitted renormalization scale $\mu$ factor, coming from the consistence of units in the dimensional regularization scheme.
\begin{eqnarray}
g_{B} =  \left(g - \frac{1}{2} \Delta Z_{A} g -  \Delta Z_{\psi} g
+  K_{2} g + \ldots \right) \mu^{\epsilon/2}= Z_{g} g \mu^{\epsilon/2}
\end{eqnarray}
Since $g_{B}$ does not depend on the scale $\mu$, one gets
\begin{eqnarray}
& & \mu \frac{d g_{B}}{d \mu} = 0 = \mu \frac{d}{d \mu} 
\left( Z_{g} g \mu^{\epsilon/2}
\right) \\
& & 0 = \frac{\epsilon}{2} Z_{g} g + \mu g \frac{d Z_{g}}{d \mu} + \mu Z_{g} \frac{d g}{ d \mu}
\end{eqnarray}

We here have the $\mu$ dependence written explicitly, so $\frac{\partial Z_{g}}{\partial \mu} = 0$. Using the expansion of $Z_{g}$ in terms of the coupling we get the expression for $\mu \frac{dg}{d \mu} $.

\begin{eqnarray}
& & Z_{g} = 1 + Z_{g}^{(2)} g^2 + \ldots \\
& & \mu \frac{d g}{ d \mu} = \frac{- \frac{\epsilon}{2} Z_{g} g}{g \frac{\partial Z_{g}}{\partial g} + Z_{g}} = 
- \frac{\epsilon}{2}g + \epsilon g^3 Z_{g}^{(2)}
\end{eqnarray}
The beta function is defined as 
\begin{eqnarray}
\beta  \left(  \mu \right) = \lim_{\epsilon \to 0} \left(  \mu \frac{d g}{ d \mu} \right)
\end{eqnarray}
Beta function expanded in terms of $g$ gives us the $\beta_{0}$ function we are interested in
\begin{eqnarray}
\beta  \left(  \mu \right) = \beta_{0} g^3 + \ldots \\
\beta_{0} = \lim_{\epsilon \to 0} \left(  \epsilon Z_{g}^{(2)} \right)
\end{eqnarray}
In our case we will calculate the $\beta_{0}$ function for $g$-coupling with help of the (\ref{ZA}), (\ref{Zpsi2}) and (\ref{Z2}).
\begin{eqnarray}
Z_{g} &=& 1 + K_{2} - \Delta Z_{\psi} - \frac{1}{2} \Delta Z_{A} + \ldots \\
\beta \left( g \right) &=&\left( - \frac{11}{3} C_{1} + \frac{4}{3} \overline{C}_{2} + \frac{1}{6} C_{2} \right) \frac{g^3}{16 \pi^2} + \ldots
\end{eqnarray}

For other couplings the renormalization constants are described by matrices with non-zero mixing terms. For a general coupling constant $f_{\alpha}$ relationships between bare and renormalised quantities can be simply written in the form:
\begin{eqnarray}
(f_{\alpha})_{B} = \sum_{\beta} \mu ^{- \omega} Z_{\alpha \beta} \times (f_{\beta}) _{R} 
\label{f_alpha}
\end{eqnarray}
where $\omega = \epsilon$ for quadrilinear coupling constant and $\omega = \frac{\epsilon}{2}$ for the Yukawa and gauge coupling. As before, with the differentiation of (\ref{f_alpha}) we will get an expression for the beta function. But now the renormalization constant depends in general on all the couplings from the model we are considering. So we obtain a more complicated result.\\
We will skip the $R$ index to make the expressions shorter.
\begin{eqnarray}
\mu \frac {d \, f_{\beta}}{d \mu} & = & \sum _{\alpha, \gamma} \left(X ^{-1} \right) _{\beta \alpha} 
\left(
- \omega Z_{\alpha \gamma} f_{\gamma} 
\right) \nonumber\\
X_{\alpha \beta} & = &  Z_{\alpha \beta} + \sum_{\gamma}\frac{\partial Z_{\alpha \gamma}}{\partial f_{\beta}} f_{\gamma}
\end{eqnarray}
Then one can expand the $Z_{\alpha \beta}$ matrix as a delta function with a small correction
\begin{eqnarray}
Z_{\alpha \beta} & = & \delta _{\alpha \beta} + \Delta Z_{\alpha \beta} \nonumber\\
X_{\alpha \beta} & = & \delta _{\alpha \beta} + \Delta Z_{\alpha \beta} + \sum_{\gamma}\frac{\partial \Delta Z_{\alpha \gamma}}{\partial f_{\beta}} f_{\gamma} 
\end{eqnarray}
For small values of $\Delta Z_{\alpha \beta} $ we can easily write the inverse matrix of $X _{\alpha \beta}$
\begin{eqnarray}
\left( X ^{-1} \right) _{\alpha \beta} & = & \delta _{\alpha \beta} - \Delta Z_{\alpha \beta} - \sum_{\gamma} \frac{\partial \Delta Z_{\alpha \gamma}}{\partial f_{\beta}} f_{\gamma} 
\end{eqnarray}
We consider first two terms expanding in $\Delta Z_{\alpha \beta}$ and $\epsilon$ with the result of
\begin{eqnarray}
\beta (f_{\beta}) = \lim_{\epsilon \to 0} \mu \frac {d \, f_{\beta}}{d \mu}  =  
\lim_{\epsilon \to 0} \left(
- \omega f_{\beta} + \sum_{\gamma ,\mu}\omega \frac{\partial \Delta Z_{\beta \gamma}}{\partial f_{\mu}}  f_{\gamma} f_{\mu} \right)
\end{eqnarray}

We will first consider $\beta (\kappa^{i}_{m n})$. Using \ref{ZetKappa} one can write
\begin{eqnarray}
\beta (\kappa ^{i}_{m n}) = 
\frac{\epsilon}{2} \sum _{i' n' m'} \sum _{i'' n'' m''} 
\frac{\partial \Delta \tilde{Z}_{i m n}^{i' m' n'}}{\partial \kappa ^{i''}_{m'' n'' }} \kappa ^{i'}_{m' n'} \kappa ^{i''}_{m'' n''} 
+ \frac{\epsilon}{2} \sum _{i' n' m'} 
\frac{\partial \Delta \tilde{Z}_{i m n}^{i' m' n'}}{\partial g} \kappa ^{i'}_{m' n'} g \nonumber\\
\end{eqnarray}
where
\begin{eqnarray}
\tilde{Z}_{i m n}^{i' m' n'} &=&
\sum _{a b c} 
\left(  Z^{-1/2}_{\psi} \right)^{*} _{b m}
\left(  Z^{-1/2}_{\psi} \right)_{c n}
\left(  Z^{-1/2}_{\phi} \right)_{a i} 
(\delta_{a i'} \delta_{b n'} \delta_{c m'}  + K_{a b c}^{i' m' n'}) \nonumber\\
&=&  \delta_{i i'} \delta_{n n'} \delta_{m m'} + K_{i m n}^{i' m' n'}
 - \frac{1}{2} \left( \Delta Z _{\psi} \right)_{n' n} \delta_{m m'} \delta_{i i'} \nonumber\\
& & - \frac{1}{2} \left( \Delta Z _{\psi} \right)^{*}_{m' m} \delta_{n n'} \delta_{i i'}
- \frac{1}{2} \left( \Delta Z _{\phi} \right)_{i i'} \delta_{n n'} \delta_{m m'} + \ldots
\end{eqnarray}
Below summation over repeated indices is assumed.
\begin{eqnarray}
&& \Delta \tilde{Z}_{i m n}^{i' m' n'} \kappa ^{i'}_{m' n'} =
\left( \overline{\textbf{T}}^a  \kappa ^{i} \overline{\textbf{T}}^a \right) _{m n}
\frac{2 g^2 (-4 + \eta)}{16 \pi^2 \epsilon}
+  \left( \kappa ^{j} \kappa ^{i} \kappa ^{j} \right) _{m n}
\frac{2}{16 \pi ^2 \epsilon} \nonumber\\
& & 
+ 
\left( \kappa ^{j} _{m n'} \overline{\textbf{T}}^{a} _{n' n}  -  \overline{\textbf{T}}^{a} _{m n'} \kappa ^{j} _{n' n}
\right) \textbf{T}^{a}_{ij} \frac{2 g^2 (- 1 + \eta)}{16 \pi^2 \epsilon}
- \frac{1}{2} \left( - \frac{2 g^2 \overline{C}_{3} (1-\eta)}{16 \pi^2 \epsilon} \kappa ^{i}_{m n} \right)
\nonumber\\
& & 
- \frac{1}{2} \left( - \frac{\left(\kappa^{j} \kappa^{j} \right)_{n' n} }{16 \pi^2 \epsilon} \kappa^{i}_{m n'} \right)
- \frac{1}{2} \left( - \frac{2 g^2 \overline{C}_{3} (1-\eta)}{16 \pi^2 \epsilon} \kappa ^{i}_{m n} \right)
- \frac{1}{2} \left( - \frac{\left(\kappa^{j} \kappa^{j}\right)_{m m'}}{16 \pi^2 \epsilon} \kappa^{i}_{m' n} \right)
\nonumber\\
& & 
- \frac{1}{2} \left( - \frac{4 \textbf{Tr} ( \kappa^{i} \kappa^{i'} )}{16 \pi^2 \epsilon} \kappa^{i'} _{m n} \right)
- \frac{1}{2} \left( \frac{g^2 (4+ 2\eta) C_{3}}{16 \pi^2 \epsilon} \kappa^{i} _{m n} \right)
\nonumber\\
\end{eqnarray}
\begin{eqnarray}
&& 16 \pi^2 \beta (\kappa ^{i}_{m n}) = 
2 g^2 (-4 + \eta) \left( \overline{\textbf{T}}^a  \kappa ^{i} \overline{\textbf{T}}^a \right) _{m n}
+ 2 \left( \kappa ^{j} \kappa ^{i} \kappa ^{j} \right) _{m n}
+ 2 g^2 \overline{C}_{3} (1-\eta) \kappa ^{i}_{m n} \nonumber\\
& & 
+ 2 g^2 (- 1 + \eta)
\left( \kappa ^{j} _{m n'} \overline{\textbf{T}}^{a} _{n' n}  -  \overline{\textbf{T}}^{a} _{m n'} \kappa ^{j} _{n' n}
\right) \textbf{T}^{a}_{i j} 
+ \frac{1}{2} \left( \kappa^{i} \kappa^{j} \kappa^{j} +  \kappa^{j} \kappa^{j} \kappa^{i} \right)_{m n}
\nonumber\\
& &
+ 2 \textbf{Tr} ( \kappa^{i} \kappa^{i'} ) \kappa^{i'} _{m n}
- g^2 (2+ \eta) C_{3} \kappa^{i} _{m n}
\label{row1}
\end{eqnarray}
Now using (\ref{tozsamosc_kapp}) one can find the following two relations
\begin{eqnarray}
& & 
C_{3} \delta _{i j} \kappa^{j} _{m n} = \textbf{T}^{a}_{i i'} \textbf{T}^{a}_{i' j} \kappa^{j} _{m n} =
 2 \overline{C}_{3} \kappa^{i} _{m n} - 2 \left( \overline{\textbf{T}}^a  \kappa ^{i} \overline{\textbf{T}}^a \right) _{m n}
\\ & &
\left( \kappa ^{j} _{m n'} \overline{\textbf{T}}^{a} _{n' n}  -  \overline{\textbf{T}}^{a} _{m n'} \kappa ^{j} _{n' n}
\right) \textbf{T}^{a}_{i j} =  2 \overline{C}_{3} \kappa^{i} _{m n} - 2 \left( \overline{\textbf{T}}^a  \kappa ^{i} \overline{\textbf{T}}^a \right) _{m n}
\end{eqnarray}
Substituting those results to (\ref{row1}) one can get
\begin{eqnarray}
16 \pi^2 \beta (\kappa ^{i}_{m n}) &=& 
- 6 g^2 \overline{C}_{3} \kappa ^{i}_{nm} 
+ 2 \left( \kappa ^{j} \kappa ^{i} \kappa ^{j} \right) _{n m}  \nonumber\\
& & 
+ \frac{1}{2} \left( \kappa^{i} \kappa^{j} \kappa^{j} +  \kappa^{j} \kappa^{j} \kappa^{i} \right)_{n m}
+ 2 \textbf{Tr} ( \kappa^{i} \kappa^{j} ) \kappa^{j} _{n m}
\label{kappa_result}
\end{eqnarray}

Now we will consider $\beta (h_{i j k l})$. Using \ref{ZetH} one can write: \nopagebreak
\begin{eqnarray}
\beta (h_{i j k l}) &=& 
\epsilon \sum _{i' j' k' l'} \sum _{i'' j'' k'' l''} 
\frac{\partial \Delta \tilde{Z}_{i j k l}^{i' j' k' l'}}{\partial h_{i'' j'' k'' l''}} h_{i' j' k' l'} 
h_{i'' j'' k'' l''}
\nonumber\\ & & 
+
\frac{\epsilon}{2} \sum _{i' n' m'} \sum _{i'' n'' m''} 
\frac{\partial \Delta \tilde{Z}_{i j k l}^{i' j' k' l'}}{\partial \kappa ^{i''}_{m'' n'' }} \kappa ^{i''}_{m'' n''}  \, h_{i' j' k' l'}  
+ \frac{\epsilon}{2} \sum _{i' n' m'} 
\frac{\partial \Delta \tilde{Z}_{i j k l}^{i' j' k' l'}}{\partial g}   g  h_{i' j' k' l'} \nonumber\\
\end{eqnarray}
where
\begin{eqnarray}
\tilde{Z}_{i j k l}^{i' j' k' l'}&=&
\sum _{a b c d} 
\left(  Z^{-1/2}_{\phi} \right)_{i a}
\left(  Z^{-1/2}_{\phi} \right)_{j b}
\left(  Z^{-1/2}_{\phi} \right)_{k c} 
\left(  Z^{-1/2}_{\phi} \right)_{l d} 
(\delta_{a i'} \delta_{b j'} \delta_{c k'} \delta_{d l'} + L_{a b c d}^{i' j' k' l'}) = \nonumber\\
& = &  
\delta_{i i'} \delta_{j j'} \delta_{k k'} \delta_{l l'} + L_{i j k l}^{i' j' k' l'}
- \frac{1}{2} \left( \Delta Z _{\phi} \right)_{i i'} \delta_{j j'} \delta_{k k'} \delta_{l l'}
- \frac{1}{2} \left( \Delta Z _{\phi} \right)_{j j'} \delta_{i i'} \delta_{k k'} \delta_{l l'}
\nonumber\\ & &
- \frac{1}{2} \left( \Delta Z _{\phi} \right)_{k k'} \delta_{i i'} \delta_{j j'} \delta_{l l'}
- \frac{1}{2} \left( \Delta Z _{\phi} \right)_{l l'} \delta_{i i'} \delta_{j j'} \delta_{k k'}
+ \ldots
\end{eqnarray}
Below summation over repeated indices is assumed
\begin{eqnarray}
& & \Delta \tilde{Z}_{i j k l}^{i' j' k' l'} h_{i' j' k' l'} =
\frac{3 g^4 B_{ijkl}}{16 \pi ^2 \epsilon} 
- \frac{8 A_{ijkl}}{16 \pi ^2 \epsilon} 
- \frac{4 g^2 (1 - \eta) C_{3} }{16 \pi^2 \epsilon} h_{i j k l}
\nonumber\\ & &
+ \frac{ \dot{\imath} }{16 \pi ^2 \epsilon }
\left( h_{i j i' j'} h_{i' j' k l} + h_{i k i' j'} h_{i' j' j l}+ h_{k j i' j'} h_{i' j' i l} \right)
\nonumber\\ & &
- \frac{1}{2} \frac{(- 4)}{16 \pi ^2 \epsilon} 
(   
\textbf{Tr}( \kappa^{i} \kappa^{i'} ) h_{i' j k l} +
\textbf{Tr}( \kappa^{j} \kappa^{j'} ) h_{i j' k l} +
\textbf{Tr}( \kappa^{k} \kappa^{k'} ) h_{i j k' l} +
\textbf{Tr}( \kappa^{l} \kappa^{l'} ) h_{i j k l'} 
)
\nonumber\\ & &
- \frac{1}{2} \frac{g^2 }{16 \pi^2 \epsilon} \left( 4 + 2 \eta \right) C_{3} h_{i j k l} \times 4
+ \ldots
\end{eqnarray}
Then one gets
\begin{eqnarray}
&& 16 \pi^2 \beta ( h_{i' j' k' l'} ) =
3 g^4 B_{ijkl} 
- 8 A_{ijkl}
+ \left( h_{i j i' j'} h_{i' j' k l} + h_{i k i' j'} h_{i' j' j l}+ h_{k j i' j'} h_{i' j' i l} \right)
\nonumber\\ & &
+ 2 (   
\textbf{Tr}( \kappa^{i} \kappa^{i'} ) h_{i' j k l} +
\textbf{Tr}( \kappa^{j} \kappa^{j'} ) h_{i j' k l} +
\textbf{Tr}( \kappa^{k} \kappa^{k'} ) h_{i j k' l} 
+ \textbf{Tr}( \kappa^{l} \kappa^{l'} ) h_{i j k l'} 
) \nonumber\\ & &
- 12 g^2 C_{3} h_{i j k l}
\end{eqnarray}

Beta functions we have calculated are expressed in terms of general group theory factors. To derive the expressions in particular models further analysis is required. For example, if the gauge group is a group product, like in the Standard Model, it is necessary to modify the results. 

The beta function found here were published for example in \cite{cheng}. We confirm the result and point out the misprint: in equation (2.7) in \cite{cheng} the group theory factor $S_2(S)$ (which in our notation is $C_{3}$) should be replaced by $S_2(F)$ (which in our notation stands for $\overline{C}_{3}$).

\section{Beta functions for the Standard Model and its extension} 

We would like to apply our general result to the Standard Model and the Minimal Standard Model (MSM) cases.

\subsection{Standard Model result} 
   
The SM\footnote{A Lagrangian for the Standard Model can be found in many places in literature, see for example \cite{bailin}, \cite{pokorski}.} has a $U(1) \times SU(2) \times SU(3)$ gauge symmetry. Following \cite{machacek3}, if a gauge group is a direct product $G_{1} \times ... \times G_{N}$ of simple groups with corresponding gauge constants $g_{1},...,g_{N}$ then the group factors we used in our general theory should be replaced as follows:
\begin{eqnarray}
g^2 C_{i}(R) &\longrightarrow & \sum_{n} g_{n}^2 C_{i}(R) \\
g^4 B_{ijkl} &\longrightarrow & \sum_{n,m} g_{n}^2 g_{m}^2 \tilde{B}^{n m}_{ijkl}
\end{eqnarray}

The factor $\tilde{B}^{n m}_{ijkl}$ is expressed by the group generators of different simple groups $\textbf{T}^a_{n},\textbf{T}^b_{m}$ ($n,m$ - simple group indices, $a,b$ - indices numbering the generators of each group)
\begin{eqnarray}
\tilde{B}^{n m}_{ijkl} &=& \sum \lbrace \textbf{T}^a_{n} , \textbf{T}^b_{m} \rbrace _{i,j} \lbrace \textbf{T}^a_{n} , \textbf{T}^b_{m} \rbrace _{k,l} +  \lbrace \textbf{T}^a_{n} , \textbf{T}^b_{m} \rbrace _{i,k} \lbrace \textbf{T}^a_{n} , \textbf{T}^b_{m} \rbrace _{j,l} + \nonumber\\
& &  \lbrace \textbf{T}^a_{n} , \textbf{T}^b_{m} \rbrace _{i,l} \lbrace \textbf{T}^a_{n} , \textbf{T}^b_{m} \rbrace _{j,k}
\end{eqnarray}

Second problem that occurs while adapting the general result to the Standard Model case is that left- and right-handed fermion fields attribute to different gauge group representations. In the SM couplings we have an additional operator $P_{L}$ or $P_{R}$ of chiral projections. If we'd like to repeat our calculations in the case of right- or left-handed fields, then there occur some additional factors. For example while integrating over a closed fermion loop, there is an additional factor $\frac{1}{2}$ from the projections, so one has to be very careful.

While calculating the final result, we will confine ourselves to the most relevant SM constants: gauge couplings $g_1, g_2, g_3$, top quark Yukawa coupling $y_t$ and quadrilinear Higgs coupling $\lambda$. We will also skip parts of the beta functions calculations, analysing only the group theory factors. 

For the SU(N) we have $C_{1}^{SU(N)} = N$ and  $C_{2}^{SU(N)}(R_{F}) = \frac{1}{2}$ for a fundamental representation $R_{F}$. For U(1) the $C_{1}^{U(1)} = 0$. To calculate  $C_{2}^{U(1)}$ one needs to add the squares of scalar hypercharges, and for the $\overline{C}_{2}^{U(1)}$  the fermion hypercharges. In all these calculations we need to remember that there are 3 generations of fermions and 3 colours of quarks. 

The $C_{3}$ factor in the beta function for the quartic coupling contributes only from U(1) and SU(2). Once again we add the squares of hypercharges in a case of U(1) symmetry, and the $\textbf{T}^a \textbf{T}^a$ for the SU(2), where the generators are half the Pauli matrices $\textbf{T}^a = \frac{1}{2} \sigma^{a}$
\begin{eqnarray}
C_{3}^{U(1)} = \frac{1}{4}, \, \,
C_{3}^{SU(2)} = \frac{3}{4}
\end{eqnarray}

The $\overline{C}_{3}$ factor in the beta function for the Yukawa coupling can be easily calculated in case of SU(2) and SU(3).
\begin{eqnarray}
\overline{C}_{3}^{SU(2)} = \frac{3}{4}, \, \,
\overline{C}_{3}^{SU(3)} = \frac{4}{3}
\end{eqnarray}
For the U(1) gauge symmetry one has to consider only the hypercharges of the top quark left- and right- handed part, which give a result
\begin{eqnarray}
\overline{C}_{3}^{U(1)} = \frac{1}{2} \left(\left(\frac{1}{6}\right)^2 + \left(\frac{2}{3}\right)^2 \right) = \frac{1}{6} \frac{17}{12}
\end{eqnarray}

Now we can present final expressions for the SM 1-loop beta functions:
\begin{eqnarray}
16 \pi^2 \beta(\lambda) & = & \frac{3}{8}  g_{1}^4 + \frac{9}{8} g_{2}^4 + \frac{3}{4} g_{1}^2 g_{2}^2 
- 6 y_{t}^4 + 24 \lambda^2
+12 y_{t}^2 \lambda -3 g_{1}^2 \lambda - 9 g_{2}^2 \lambda \nonumber\\ \\
16 \pi^2 \beta(g_{1}) & = & \frac{41}{6} g_{1}^3 \\
16 \pi^2 \beta(g_{2}) & = & -\frac{19}{6} g_{2}^3 \\
16 \pi^2 \beta(g_{3}) & = & -7 g_{3}^3 \\
16 \pi^2 \beta(y_{t}) & = & \left( -\frac{17}{12} g_{1}^2 -\frac{9}{4} g_{2}^2 - 8 g_{3}^2 \right) y_{t} + \frac{9}{2} y_{t}^3
\end{eqnarray}

The results agree with those from the literature, see e.g. \cite{pirogov}

\subsection{Standard Model plus scalar singlets}

We'd like to consider now a model with additional scalar singlet fields. The general scalar potential with the SM doublet of scalars $H$ and $N_{\phi}$ scalar singlets $\phi_{i}$ is:
\begin{eqnarray}
V(H, \phi_{n}) = -m^2 H ^\dagger H + \lambda (H ^\dagger H)^2 + \frac{1}{2}  \sum _{i} ^{N_{\phi}} \mu ^{i}_{\phi} \phi _ i ^2 + \sum _{i,j}^{N_{\phi}} \lambda_{\phi}^{ij} \phi ^2 _{i} \phi ^2 _{j} +
\sum _{i}^{N_{\phi}} \lambda_{x}^{i} (H ^\dagger H) \phi _{i} ^ 2 \nonumber\\
\end{eqnarray}

All the previously mentioned problems occur here as well. Additional calculations to make are rather simple and do not require a special comment.
Resulting scalar sector beta functions for the SM with $N_{\phi}$ scalar singlets are:
\begin{eqnarray}
16 \pi^2 \beta(\lambda) & = & \frac{3}{8}  g_{1}^4 + \frac{9}{8} g_{2}^4 + \frac{3}{4} g_{1}^2 g_{2}^2 
- 6 y_{t}^4 + 24 \lambda^2
+12 y_{t}^2 \lambda -3 g_{1}^2 \lambda - 9 g_{2}^2 \lambda \nonumber\\
& & + 2 N_{\phi} \lambda_{x}^2 \\
16 \pi^2 \beta(\lambda_{\phi}) & = &  (64 + 8 N_{\phi}) \lambda_{\phi} ^2 + 2 \lambda_{x} ^2 \\
16 \pi^2 \beta(\lambda_{x}) & = &  12 \lambda \lambda_{x} + 24 \lambda_{\phi} \lambda_{x} + 8 \lambda_{x}^2 + 6 y_{t}^2 \lambda_{x} - \frac{3}{2} g_{1}^2 \lambda_{x} - \frac{9}{2} g_{2}^2 \lambda_{x}
\end{eqnarray}

The above results agree with \cite{davoudiasl}.

\subsection{Right-handed neutrinos} \label{right_neutrinos_sec}
 
After adding singlet scalar fields to the theory, it is very natural to include also right-handed Majorana  neutrino singlets (see \cite{casas} or \cite{akhmedov}) and their couplings to scalar singlets:

\begin{eqnarray}
L_{\nu} = - \frac{1}{2} \overline{(\nu_{R})^{c}} Y_{\phi} \nu_{R} \phi  + h.c.
\end{eqnarray}
where $(\,)^c$ denotes the charge conjugation operator acting on a fermion field. 

The coupling $Y_{\phi}$ contributes to the $\beta \left( \lambda_x \right)$ and $\beta \left( \lambda_{\phi} \right)$. To calculate those corrections we need to consider a scalar singlet propagator correction from right neutrinos.

\renewcommand{\figurename}{Diagram}
\begin{figure}
  \centering
  \includegraphics[height = 2 cm]{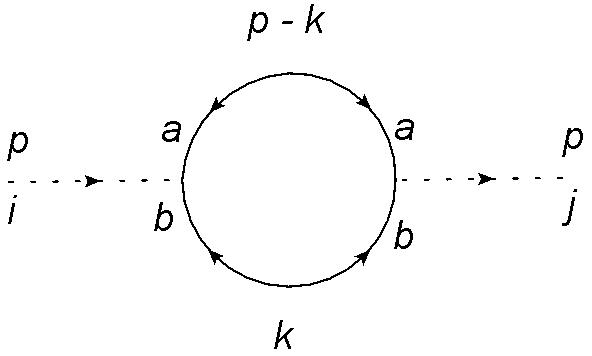}
  \caption{Self energy corrections to the scalar propagator from Majorana fermion}
  \label{2point_scalar_majorana_diag}
\end{figure} 
\renewcommand{\figurename}{Figure}

For diagram \ref{2point_scalar_majorana_diag} we need to include a standard combinatorial factor 1/2 for such loop with self-conjugate particles. A $(-1)$ factor originates from a fermion loop. Feynman rules for the Majorana neutrinos can be found in Appendix C.

\begin{eqnarray}
\texttt{diagram \ref{2point_scalar_majorana_diag}} &=& - \frac{1}{2} \sum_{a b}\int \frac{d^4 k}{(2 \pi)^4} Tr \left(
  \left( - \tilde{S}_{F} (p - k) \hat{C}  \right) (-i) \hat{C}  Y_{i} ^{a b}
 \left( - \tilde{S}_{F} (k) \hat{C}  \right) (- i ) \hat{C} Y_{j} ^{a b}
 \right) \nonumber\\
&=&  \frac{2 i p^2 Tr(Y_{i} Y_{j})}{16 \pi^2 \epsilon}
\end{eqnarray}
where we use the fact that $\hat{C}^2 = 1$. \\

This diagram contributes to the general beta function formula in the following way:
\begin{eqnarray}
16 \pi^2 \beta(h_{ijkl}) &=& \ldots + \left( Tr( Y_{i} Y_{i'} ) h_{i'jkl} +
 Tr(Y_{j} Y_{j'}) h_{ij'kl}+  Tr(Y_{k} Y_{k'}) h_{ijk'l} \right. \nonumber\\ && \left.
 +  Tr(Y_{l} Y_{l'}) h_{ijkl'}
\right)
\end{eqnarray}

Now one can calculate the contribution to the $\lambda_x$ and $\lambda_\phi$ beta function for the one singlet SM extension (we also include top-quark Yukawa interaction contribution for comparison)
\begin{eqnarray}
& & 16 \pi^2 \beta ( \lambda_x )=
4 (   
3 \times \textbf{Tr}  ( Y_{t} Y_{t}  )   \lambda_x ) 
+ 2 \textbf{Tr}  ( Y_{\phi} Y_{\phi} ) \lambda_x
+ \ldots \\
& & 16 \pi^2 \beta(\lambda_{\phi}) = 4 \textbf{Tr} ( Y_{\phi} Y_{\phi} ) \lambda_{\phi} + \ldots
\end{eqnarray}

To have a full $\beta(\lambda_{\phi})$ from the right neutrino coupling one has to consider also 1-loop correction to the $\phi^4$ vertex. As we do not need the $\beta(\lambda_{\phi})$ for our purposes, we will skip this calculation.

\section{1-loop quadratic divergences in a generic gauge
   theory}   
   
\renewcommand{\figurename}{Figure}
\begin{figure}
  \centering
  \includegraphics[height = 3 cm]{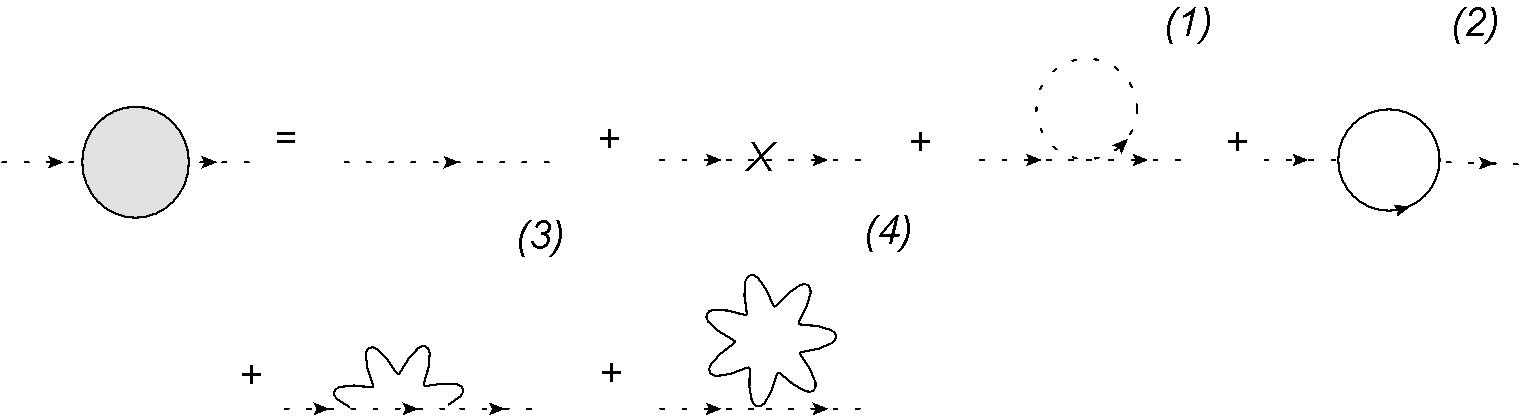}
  \caption{Scalar particle 1-loop self-energy corrections for a generic gauge theory with fermions, which contain quadratic divergence.}
  \label{1loop_propagator}
\end{figure}

In this section we will find the quadratically divergent contributions to scalar 2-point Green's function in a general gauge theory with scalar and fermion fields (as introduced in section \ref{general_theory}). We will adopt the cut-off regularization (see the Appendix B for necessary integrals). For all the loops we assume the same cut-off $\Lambda$ and we keep the $\Lambda$ contributions and $\log (\Lambda)$ for scalar loops, as they will be relevant in later discussion. Below there are mentioned only the diagrams that contribute in this regularization.

Below we list all the contributions from diagrams in figure \ref{1loop_propagator}.\\

$\texttt{Diagram 1}$: symmetry factor $\frac{1}{2}$
\begin{eqnarray}
- \frac{1}{2}  \int \frac{d^4 k}{(2 \pi)^4} \frac{i}{k^2 - m_{i'}^2} i h_{i j i' i'} = 
- \frac{1}{2} h_{i j i' i'} \frac{i}{16 \pi^2} \left( \Lambda ^2 - m_{i'}^2 \log \left( \frac{m_{i'}^2 + \Lambda^2}{m_{i'}^2}  \right)\right) \nonumber\\
\end{eqnarray}

$\texttt{Diagram 2}$: symmetry factor  1, (-1) factor from a fermion loop
\begin{eqnarray}
(-1) \int \frac{d^4 k}{(2 \pi)^4} 
Tr \left( i \kappa ^i _{m n} \tilde{S}_{F} (p + k) i \kappa ^j _{n m} \tilde{S}_{F} (k) \right) = Tr (\kappa^i \kappa^j) \frac{4 i \Lambda ^2}{16 \pi^2}
\end{eqnarray}

$\texttt{Diagram 3}$: symmetry factor 1, summing over gauge fields
\begin{eqnarray}
- g^2 T^a_{i i'}  T^a_{i' j} 
\int \frac{d^4 k}{(2 \pi)^4} (2p - k)_{\mu}  \tilde{D}_{F} (p-k)  (2p - k)_{\nu} \tilde{D}_{F}^{\mu \nu} (p-k) =
g^2 ( T^a T^a ) _{i j} \frac{i \Lambda ^2}{16 \pi^2} (1 - \eta) \nonumber\\
\end{eqnarray}

$\texttt{Diagram 4}$: symmetry factor $\frac{1}{2}$, 
\begin{eqnarray}
\frac{1}{2}  \int \frac{d^4 k}{(2 \pi)^4} i g^2 g_{\mu \nu} 2 (T^a T^a)_{i j} \tilde{D}_{F}^{\mu \nu}
= g^2 ( T^a T^a ) _{i j} \frac{i \Lambda ^2}{16 \pi^2} (-4 + \eta) 
\end{eqnarray}

Now one can write an expression for a 1-loop correction to the scalar particle mass in generic gauge theory (summing over primed indices):
\begin{eqnarray}
\delta m^2 _{ij} =
\frac{1}{16 \pi^2} \left(
- \frac{1}{2} h_{i j i' i'}  \left( \Lambda ^2 - m_{i'}^2 \log \left( \frac{m_{i'}^2+\Lambda^2}{m_{i'}^2}  \right)\right)  + 4 Tr (\kappa^i \kappa^j) \Lambda^2 - 3 g^2 ( T^a T^a ) _{i j} \Lambda ^2
\right) \nonumber\\
\label{general_1loop}
\end{eqnarray}

\subsection{Standard Model with scalar singlets case}

We'd like to calculate a 1-loop correction to the Higgs mass in a case of a SM Higgs doublet and $N_{\phi}$ singlet scalar fields (for the potential see equation (\ref{scalar_potential})) with the common mass $m_{\phi}$.

Using  $m_{h}^2 = - \mu^2 + 3 \lambda v^2 = 2 \mu^2 $ (where $v$ is the vacuum expectation value of the Higgs field) and (\ref{general_1loop}) one can calculate the Higgs boson mass correction
\begin{eqnarray}
& & \delta m^2_{h} = 
\frac{\Lambda ^2}{16 \pi^2} \left(
12 \lambda + 2 N_{\phi} \lambda_x - 12 y_t^2 + \frac{3}{2} g_1^2 + \frac{9}{2} g_2^2
\right) \nonumber\\
& & - \frac{1}{16 \pi^2} \left(
6 \lambda m_{h}^2 \log \left( \frac{m_{h}^2+\Lambda^2}{m_{h}^2} \right)+
2 \lambda \sum_{I = 1,2,3} m_{I}^2 \log \left( \frac{m_{I}^2+\Lambda^2}{m_{I}^2}  \right) 
\right. \nonumber\\ && \left. 
+ 2 \lambda_{x} N_{\phi} m_{\phi}^2 \log  \left( \frac{m_{\phi}^2+\Lambda^2}{m_{\phi}^2}  \right) \right)
\end{eqnarray}
where $m_{I}$ stands for the masses of Goldstone bosons and $m_{\phi}$ is the mass of singlet scalar fields 
$m_{\phi}^2 = \mu _{\phi} ^2 + \lambda_{x} v ^2$

\section{Leading quadratic divergences in higher orders}

\renewcommand{\figurename}{Figure}
\begin{figure}
  \centering
  \includegraphics[height = 3 cm]{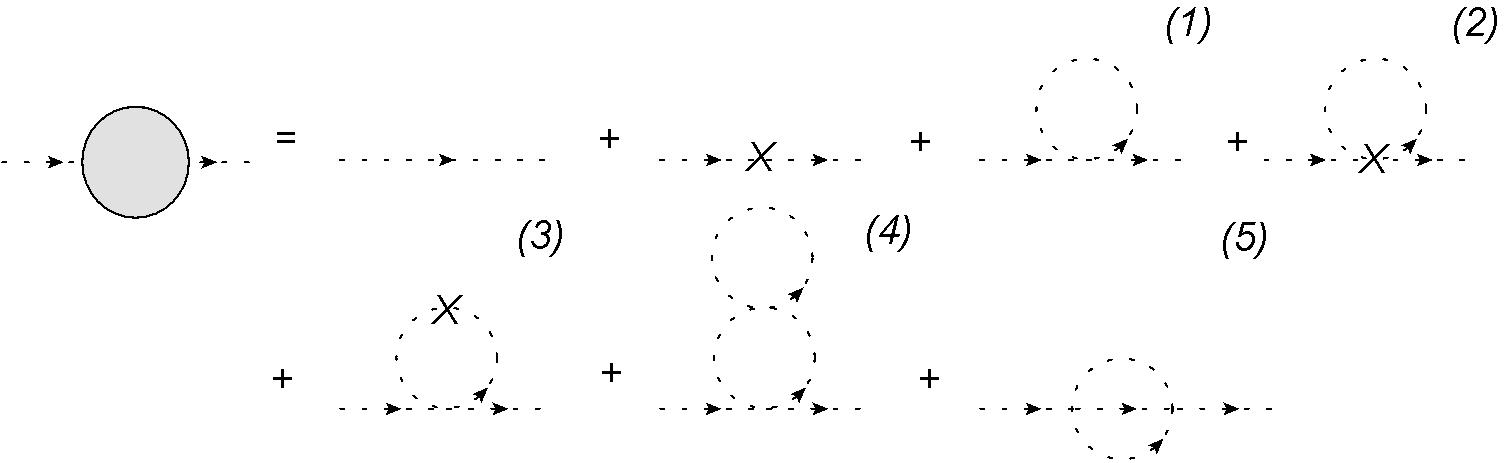}
  \caption{Scalar particle 2-loop self-energy corrections from scalar quartic couplings that contribute to quadratic divergences. The cross stands for the 1-loop counterterm.}
  \label{2loop_propagator}
\end{figure}

In this chapter we'd like to show how to calculate quadratic divergences in two ways. As in previous chapter, we're interested in divergences within general gauge theory with scalar and fermion fields, in a cut-off regularization scheme. We mention only the diagrams that contribute in cut-off regularization scheme.

\subsection{2-loop Higgs effects in a generic theory}

The most common approach to calculate 2-loop divergences is a straightforward computation with help of Feynman diagrams. In the figure \ref{2loop_propagator} we drew contributing diagrams in a 2-loop calculation that originate quartic scalar coupling.

$\texttt{Diagram 1}$: symmetry factor $\frac{1}{2}$\\
\begin{eqnarray}
\frac{1}{2}  \int \frac{d^4 k}{(2 \pi)^4} \frac{i}{k^2} i h_{i j i' i'}
= \frac{i \Lambda ^2}{16 \pi^2} \frac{1}{2} h_{i j i' i'}  
\end{eqnarray}

$\texttt{Diagram 2}$: symmetry factor $\frac{1}{2}$\\
\begin{eqnarray}
\frac{1}{2}  \int \frac{d^4 k}{(2 \pi)^4} \frac{i}{k^2} \left(i \Delta h_{i j i' i'} \right) 
= \frac{i \Lambda ^2}{16 \pi^2} \frac{1}{2}  \left(i \Delta h_{i j i' i'} \right) 
\end{eqnarray}

$\texttt{Diagram 3}$: symmetry factor $\frac{1}{2}$\\
\begin{eqnarray}
\frac{1}{2}  \int \frac{d^4 k}{(2 \pi)^4} \frac{i}{k^2} \left( i \Delta Z_{i' j'} \right) \frac{i}{k^2} i h_{i j i' j'}
= \frac{i}{16 \pi^2} \frac{1}{2} \Delta Z_{i' j'} h_{i j i' j'} \log  (\frac{\Lambda ^2}{m_{i} m_{j}})
\end{eqnarray}

$\texttt{Diagram 4}$: symmetry factor $\frac{1}{4}$\\
\begin{eqnarray}
\frac{1}{4}  \int \frac{d^4 k}{(2 \pi)^4} \int \frac{d^4 q}{(2 \pi)^4} \frac{i}{k^2} \frac{i}{k^2} \frac{i}{q^2}
i h_{i j i' j'} i h_{i' j' i'' i''}
= \frac{i}{(16 \pi^2)^2} \frac{1}{4}  h_{i j i' j'}  h_{i' j' i'' i''} \Lambda ^2 \log  (\frac{\Lambda ^2}{m_{i'} m_{j'}})  \nonumber\\
\end{eqnarray}

$\texttt{Diagram 5}$: symmetry factor $\frac{1}{6}$\\
\begin{eqnarray}
\frac{1}{6}  \int \frac{d^4 k}{(2 \pi)^4} \int \frac{d^4 q}{(2 \pi)^4} \frac{i}{k^2} \frac{i}{(k - q)^2} \frac{i}{q^2}
i h_{i i' j' k'} i h_{j i' j' k'}
= \frac{1}{(16 \pi^2)^2} \frac{i \Lambda ^2}{3} h_{i i' j' k'}  h_{j i' j' k'} \nonumber\\
\end{eqnarray}

From the results above we can determine the 1-loop counterterms:
\begin{eqnarray}
& & \Delta Z_{i j} = - \frac{1}{2} h_{i j i' i'} \frac{\Lambda ^2}{16 \pi^2}  \\
& & \Delta h_{i j k l} = - \frac{1}{2} \frac{1}{16 \pi^2} \log  (\frac{\Lambda ^2}{m_{i} m_{j}}) 
\left( h_{i j i' j'} h_{k l i' j'} + h_{i k i' j'} h_{j l i' j'} + h_{i l i' j'} h_{j k i' j'}
\right)\nonumber\\
\end{eqnarray}

And write the final result of the mass correction leading scalar contributions
\begin{eqnarray}
\texttt{1-loop correction} &=& - \frac{\Lambda ^2}{16 \pi^2} \frac{1}{2} h_{i i i' i'}  \label{result_1loop} \\
\texttt{2-loop correction} &=&  - \frac{1}{(16 \pi^2)^2} \Lambda ^2 \log  \left( \frac{\Lambda^2}{m^2} \right)
\frac{1}{4} \left( 
h_{i j i' j'} h_{i' j' k' k'} + 2 h_{i i' j' k'} h_{j i' j' k'}
\right) \nonumber\\
& & + 
\frac{\Lambda ^2}{(16 \pi ^2)^2} \frac{1}{3}  h_{i i' j' k'} h_{j i' j' k'} \label{2_loop_scalar_coefficent}
\end{eqnarray}

One can neglect the result proportional to the $\frac{\Lambda ^2 h^2}{(16 \pi ^2)^2}$ as small in comparison to the 1-loop term\footnote{In the SM with singlets the ratio of the $\Lambda ^2$ (no $\log  (\Lambda)$ ) term in 2-loop correction and the 1-loop correction is $\frac{0.15 \lambda^2 + 0.05 \lambda_{x}^2}{6 \lambda + \lambda_{x}}$ which for $\lambda \sim 1$ and $\lambda_{x} < 5$ is negligible.}.

Alternatively, one can obtain the leading higher order quadratic divergences indirectly, with some help of beta functions. Following \cite{einhorn}, in a theory with many couplings $\lambda_{i}$ the leading (containing the highest power of $\log  (\Lambda)$ ) quadratic divergences can be written as
\begin{eqnarray}
\delta m^2 = \Lambda^2 \sum_{n = 0} ^{\infty} f_{n} (\lambda_{i}) \log ^{n} \left( \frac{\Lambda}{\mu}\right) + \ldots
\label{einhorn_eq}
\end{eqnarray}
where $n+1$ is the number of loops considered, $\mu$ is the renormalization scale and the coefficients $f_{n}$ satisfy
\begin{eqnarray}
(n+1)f_{n+1} = \mu \frac{\partial}{\partial \mu} f_{n} = \beta_{i} \frac{\partial}{\partial \lambda_{i}} f_{n}
\label{recursion}
\end{eqnarray}

This method allows to determine only terms proportional to the $\Lambda^2 \log ^{n} \left( \frac{\Lambda}{\mu}\right)$. Terms with the logarithm power less than $n$ are not controlled within this method. The results (\ref{result_1loop}) and (\ref{2_loop_scalar_coefficent}) could be used to verify the recursion  (\ref{recursion}). With $f_{0}$ from (\ref{result_1loop}) and the beta function for $h_{ijkl}$ we get:
\begin{eqnarray}
& & f_{0} = - \frac{1}{16 \pi^2} \frac{1}{2} h_{iji'i'} + ... \\
& & \beta (h_{ijkl}) = ... + \frac{1}{16 \pi^2}( h_{iji'j'}h_{kli'j'} + h_{iki'j'}h_{jli'j'} + h_{ili'j'}h_{jki'j'}) + ... \\
& & f_{1} = \beta (h_{a b c d}) \frac{\partial}{\partial h_{a b c d}} f_{0} = 
- \frac{1}{(16 \pi^2)^2} \frac{1}{2}( h_{ijj'k'}h_{i'i'j'k'} + 2 h_{i i' j' k'}h_{j i' j' k'}) + ...
\end{eqnarray}

which is the same as in (\ref{2_loop_scalar_coefficent}) (watch the form of the logarithm).
In (\ref{2_loop_scalar_coefficent}) and (\ref{recursion}) we use interchangeably $m \longleftrightarrow \mu$ in the logarithm, because the difference from this change is sub-leading.

\subsection{2-loop Higgs mass corrections in the scalar singlets case}
For the SM with a single scalar extension we have
\begin{eqnarray}
f_{0} = \frac{1}{16 \pi ^2} \left( 12\lambda + 2\lambda _{x} - 12 y_{t}^2 
+ \frac{3}{2} g_{1}^2 + \frac{9}{2} g_{2}^2  \right)
\end{eqnarray}

That let us calulate the $f_{1}$ coefficient
\begin{eqnarray}
f_{1} =  \frac{1}{16 \pi ^2} \left(
12 \beta(\lambda)+ 2 \beta( \lambda _{x} ) - 24 y_{t} \beta(y_{t}) + 3 g_{1} \beta(g_{1}) 
+ 9 g_{2} \beta(g_{2}) \right)
\end{eqnarray}

Inserting the beta functions from chapter 2, we obtain:
\begin{eqnarray}
f_{1} &=& \frac{1}{(16 \pi ^2)^2} 
( 25 g_1^4 + 9 g_1^2 g_2^2 - 15 g_2^4 + 34 g_1^2 y_t^2 + 54 g_2^2 y_t^2 + 
 192 g_3^2 y_t^2 - 180 y_t^4 
\nonumber\\ & &
 - 36 g_1^2 \lambda - 108 g_2^2 \lambda + 
 144 y_t^2 \lambda + 288 \lambda^2 - 3 g_1^2 \lambda_x - 
 9 g_2^2 \lambda_x + 12 y_t^2 \lambda_x 
 \nonumber\\& & 
+ 24 \lambda \lambda_x + 40 \lambda_x^2 + 48 \lambda_x \lambda_{\phi}+
4 \lambda_x \textbf{Tr}\left(  Y_{\phi}  Y_{\phi}  \right) )+ \ldots
\label{f1}
\end{eqnarray}
Standard Model result can be easily reproduced by putting all the singlet parameters to zero.

\section{2-loop fine-tuning in the Standard Model}

There are several classical theoretical constraints on the Higgs boson mass: unitarity, triviality, vacuum stability and fine-tuning. For a summary discussion of all these constraints see \cite{kolda}, here we will concentrate on the triviality and the fine-tuning.

\subsection{Triviality bound} \label{Triviality}
A constraint traditionally called 'triviality', is basically a constraint coming from the scale $\Lambda_{\infty}$ at which the value of a theory running parameter tends to infinity. If couplings increase monotonically with the momentum scale (running constants), the theory becomes non-perturbative near the pole (Landau Pole). The name of this effect comes from the fact, that only trivial (non-interacting) theory with vanishing quartic interactions is allowed if one tries to shift location of the pole to infinity. Similar effect is also present in QED.
If the only allowed value for the renormalized charge is zero, theory is called non-interacting or 'trivial'.

While the triviality problem in QED can be considered minor because the Landau pole scale is far beyond any observable energies, the Higgs boson's Landau pole appears for much smaller energies and an acceptable solution is to make sure that the pole is above the value of the SM cut-off. This is used to set the 'triviality bound' on the Higgs mass and the energy scale allowed for the SM. 

To evaluate location of the pole as a function of the Higgs mass, we will use the beta functions for the SM. In general, one has to solve the set of equations for all of the parameters in the SM. For our purposes, we will approximate the result by considering only the evolution of $\lambda$.
\begin{eqnarray} 
\mu \frac{d \lambda}{d \mu} &=& 
\frac{3}{8}  g_{1}^4 + \frac{9}{8} g_{2}^4 + \frac{3}{4} g_{1}^2 g_{2}^2 
- 6 y_{t}^4 + 24 \lambda^2 + 12 y_{t}^2 \lambda -3 g_{1}^2 \lambda - 9 g_{2}^2 \lambda
\label{triv_eq}
\end{eqnarray}  

We need also a specification of the initial conditions and we assume a given value of $\lambda$ at the energy scale $80$ GeV. 
\begin{eqnarray} 
\lambda(\mu = 80 \, \mathrm{GeV}) = \lambda_{0} \label{initial}
\end{eqnarray} 

\begin{figure}
  \centering
  \includegraphics[height = 7.1 cm]{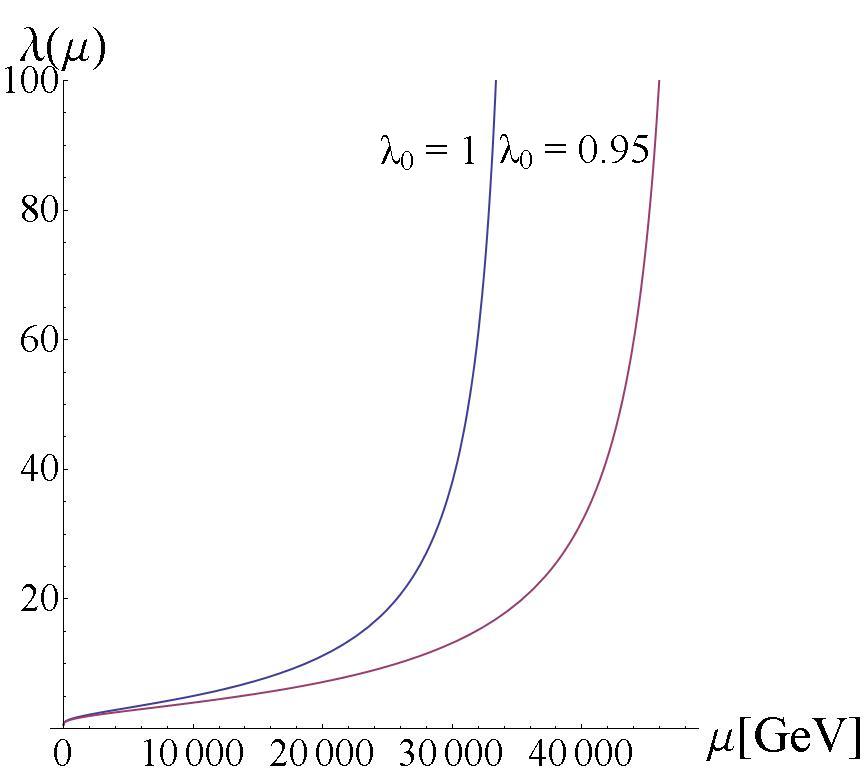} 
  \includegraphics[height = 7.1 cm]{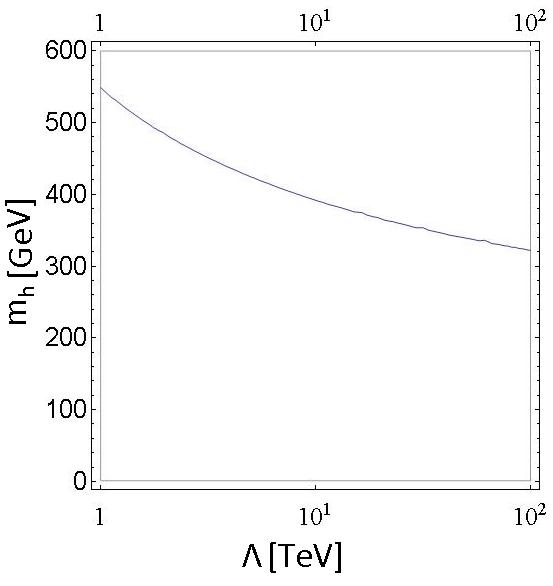}
  \caption{The left panel shows the running of $\lambda$ for $\lambda_{0} = 1$ and $\lambda_{0} = 0.95$. The right panel illustrates the triviality constraint on Higgs mass as a function of the cut-off $\Lambda$ (location of the pole) using 1-loop beta function for $\lambda$.
  }
  \label{SM_running}
\end{figure}

Solutions of (\ref{triv_eq}) for specific values of $\lambda_
0$ are shown in the left panel of fig. \ref{SM_running}.

The condition for the Landau pole $\Lambda_{\infty}$ is the following:
\begin{eqnarray} 
\lambda(\mu) |_{\mu \rightarrow \Lambda_{\infty}} \rightarrow \infty \label{landau_pole}
\end{eqnarray}  

Equation (\ref{landau_pole}) can be solved with respect to $\lambda_{0}$ and then the function $\lambda_{0} \left( \Lambda_{\infty} \right)$ leads to the triviality bound. For each $\lambda_{0}$ we want the Landau pole to be above the value of the SM cut-off, so the values of $\Lambda$ beyond $\Lambda_{\infty}$ are forbidden. The result is shown in the right panel of fig. \ref{SM_running} in terms of the Higgs mass $m_{h} = v \sqrt{2 \lambda_{0}}$.

We obtained the solution $\lambda_{0} \left( \Lambda_{\infty} \right)$ shown in fig. \ref{SM_running} using numerical solving of the differential equation (\ref{triv_eq}) with initial condition (\ref{initial}) in \textit{Wolfram Mathematica 7}. The numerical solution procedure builds a so-called Interpolating Function Grid (see \cite{wolfram}) - a grid of points at which data is specified while solving the differential equation. The algorithm for a sufficiently large sampling range breaks down at a certain value, which in our case is the pole of the function $\lambda(\mu)$. We can extract the value of the breaking point $\Lambda_{\infty}$ from the Interpolating Function Grid for each initial parameter $\lambda_{0}$, which gives us $\lambda_{0} \left( \Lambda_{\infty} \right)$ - the triviality bound. In the language of \textit{Mathematica}, the function looks as follows:

\begin{eqnarray}
& &\texttt{Needs["DifferentialEquations`InterpolatingFunctionAnatomy`"]}; \nonumber\\
& &\Lambda \texttt{infinity[}\lambda_0 \texttt{]} :=
\texttt{Last[InterpolatingFunctionGrid[First[}
\lambda /. \, \texttt{NDSolve[} \lbrace \nonumber\\
& & \,\,\,\,\,\,\,\,\,\,\,\,\,\,\,
\beta \, \texttt{[} \lambda\texttt{[}\mu\texttt{]]} == \mu \, \lambda'\texttt{[}\mu\texttt{]]}, \nonumber\\
& & \,\,\,\,\,\,\,\,\,\,\,\,\,\,\, 
\lambda \, \texttt{[}80\texttt{]} == \lambda_0 
\rbrace, \nonumber\\
& & \,\,\,\,\,\,\,\,\,\,\,\,\,\,\,
\lambda, \lbrace 
\mu, 1,1000000 \rbrace
\texttt{]]]]} \, \texttt{[[}1\texttt{]]};
\end{eqnarray}  
where the number $1000000$ corresponds to the optional value of an upper bound of the sampling range in GeV, $\beta \, \texttt{[} \lambda\texttt{[}\mu\texttt{]]}$ is defined as the RHS of (\ref{triv_eq}). The function $\Lambda \texttt{infinity[}\lambda_0 \texttt{]}$ has to be inverted.

\renewcommand{\figurename}{Figure}

\newpage
\subsection{The fine-tuning}   
As we have mentioned before in the introduction, the fine-tuning is a very precise adjustment of parameters and we would like our theory not to require such procedures. 

The mass of Higgs boson has quadratically divergent corrections. For a large SM cut-off $\Lambda$, the mass of the Higgs particle should be of an order of $\Lambda$. To get an acceptable Higgs masses not larger than $1$ TeV, the self-energy corrections should be cancelled by the counterterms to a relatively small value of the Higgs boson mass\footnote{Radiative corrections for fermion and vector boson masses do not contain quadratic divergences}. If $\Lambda$ is large, the fine-tuning between counterterms and quadratically divergent terms is needed. We would like to avoid such a fine-tuning.

A solution for this problem was proposed at first by Veltman (see \cite{veltman} or \cite{kundu}). If the corrections to the Higgs self-energy at 1-loop accuracy are zero, the fine-tuning problem vanishes at the 1-loop order:  
\begin{eqnarray} 
m_{h}^2 + m_{Z}^2 + 2m_{W}^2 - 4 m_{t}^2 \simeq 0 
\label{veltman_condition}
\end{eqnarray}   

By presenting such condition we assume an underlying theory that can explain the zero value of the divergence coefficient. Such theory may include an additional symmetry and should explain the relationship between the Higgs mass and masses of other particles obtained from (\ref{veltman_condition}).

We'd like to estimate the cut-off $\Lambda$ by requiring the following:
\begin{eqnarray} 
\left| \frac{\delta m_{h}^2}{m_{h}^2} \right| \leq \Delta_{h}
\label{D_SM_1loop}
\end{eqnarray} 

\begin{figure}
  \centering
  \includegraphics[height = 7.5 cm]{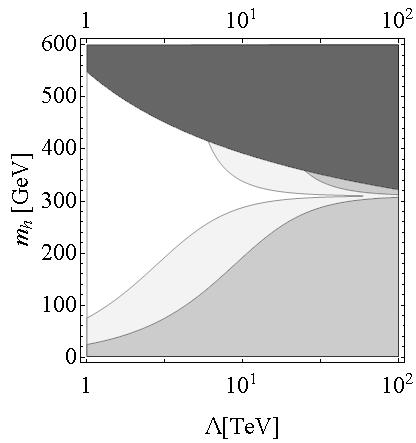}
  \caption{1-loop constraints on $m_{h}$ and $\Lambda$: triviality (black region is excluded) and 1-loop fine-tuning.
  For the fine-tuning $\Delta_{h} = 10$ the white region is allowed.
  For the fine-tuning $\Delta_{h} = 100$ the white and light grey regions are allowed.
  Dark grey region corresponds to $\Delta_{h} > 100$.}
  \label{FT_SM_1loop}
\end{figure}

Knowing the expression for $\delta m_{h}$ at 1-loop accuracy (here we take only the leading $\Lambda^2$ part)
\begin{eqnarray} 
\delta m_{h \, 1loop \, SM}^{2} = \frac{\Lambda^2}{16 \pi^2} 
\left(
12 \lambda - 12 y_{t}^2 + \frac{3}{2} g_{1}^2 + \frac{9}{2} g_{2}^2 
\right)
\end{eqnarray} 
one can impose the condition (\ref{D_SM_1loop}) which gives us a fine-tuning allowed region in a plane $(\lambda, \Lambda)$ for specified values of $\Delta_{h}$. The plot shown in fig. \ref{FT_SM_1loop} was obtained with the help of a simple RegionPlot function (see \cite{wolfram2}) in \textit{Wolfram Mathematica 7}
\begin{eqnarray}
\texttt{RegionPlot[FineTuning[}m_{H}, \Lambda\texttt{]} \leq \Delta_{h} , \lbrace\Lambda, 1000, 100000\rbrace, \lbrace m_{H}, 1, 600 \rbrace \texttt{]}
\end{eqnarray}  
where the numbers $\lbrace\Lambda, 1000, 100000\rbrace$ correspond to the $\Lambda$ range in GeV, $\lbrace m_{H}, 1, 600 \rbrace$ is the $m_{h}$ range also in GeV and the function $\texttt{FineTuning[}m_{H}, \Lambda\texttt{]}$ is the LHS of (\ref{D_SM_1loop}).

The $\Delta_{h} = 0$ is fulfilled for $m_{h} \sim 310$ GeV. One can assume the fine-tuning cancellation to be very precise ($\Delta_{h} \sim 0.1$) or just quite good ($\Delta_{h} \sim 100$). Even the assumption of $\Delta_{h} \sim 100$ is very useful, because it reduces the arbitrariness of Higgs mass choice.

The Veltman condition is sufficient to cancel quadratically divergent contributions to the Higgs mass only at the 1-loop order. A general form of leading higher order contributions, as in equation (\ref{einhorn_eq}) is
\begin{eqnarray} 
m_{h}^2 \longrightarrow m_{h}^2 + \Lambda^2 \sum_{n = 0}^{\infty} f_{n}(\lambda_{i}) \log ^n \left(\frac{\Lambda}{\mu} \right)
\end{eqnarray}  

\begin{figure}
  \centering
  \includegraphics[height = 7.5 cm]{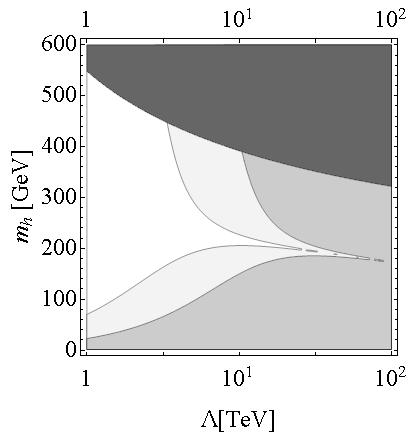}
  \caption{Constraints on $m_{h}$ and $\Lambda$: triviality (black region is excluded) and 2-loop fine-tuning.
  For the fine-tuning $\Delta_{h} = 10$ the white region is allowed.
  For the fine-tuning $\Delta_{h} = 100$ the white and light grey regions are allowed.
  Dark grey region corresponds to $\Delta_{h} > 100$.}
  \label{FT_SM_2loop}
\end{figure}

The coefficient $f_{1}$ for Standard Model can be deduced from (\ref{f1}). We will concentrate on the 2-loop accuracy corrections, because 3-loop corrections are not relevant up to $\sim 50$ TeV scale. 
\begin{eqnarray} 
\delta m^2_{h \,2loops \, SM} &=& \frac{\Lambda ^2}{(16 \pi ^2)^2} \log  \left( \frac{\Lambda}{\mu} \right)
( 25 g_1^4 + 9 g_1^2 g_2^2 - 15 g_2^4 + 34 g_1^2 y_t^2 + 54 g_2^2 y_t^2 + 
 192 g_3^2 y_t^2 \nonumber\\ 
 & & - 180 y_t^4  \lambda - 36 g_1^2 - 108 g_2^2 \lambda + 
 144 y_t^2 \lambda + 288 \lambda^2 
 )
\end{eqnarray}  
where we put the renormalization scale to be the vacuum expectation value for the Higgs field $\mu = v = 246$ GeV.
As before we can use the estimation of corrections for different $\Delta_{h}$
\begin{eqnarray} 
\left| \frac{\delta m_{h \, 1loop \, SM}^2 + \delta m_{h \, 2loops \, SM}^2}{m_{h}^2} \right| \leq \Delta_{h}
\label{D_SM_2loop}
\end{eqnarray} 
As a result of this constraint we have a forbidden region in a plane $(\lambda,\Lambda)$ (or $(m_{h},\Lambda)$), which one can see in fig. \ref{FT_SM_2loop}. The plot was obtained in the same way as fig. \ref{FT_SM_1loop}.

\renewcommand{\figurename}{Figure}

\renewcommand{\figurename}{Figure}

\section{ 2-loop fine-tuning in the scalar singlet extension of the Standard Model}

So far we presented the SM extension with $N_{\phi}$ singlet scalar fields and singlet right-handed massive neutrinos. We would like now to show, why this particular SM extension is a useful idea to the particle physics and, as in the previous chapter, discuss classical Higgs mass constraints: triviality and fine-tuning.

The model with one singlet was presented in \cite{grzadkowski}. It is the most economic extension of the SM for which the fine-tuning problem is improved while preserving all the successes of the SM. Other advantages of the model are the presence of the Dark Matter candidate, neutrino masses and mixing or possible lepton asymmetry, however in this work we concentrate only on moderating the quadratic divergences of the Higgs mass. The Lagrangian for the model with a single new scalar field $\phi$ with the gauge invariant coupling to the Higgs doublet and three singlet right-handed Majorana neutrinos reads:
\begin{eqnarray} 
L &=& L_{SM} + \frac{1}{2} \partial_{\mu} \phi  \partial^{\mu} \phi + \frac{1}{2} \mu _{\phi} \phi ^2 + \lambda_{\phi} \phi ^4 + \lambda_{x} (H ^\dagger H) \phi ^ 2 + \nonumber\\
& &  + \overline{\nu_{R}} i \slashed{\partial} \, \nu_{R}  
- \frac{1}{2} \left( \overline{(\nu_{R})^{c}} M \nu_{R} + h.c.\right)
 - \frac{1}{2}  \left( \overline{(\nu_{R})^{c}} Y_{\phi} \nu_{R} \phi  + h.c. \right) \label{lagrangian_scalarandneutrinos}
\end{eqnarray}

Through this renormalizable extension, we would like to generate additional radiative corrections to the Higgs boson mass that can soften the little hierarchy problem. The SM contributions to the quartic divergence are dominated by the top quark. Therefore introducing an extra scalar (different statistics) can suppress the SM result leading to a theory with ameliorated hierarchy problem. We will show that this leads also to constraints for the mass of the Higgs boson.

\subsection{The triviality bound}
As mentioned before in section \ref{Triviality}, for the full triviality constraint, one has to solve the set of equations for all of the parameters in the SM extension. For our purposes, we will approximate the result by considering only the evolution of $\lambda$ and $\lambda_x$.

\begin{eqnarray} 
\mu \frac{d \lambda}{d \mu} &=& 
\frac{3}{8}  g_{1}^4 + \frac{9}{8} g_{2}^4 + \frac{3}{4} g_{1}^2 g_{2}^2 
- 6 y_{t}^4 + 24 \lambda^2 + 12 y_{t}^2 \lambda -3 g_{1}^2 \lambda - 9 g_{2}^2 \lambda + 2\lambda_{x}^2 \nonumber\\ \label{diff_set1}\\
\mu \frac{d \lambda_{x}}{d \mu} &=& 12 \lambda \lambda_{x} + 24 \lambda_{\phi} \lambda_{x} + 8 \lambda_{x}^2 + 6 y_{t}^2 \lambda_{x} - \frac{3}{2} g_{1}^2 \lambda_{x} - \frac{9}{2} g_{2}^2 \lambda_{x} + 2 \textbf{Tr}  ( Y_{\phi} Y_{\phi} ) \lambda_x 
\nonumber\\ \label{diff_set2}
\end{eqnarray}  

The solution for this set of differential equations, with initial conditions 
\begin{eqnarray} 
\lambda(\mu = 80 \, \mathrm{GeV})&=& \lambda_0 \label{initial1}\\
\lambda_{x}(\mu = 80 \, \mathrm{GeV})&=& \lambda _{x \, 0} \label{initial2}
\end{eqnarray}
are functions $\lambda(\mu)$ and $\lambda_{x}(\mu)$ that have a pole for a specific value $\Lambda_{\infty}$ depending on (\ref{initial1}) and (\ref{initial2}). As in the previous chapter, if we want to make sure that the Landau pole is above the SM cut-off, then we receive a constraint on $m_{h}$ and $\Lambda$. The region in $(m_{h},\Lambda)$ plane, forbidden due to this constraint, depends on the initial parameter $\lambda_{x \, 0}$ and the matrix $Y_{\phi}$ in (\ref{lagrangian_scalarandneutrinos}).

We assume $\lambda_{\phi} \sim 0.1$ and therefore $\lambda_{\phi}$ effects do not influence the result much. We will also assume the form of $Y_{\phi}$ matrix as it is in \cite{grzadkowski} (which is a consequence of the $Z_{2}$ symmetry of the singlet scalar field):
\begin{eqnarray} 
Y_{\phi} = \left( \begin{array}{ccc}
0 & 0 & b_1 \\
0 & 0 & b_2 \\
b_1 & b_2 & 0 \\
\end{array} \right)
\end{eqnarray}

We will assume $b_1 = b_2 = b$ and choose $b$ such that the 1-loop corrections to the singlet scalar mass $m_{\phi}$ cancel assuming small $\lambda_{\phi}$ (see \cite{grzadkowski} and \cite{grzadkowski2} for details). From (\ref{general_1loop}) we can determine the correction to the scalar singlet mass
\begin{eqnarray} 
\delta m_{\phi}^2 = \frac{\Lambda^2}{16 \pi^2 } \left( - \frac{\Lambda^2}{2} h_{\phi \phi i i} +
2 \textbf{Tr}( Y_{\phi} Y_{\phi} ) \right) = \frac{1}{16 \pi^2 } \left( - 4 \lambda_{x} - 12\lambda_{\phi} + 8 b^2 \right) \simeq 0
\end{eqnarray} 
which gives us $b \simeq \sqrt{\frac{\lambda_{x}}{2}}$.

\begin{figure}[tp]
  \centering
\includegraphics[height=9 cm]{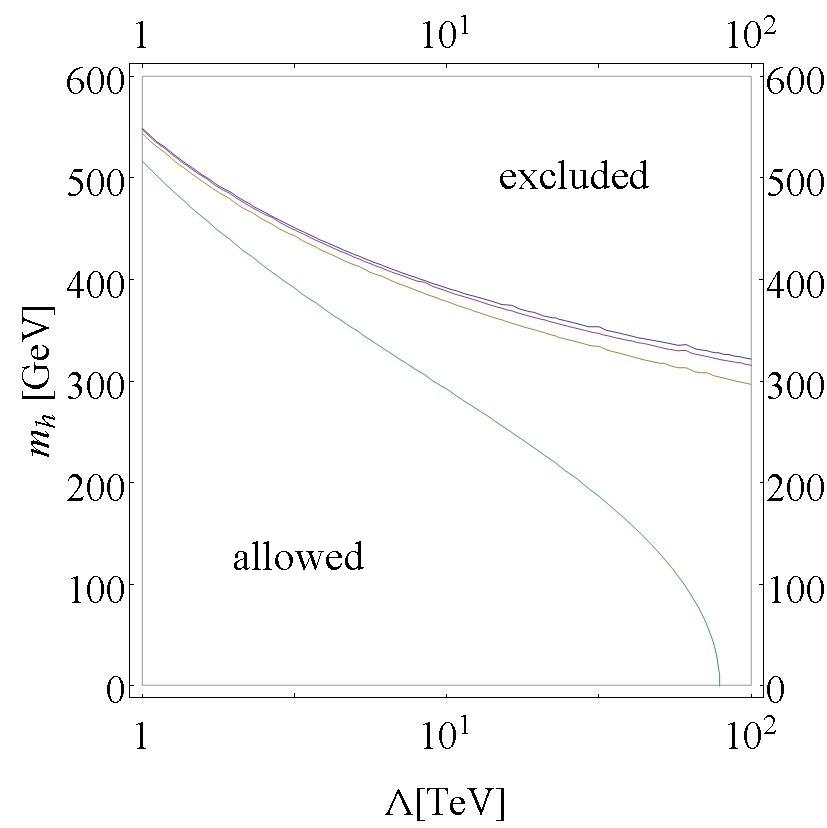}
  \caption{The "Triviality bound" dependence on $\Lambda$ for fixed values $\lambda_{x \, 0} = 0.1, 1, 2, 5$ (starting with the upper most). The region above each curve is forbidden by the triviality constraint for the specific set of parameters.}
  \label{triv_lambdax}
\end{figure}

\begin{figure}[tp]
  \centering
\includegraphics[height=9 cm]{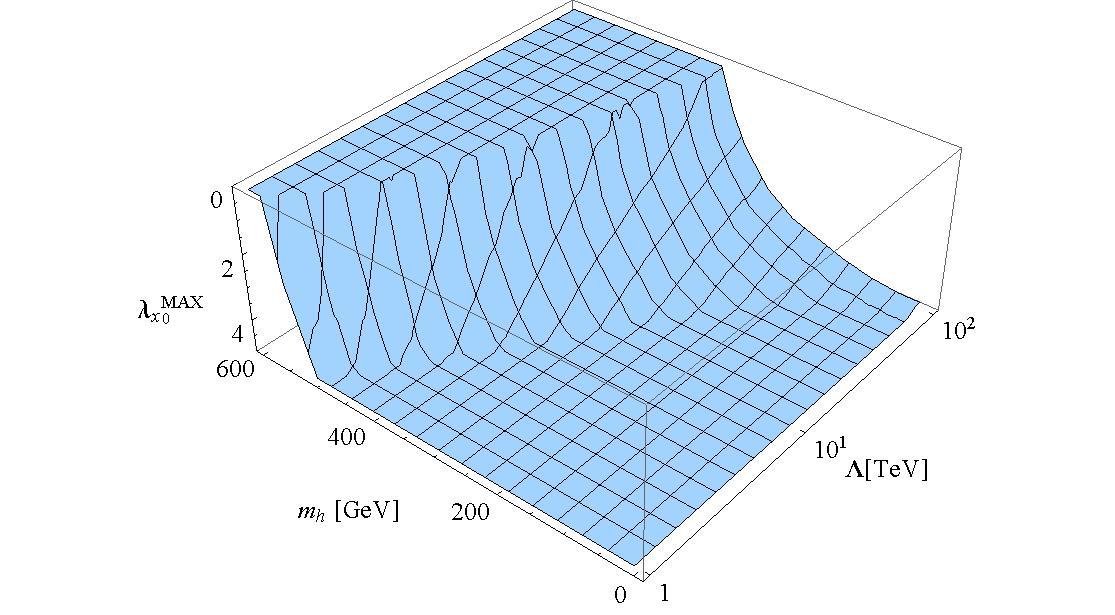}
  \caption{Maximum $\lambda_{x \, 0}$ allowed by the triviality bound, as a function of $m_h$ and $\Lambda$. We assume also $0.1 \leq \lambda_{x \, 0} \leq 5$.}
  \label{lambdaMAX}
\end{figure}

The triviality bound on $m_{h}$ as a function of $\Lambda$, for different values of the initial parameter $\lambda_{x \, 0}$, can be seen in fig. \ref{triv_lambdax}. As one can see, a point $(m_{h},\Lambda)$ that is prohibited for $\lambda_{x \, 0} = 5$ can be allowed if $\lambda_{x \, 0} = 0$. The allowed region shrinks as $\lambda_{x \, 0}$ grows. Therefore, we will take the intersection of the prohibited regions as the triviality bound for $m_{h}$, which corresponds to the $\lambda_{x \, 0} = 0$. 
We should not forget that also the $\lambda_{x}(\mu)$ function has the Landau divergence. Location of the pole depends on the initial values $\lambda_{x \, 0}$ and $\lambda_{0}$. Growing $\lambda_{x \, 0}$ implies a shift of the pole position towards smaller energies.
For every initial condition $\lambda_{0}$ we should specify a certain range of $\lambda_{x \, 0}$ that the Landau pole of $\lambda_{x}(\mu)$ is above a given value of $\Lambda$. Therefore, not every value of $\lambda_{x \, 0}$ parameter is allowed for each Higgs mass and the cut-off $\Lambda$. The maximum $\lambda_{x \, 0}$ one can see in the figure \ref{lambdaMAX}.


The results in figures \ref{triv_lambdax} and \ref{lambdaMAX}, were both obtained through the same numerical procedure in \textit{Mathematica} as introduced before in section \ref{Triviality}.

\newpage 
\subsection{The fine-tuning}

To discuss the fine-tuning in the SM extension with a singlet scalar field and right singlet neutrinos, we need the full result for 1-loop and 2-loops corrections to the Higgs mass:
\begin{eqnarray} 
\delta m^2_{h \,1loop} &=& \frac{\Lambda ^2}{16 \pi^2} \left(
12 \lambda + 2 \lambda_x - 12 y_t^2 + \frac{3}{2} g_1^2 + \frac{9}{2} g_2^2 \right) \nonumber\\
& & - \frac{1}{16 \pi^2} \left(
6 \lambda m_{h}^2 \log  \left( \frac{m_{h}^2+\Lambda^2}{m_{h}^2}  \right) +
2 \lambda_x m_{\phi}^2 \log  \left( \frac{m_{\phi}^2+\Lambda^2}{m_{\phi}^2}  \right)
\right)\\
\delta m^2_{h \,2loops} &=& \frac{\Lambda ^2}{(16 \pi ^2)^2} \log  \left( \frac{\Lambda}{\mu} \right)
\left( 25 g_1^4 + 9 g_1^2 g_2^2 - 15 g_2^4 + 34 g_1^2 y_t^2 + 54 g_2^2 y_t^2 \right.
 \nonumber\\
& &
 +  192 g_3^2 y_t^2 - 180 y_t^4  \lambda - 36 g_1^2 - 108 g_2^2 \lambda 
+  144 y_t^2 \lambda + 288 \lambda^2
 \nonumber\\
& & 
- 3 g_1^2 \lambda_x -  9 g_2^2 \lambda_x + 12 y_t^2 \lambda_x + 24 \lambda \lambda_x +  40 \lambda_x^2 + 48 \lambda_x \lambda_{\phi} 
 \nonumber\\
& &
\left. + 4 \lambda_x \textbf{Tr}\left(  Y_{\phi}  Y_{\phi}  \right) \right.)
\end{eqnarray} 
where the logarithmic terms in the 1-loop correction are kept as relevant because of the high value of $m_{\phi}$ parameter.

As before, the corrections should be relatively small in comparison with the Higgs mass, so we again introduce the fine-tuning parameter $\Delta_{h}$

\begin{eqnarray} 
\left| \frac{\delta m_{h \, 1loop}^2 + \delta m_{h \, 2loops}^2}{m_{h}^2} \right| \leq \Delta_{h}
\label{D_sinlet_2loop}
\end{eqnarray} 

We would like to repeat the assumptions from the previous section: $\lambda_{\phi} \sim 0.1$, $Y_{\phi}$ should roughly cancel the 1-loop correction to the scalar singlet mass $m_{\phi}$. Higgs coupling to the singlet scalar has to satisfy the following condition for every  $m_{h}$ and $\Lambda$
\begin{eqnarray} 
0.1 \leq \lambda_{x \, 0} \leq \lambda_{x \, 0}^{MAX} (m_{h},\Lambda) \leq 5 \label{lambda_condition}
\end{eqnarray} 
where $\lambda_{x \, 0}^{MAX}(m_{h},\Lambda)$ is the triviality constraint (see fig. \ref{lambdaMAX}). We would like the singlet scalar mass $m_{\phi}$ to be in a range 500 - 5000 GeV and, in order to satisfy $<\phi>=0$, it must also fulfil the inequality 
\begin{eqnarray} 
m_{\phi}^2 - \lambda_{x \, 0} v^2 > 0 \label{mphi_condition}
\end{eqnarray}
where $v = 246$ GeV is the Higgs field vacuum expectation value (see \cite{grzadkowski} for details). With all these assumptions we can now consider allowed values of $m_{h}$ and $\Lambda$ for different $\Delta_{h}$. 

\begin{figure}[tp]
\centering
\includegraphics[height=0.3 cm]{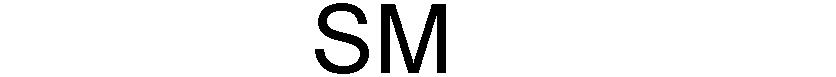}  \, \, \, \, \, \, \, \,  \, \, \, \, \, \, \ \includegraphics[height=0.3 cm]{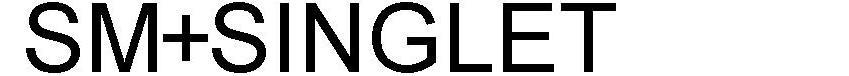} \\
\includegraphics[height=4cm]{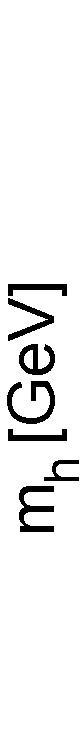}
\includegraphics[height=6cm]{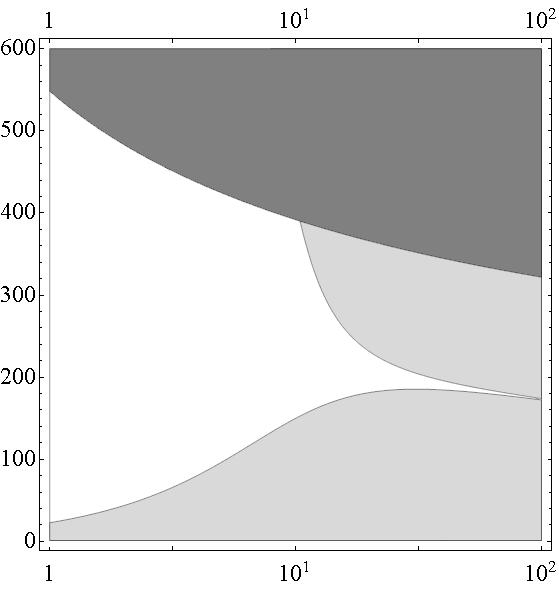}
\includegraphics[height=6cm]{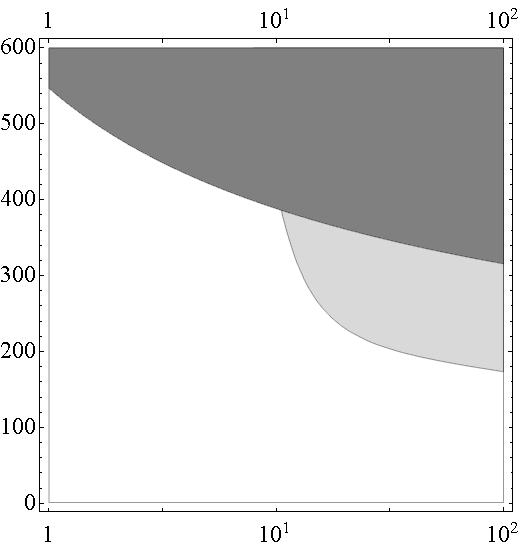} 
\includegraphics[height=4cm]{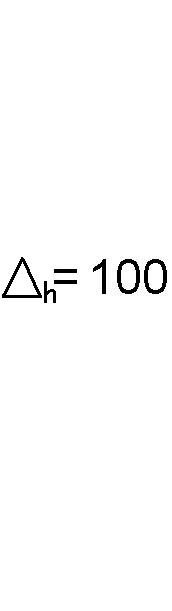}\\
\includegraphics[height=4cm]{Graph/m_h_GeV.jpg}
\includegraphics[height=6cm]{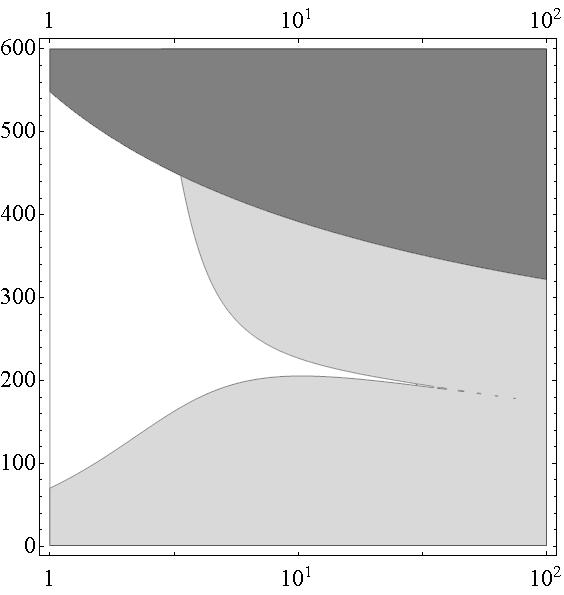}
\includegraphics[height=6cm]{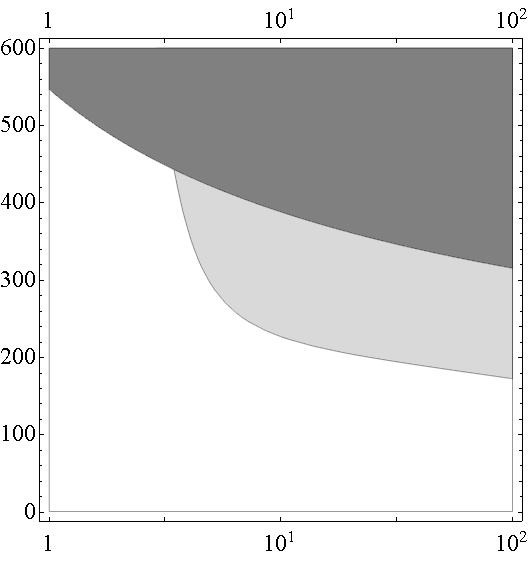} 
\includegraphics[height=4cm]{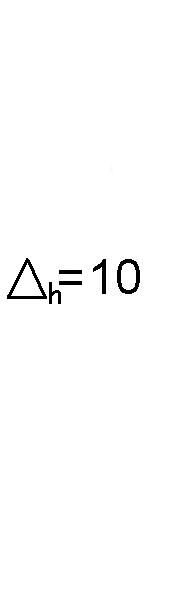}\\
\includegraphics[height=0.4 cm]{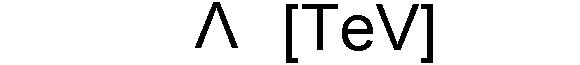}  \, \, \, \, \, \, \, \, \, \includegraphics[height=0.4 cm]{Graph/Lambda_TeV.jpg}
  \caption{Allowed regions (white) for $m_h$ and $\Lambda$ resulting from the fine-tuning in the SM and the SM singlet extension for $\Delta_{h} = 100$ and $\Delta_{h} = 10$. Dark grey region is excluded by the triviality argument for any value of $\lambda_{x \, 0}$ in the range $0.1 \leq \lambda_{x \, 0} \leq 5$.}
  \label{wszystkieNaRaz1}
\end{figure}

\begin{figure}[tp]
\centering
\includegraphics[height=0.3 cm]{Graph/SM_napis.jpg}  \, \, \, \, \, \, \, \,  \, \, \, \, \, \, \ \includegraphics[height=0.3 cm]{Graph/SMS_napis.jpg} \\
\includegraphics[height=4cm]{Graph/m_h_GeV.jpg}
\includegraphics[height=6cm]{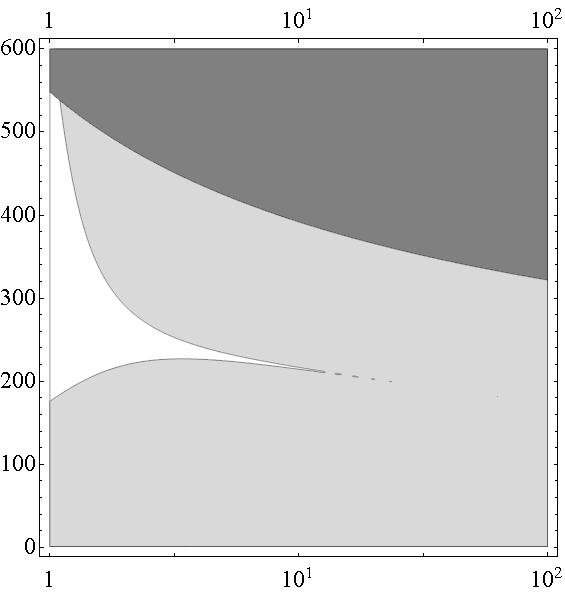}
\includegraphics[height=6cm]{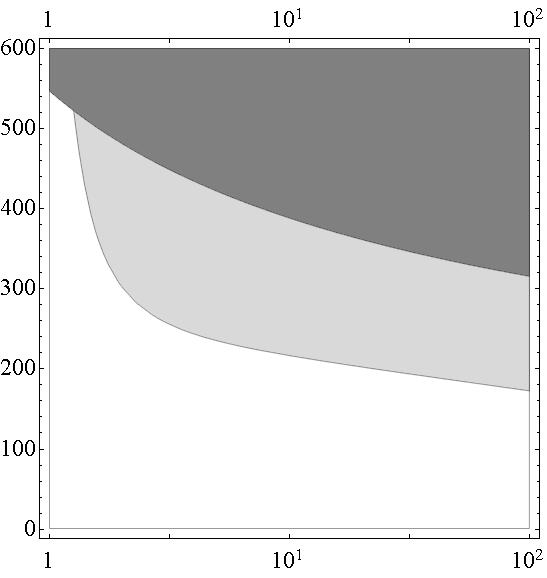} 
\includegraphics[height=4cm]{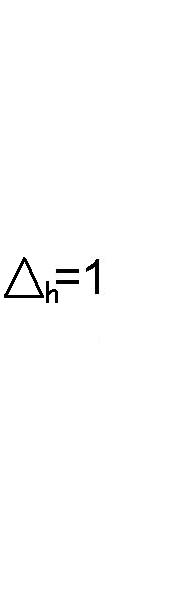}\\
\includegraphics[height=4cm]{Graph/m_h_GeV.jpg}
\includegraphics[height=6cm]{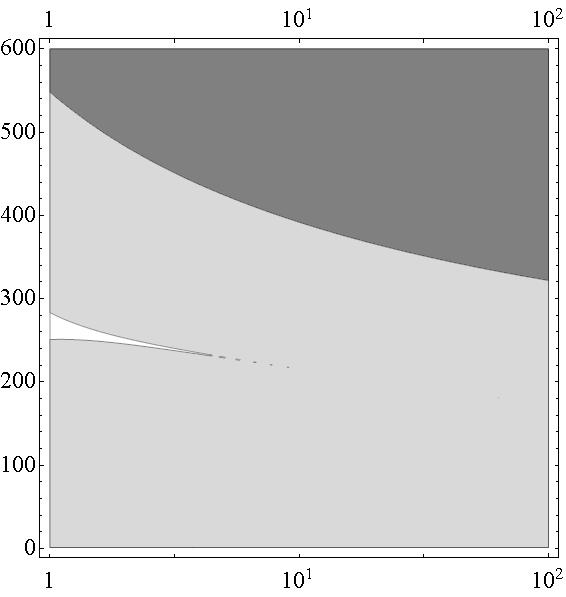}
\includegraphics[height=6cm]{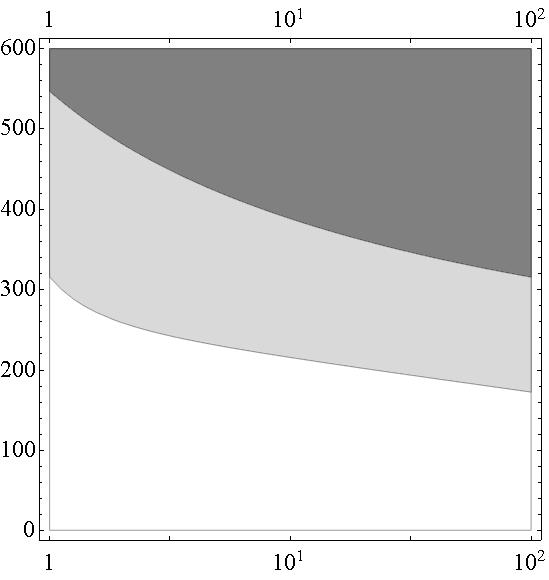} 
\includegraphics[height=4cm]{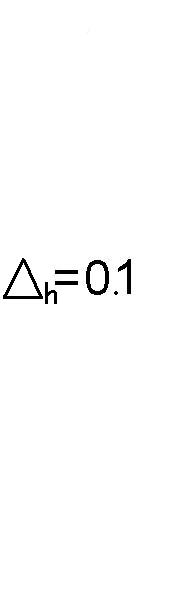}\\
\includegraphics[height=0.4 cm]{Graph/Lambda_TeV.jpg}  \, \, \, \, \, \, \, \, \, \includegraphics[height=0.4 cm]{Graph/Lambda_TeV.jpg}
  \caption{Allowed regions (white) for $m_h$ and $\Lambda$ resulting from the fine-tuning in the SM and the SM singlet extension for $\Delta_{h} = 1$ and $\Delta_{h} = 0.1$. Dark grey region is excluded by the triviality argument for any value of $\lambda_{x \, 0}$ in the range $0.1 \leq \lambda_{x \, 0} \leq 5$.}
  \label{wszystkieNaRaz2}
\end{figure}

For each point in the allowed by triviality part of the $(m_{h},\Lambda)$ plane we have a set of parameters $\lambda_{x \, 0}$ and $m_{\phi}$ such that they satisfy all of the just mentioned conditions. If there is no such a set of $\lambda_{x \, 0}$ and $m_{\phi}$ that the fine-tuning inequality (\ref{D_sinlet_2loop}) is fulfilled for a specified value of $\Delta_{h}$, then the point $(m_{h},\Lambda)$ belongs to the forbidden by $\Delta_{h}$ fine-tuning region. We can solve these numerically using \textit{Mathematica}. A simplified program that minimizes the LHS of (\ref{D_sinlet_2loop}) in terms of allowed $\lambda_{x \, 0}$ and $m_{\phi}$ and compares it with $\Delta_{h}$,
obtaining plots such as in figs. \ref{wszystkieNaRaz1} and \ref{wszystkieNaRaz2}, is the following:
\begin{eqnarray} 
& &\texttt{Figure[}\Delta_{h}\texttt{] := RegionPlot[First[NMinimize[}\lbrace        \nonumber\\
& & \,\,\,\,\,\,\,\,\,\,\,\,\,\,\, \texttt{FineTuning[}m_{H}, \lambda_{x \, 0}, b, \Lambda, m_{\phi}\texttt{]},       \nonumber\\
& & \,\,\,\,\,\,\,\,\,\,\,\,\,\,\, 0.1 \leq  \lambda_{x \, 0} < \texttt{LambdaXMAX[}m_{h}, \Lambda\texttt{]},          \nonumber\\
& & \,\,\,\,\,\,\,\,\,\,\,\,\,\,\, 500 < m_{\phi} < 5000,           \nonumber\\
& & \,\,\,\,\,\,\,\,\,\,\,\,\,\,\, b == \sqrt{(  \lambda_{x \, 0} / 2)},      \nonumber\\
& & \,\,\,\,\,\,\,\,\,\,\,\,\,\,\, m_{\phi}^2 -  \lambda_{x \, 0} v^2 > 0 \rbrace,        \nonumber\\
& & \,\,\,\,\,\,\,\,\,\,\,\,\,\,\, \lbrace   \lambda_{x \, 0}, m_{\phi} \rbrace \texttt{]]} > \Delta_{h} , \lbrace \Lambda, 1000, 100000 \rbrace, \lbrace m_{h}, 1, 600 \rbrace\ \texttt{]}
\end{eqnarray}
where $\texttt{FineTuning[}m_{H},  \lambda_{x \, 0}, b, \Lambda, m_{\phi}\texttt{]}$ is the LHS from (\ref{D_sinlet_2loop}), $\texttt{LambdaXMAX[}m_{h}, \Lambda\texttt{]}$ is the function from (\ref{lambda_condition}), the $m_{\phi}$ range 500 to 5000 is in GeV, such as the ranges $\lbrace \Lambda, 1000, 100000 \rbrace$ and $\lbrace m_{h}, 1, 600 \rbrace $.

In the right panel of figs. \ref{wszystkieNaRaz1} and \ref{wszystkieNaRaz2} allowed regions of $m_{h}$ and $\Lambda$ are shown in the singlet extended model in comparison with the SM results (left panel). What we can observe, is that the part for low Higgs mass which is forbidden in the SM fine-tuning plots, is allowed in the singlet scalar extension. 
This happens because, for low Higgs mass the leading contribution to the mass correction comes from the Yukawa top quark coupling. In the extended model it cancels with the contributions from the singlet scalar, as they come with opposite sings (different statistics). For large Higgs masses, the mass correction coming from the Higgs quartic coupling dominates over the correction from the top quark. As all the scalar contributions are of the same sign, they can't cancel each other. Increasing the additional couplings coming from the presence of the singlet scalar only worsen the fine-tuning condition. That is also why the upper bound difference between models is negligible - for the large Higgs masses we have $\lambda_x \simeq 0$.

\section{Summary and conclusions}  
   
There are two main results of this work.
   
First result are the derived 1-loop equations for beta functions in general gauge theory with scalars and fermions and a single gauge symmetry and the 1- and 2-loop quadratic corrections to scalar masses, including contributions from Dirac and Majorana fermions.
   
In the second part of the work we studied the theoretical constraints on the Higgs mass and new physics scale coming from triviality and fine-tuning. In the SM, the fine-tuning condition gives a significant constraint on the Higgs boson mass and on the scale of new physics beyond the SM. \textbf{However, the one scalar singlet SM extension opens a window for the low Higgs masses without significant constraint on the new physics scale.}

There are still more questions to be answered about the singlet scalar Standard Model extension. Is the new particle a good Dark Matter candidate? Can it explain the leptogenesis? What about multi-singlet SM extensions?

\newpage
\appendix
\section{Feynman rules for general gauge theory}

The Feynman rules for the propagators for the general gauge theory with scalar, gauge boson, ghost and fermion fields, with no mass for the scalar and gauge fields (for the full Lagrangian see \ref{lagrangian_general})\\
\noindent
\begin{tabularx}{\textwidth}{XX}
\includegraphics[height = 0.7 cm]{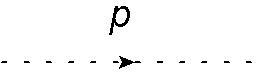} &
$\tilde{D}_{F}(p) = \frac{i}{p^2 - m^2 + i \epsilon }$ \\
\includegraphics[height = 0.7 cm]{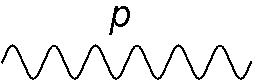} &
$\tilde{D}_{F}^{\mu \nu}(p) = \frac{i (- g^{\mu \nu} + (1- \xi) \frac{p^{\mu} p^{\nu}}{p^2} )}{p^2 - m^2 + i \epsilon }$ \\
\includegraphics[height = 0.7 cm]{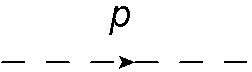} &
$\tilde{G}_{F}(p) = \frac{i }{p^2 + i \epsilon }$ \\
\includegraphics[height = 0.7 cm]{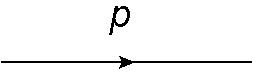} &
$\tilde{S}_{F}(p) = \frac{i (\slashed{p} + m)}{p^2 - m^2 + i \epsilon }$ \\
\end{tabularx}

Wave-function renormalization counterterms contribution to propagators:

\begin{tabularx}{\textwidth}{XX}
\includegraphics[height = 1.3 cm]{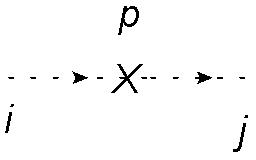} &
$ i \left( \Delta Z_{\phi} \right)_{i j} p^2 $ \\
\includegraphics[height = 1.3 cm]{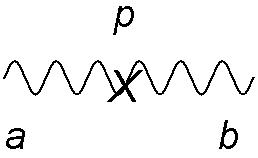} &
$ i  \Delta Z_{A} \left(- p^2 g_{\mu \nu} + p_{\mu} p_{\nu} \right)\delta_{a b} - \frac{i}{\xi} K_{\xi} p_{\mu} p_{\nu} \delta_{a b} $ \\
\includegraphics[height = 1.3 cm]{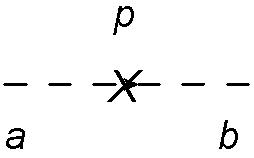} &
$ i \Delta Z_{\eta} p^2 \delta_{a b} $\\
\includegraphics[height = 1.3 cm]{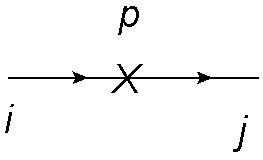} &
$ i \frac{1}{2} \left( \Delta Z_{\psi} ^{\dagger} + \Delta Z_{\psi} \right)_{n m} \slashed{p} $ \label{fermion_propagator_counter}\\
\end{tabularx}

The Feynman rules for the vertices:\\
\noindent
\begin{tabularx}{\textwidth}{XX}
\includegraphics[height = 2.5 cm]{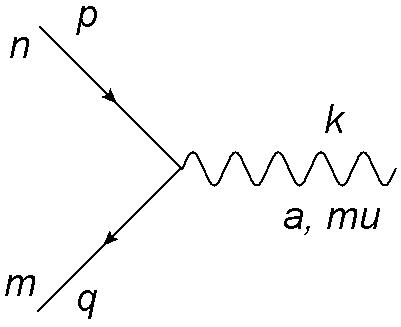} &
$-ig \overline{\textbf{T}}^a_{mn} \gamma_{\mu}$ \\
\includegraphics[height = 2.5 cm]{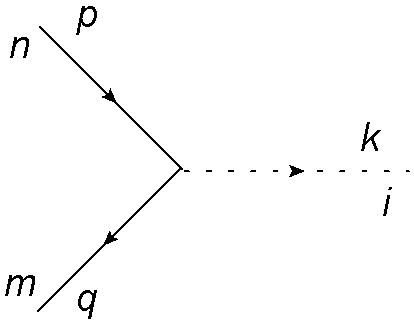} &
$-i \kappa^{i}_{mn}$ \\
\end{tabularx}\\

The following diagram is symmetric under interchanges $i, j$, which must be included in the vertex coupling. Considering the fact that $T^{a}_{i,j}$ is hermitian and imaginary, the vertex coupling simplifies to:

\begin{tabularx}{\textwidth}{XX}
\includegraphics[height = 2.5 cm]{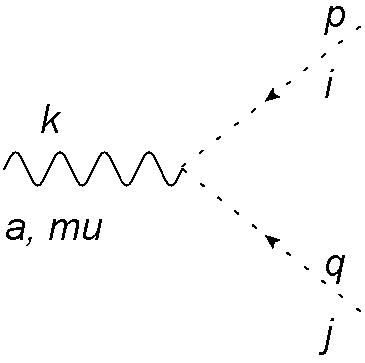} &
$ ig (p^{\mu} - q^{\mu}) \textbf{T}^a _{ij} $ \\
\end{tabularx}\\

Fol term is symmetric under interchanges $(a, \mu),(b, \nu)$ and $i, j$. To have an expression which treats all of the interacting in the vertices fields the same, we need to include all the interchanges.

\begin{tabularx}{\textwidth}{XX}
\includegraphics[height = 2.5 cm]{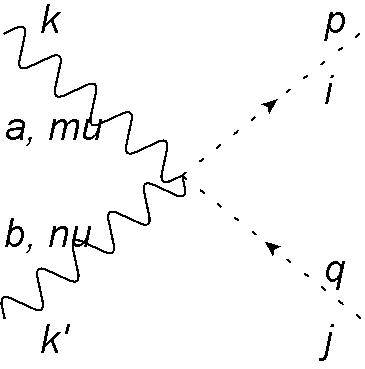} &
$ -i g^2  g^{\mu \nu} (\textbf{T}^a _{kj} \textbf{T}^b _{ki} + \textbf{T}^a _{ki} \textbf{T}^b _{kj}) $ \\
\end{tabularx}\\

The quadrilinear term is symmetric under interchanges $(a, \mu),(b, \nu),(c, \rho),(d, \sigma)$. To have an expression which treats all of the interacting in the vertices gauge fields the same, we need to include all the interchanges.

\begin{tabularx}{\textwidth}{XX}
\includegraphics[height = 2.5 cm]{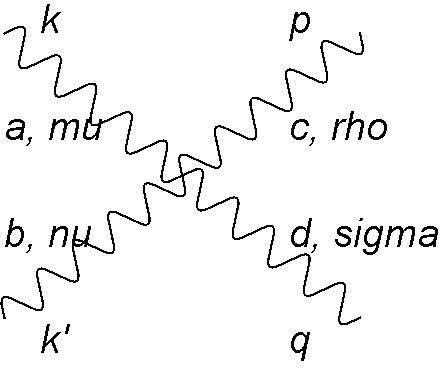} &
\begin{eqnarray} 
-i g^2 ( f_{eab} f_{ecd} (g^{\mu \rho} g^{\nu \sigma} - g^{\mu \rho} g^{\nu \sigma}) \nonumber\\
 f_{eac} f_{ebd} (g^{\mu \nu} g^{\rho \sigma} - g^{\mu \sigma} g^{\nu \rho}) \nonumber\\
 f_{ead} f_{ebc} (g^{\mu \nu} g^{\rho \sigma} - g^{\mu \rho} g^{\nu \sigma}) ) \nonumber
\end{eqnarray} \\

\end{tabularx}

The trilinear term is totally antisymmetric under interchanges $(k, \mu),(q, \nu),(p, \rho)$. To have an expression which treats all of the interacting in the vertices gauge fields the same, we need to include all the interchanges.

\begin{tabularx}{\textwidth}{XX}
\includegraphics[height = 2.5 cm]{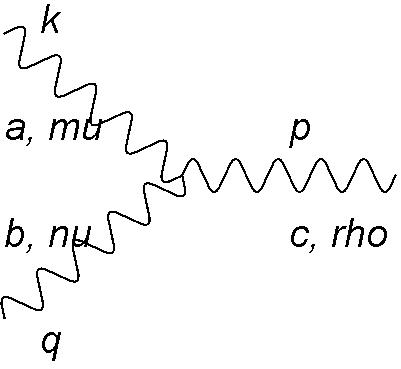} &
$ g f_{abc} (g^{\mu \nu}(k-q)^{\rho}+g^{\nu \rho}(q-p)^{\mu}+g^{\rho \mu}(p-k)^{\nu})  $ \\
\end{tabularx}

\begin{tabularx}{\textwidth}{XX}
\includegraphics[height = 2.5 cm]{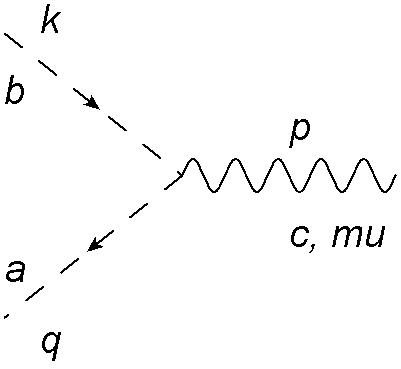} &
$ - g f_{abc} q^{\mu}$ \\
\end{tabularx}
    
\begin{tabularx}{\textwidth}{XX}
\includegraphics[height = 2.5 cm]{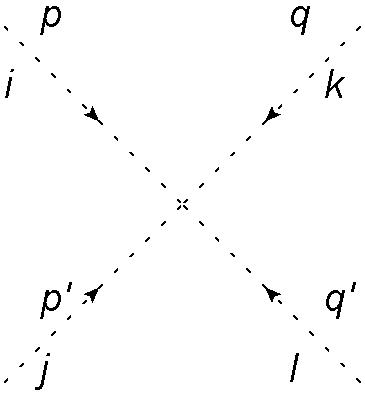} &
$ -i h _{ijkl}$ \\
\end{tabularx}
  
Feynman rules for the counterterms relevant in the work:  
  
\begin{tabularx}{\textwidth}{XX}
\includegraphics[height = 2.3 cm]{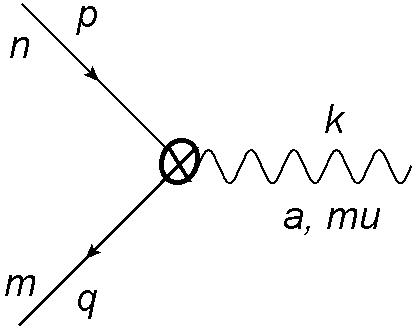} &
$ - i \gamma ^\mu \left( \Delta g \overline{\textbf{T}}^a \right)_{m n} $ \\
\end{tabularx}

\begin{tabularx}{\textwidth}{XX}
\includegraphics[height = 2.3 cm]{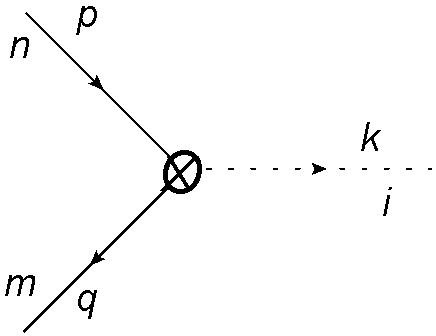} &
$ - i \Delta \kappa^{i}_{m n}$ \\
\end{tabularx}
    
\begin{tabularx}{\textwidth}{XX}
\includegraphics[height = 2.5 cm]{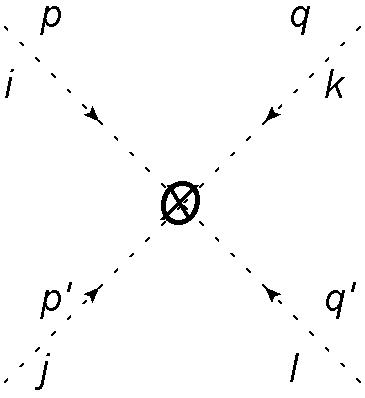} &
$ -i \Delta h _{ijkl}$ \\
\end{tabularx}  
  
\newpage
\section{Table of Integrals}  
Integrals in the dimensional regularization\\
\begin{eqnarray}
& &
\mu^{\epsilon} \int \frac{d^{d} q}{(2\pi)^{d}} \frac{1}{q^2 (p+q)^2} = \frac{2 i}{16 \pi^2 \epsilon}
\\ & &
\mu^{\epsilon} \int \frac{d^{d} q}{(2\pi)^{d}} \frac{q^{\mu}}{q^2 (p+q)^2} = - \frac{ i p^{\mu}}{16 \pi^2 \epsilon} 
\\ & &
\mu^{\epsilon} \int \frac{d^{d} q}{(2\pi)^{d}} \frac{q^{\mu} q^{\nu}}{q^2 (p+q)^2} = \frac{i}{16 \pi^2 \epsilon}
\left(
\frac{2}{3} p^{\mu} p^{\nu} - \frac{1}{6} g^{\mu \nu} p^2
\right) 
\\ & &
\mu^{\epsilon} \int \frac{d^{d} q}{(2\pi)^{d}} \frac{q^{\mu} q^{\nu}}{q^2 q^2 (p+q)^2} = \frac{i g^{\mu \nu}}{32 \pi^2 \epsilon}
\\ & &
\mu^{\epsilon} \int \frac{d^{d} q}{(2\pi)^{d}} \frac{q^{\mu} q^{\nu} q^{\alpha} q^{\beta}}{q^2 q^2 q^2 (p+q)^2} = \frac{1}{12} \frac{i}{16 \pi^2 \epsilon} (g^{\mu \nu} g^{\alpha \beta} + g^{\mu \alpha} g^{\nu \beta} + g^{\mu \beta} g^{\nu \alpha})
\end{eqnarray}

Integrals in the cut-off regularization (following \cite{inami}, \cite{varin} )\\
\begin{eqnarray}   
& &
\int \frac{d^{4} q}{(2\pi)^{4}} \frac{i}{q^2 - m^2} = \frac{1}{16 \pi^2} 
\left( \Lambda ^2 - m^2 \log  \left( \frac{m^2 + \Lambda^2}{m^2}  \right) \right) 
\\ & & 
\int \frac{d^{4} q}{(2\pi)^{4}} \frac{i}{(q^2 - m_{a}^2) (q^2 - m_{b}^2)} = \frac{1}{16 \pi^2} 
\log  \left( \frac{\Lambda ^2}{m_{a} m_{b}} \right)+ \ldots
\\ & & 
\int \int \frac{d^{4} q}{(2\pi)^{4}} \frac{d^{4} k}{(2\pi)^{4}} 
\frac{1}{(q^2 - m_{a}^2) ((q+k)^2 - m_{b}^2) (k^2 - m_{c}^2) } 
= - \frac{1}{(16 \pi^2)^2} 2 \Lambda ^2 + \ldots
\end{eqnarray}

\newpage
\section{Feynman Rules for Majorana Fermions} 

Feynman rules for Majorana neutrinos can be found for example in \cite{denner} or \cite{gluza}.

In this appendix $\psi$ denotes a Majorana fermion field and $\varphi$ a scalar field. We are interested in the following Lagrangian:
\begin{eqnarray}   
L = \overline{\psi} i \slashed{\partial} \psi  - \frac{1}{2} M \left( \overline{\psi^{c}} \psi + \overline{\psi}  \psi^{C} \right)
 - \frac{1}{2} \left( \overline{\psi^c} Y_{\varphi} \psi \varphi + \overline{\psi} Y_{\varphi} \psi^c \varphi\right)
\end{eqnarray}  
where $(\,)^{c}$ denotes the charge conjugation operator, $\overline{\psi^{c}} = \psi^{T} \hat{C}$, $\hat{C}$ is an antisymmetric charge conjugation matrix.

We define $a^{\dagger}$ and $b^{\dagger}$ as the creation operator of fermion and antifermion, respectively. Similarly $a$ and $b$ are the annihilation operators. $d^{\dagger}$ and $d$ are creation and annihilation operators of the scalar particle $\varphi$. $|(k,\lambda)\rangle$ is a state of a single Majorana fermion of momentum $k$ and helicity $\lambda$. $|k\rangle$ denotes a one scalar particle state of momentum $k$. 
\begin{eqnarray}   
& & |(k,\lambda)\rangle = a^{\dagger}_{k,\lambda} |0\rangle \\
& & |k \rangle = d^{\dagger}_{k} |0\rangle
\end{eqnarray}  
where $|0\rangle$ is the vacuum state.
\renewcommand{\figurename}{Diagram}
\begin{figure}
  \centering
  \includegraphics[height = 2 cm]{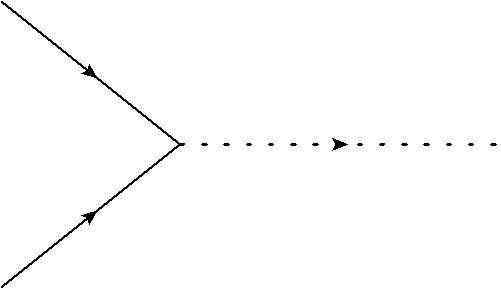}
  \caption{Yukawa interaction for Majorana fermions vertex}
  \label{majorana_yukawa}
\end{figure}

We would like to determine the Feynman rule for a Yukawa interaction vertex with two Majorana fermions, as in diagram \ref{majorana_yukawa}. Below $\textbf{T}$ denotes the time-order operator.
\begin{eqnarray}
\texttt{diagram \ref{majorana_yukawa}} &=& \langle 0 \vert a_{k_1, \lambda_1} a_{k_2, \lambda_2}
\textbf{T} \left[
\int d^4 x \left( - i \frac{1}{2} \varphi (x) \psi (x) ^T \hat{C} Y_{\varphi} \psi (x) \right) \right]
 b^{\dagger}_{k_3} \vert 0 \rangle = \nonumber\\ 
&=& 
- i v^T_{k_1, \lambda_1} \,\frac{1}{2}(\hat{C} Y_{\varphi} - Y_{\varphi} \hat{C} ^T ) \, v _{k_2, \lambda_2} 
\, \delta^4 (k_1 + k_2 - k_3)
\nonumber\\
&=& 
- i v^T_{k_1, \lambda_1} \, \hat{C} Y_{\varphi} \, v _{k_2, \lambda_2} \,
\delta^4 (k_1 + k_2 - k_3)
\end{eqnarray} 

Therefore, the Feynman rule for a Yukawa interaction for Majorana fermion vertex with fermion lines as in diagram \ref{majorana_yukawa} is simply $(-i \hat{C} Y_{\varphi})$.

The Feynman rule for Majorana fermion propagator can be obtained for example from \cite{gluza}: \\

\begin{tabularx}{\textwidth}{XX}
 \includegraphics[height = 0.2 cm]{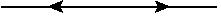}  &
$ \langle 0 \vert T \left[ \psi ^T (x) \psi (y) \right] \vert 0 \rangle = - i \left( S(x - y) \hat{C} \right) $ \\
\end{tabularx} \\

\end{document}